%% file: main.tex
\documentclass[conference]{IEEEtran}

\usepackage{cite}
\usepackage{amsmath,amssymb,amsfonts,amsthm}
\usepackage{algorithmic}
\usepackage{graphicx}
\usepackage{textcomp}
\def\BibTeX{{\rm B\kern-.05em{\sc i\kern-.025em b}\kern-.08em
    T\kern-.1667em\lower.7ex\hbox{E}\kern-.125emX}}

\usepackage[hidelinks]{hyperref}

\usepackage{xspace,balance,tabularx,multirow}
\usepackage{flushend}
\usepackage{tikz}
\usepackage{pgfplots}
\pgfplotsset{compat=1.16}
\usetikzlibrary{patterns}
\usepackage{subfig}
\usepackage[ruled, vlined, linesnumbered]{algorithm2e}
\usepackage{colortbl}
\usepackage{bbold}
\SetKwComment{Comment}{$\triangleright$\ }{}
\usepackage{enumitem}
\usepackage{tablefootnote}
\usepackage{upgreek,textgreek}
\usepackage{pifont}%
\usepackage[noabbrev]{cleveref}
\usepackage{titlecaps}
\usepackage{lipsum}

\theoremstyle{plain}
\newtheorem{theorem}{Theorem}[section]

\newtheorem{lemma}[theorem]{Lemma}

\theoremstyle{definition}
\newtheorem{definition}[theorem]{Definition}

\theoremstyle{remark}

\pgfplotsset{every tick label/.append style={font=\tiny}}

\newlength{\starsize}
\newlength{\starspread}
\tikzset{starsize/.code={\setlength{\starsize}{#1}},
         starspread/.code={\setlength{\starspread}{#1}}}
\tikzset{starsize=1mm,
         starspread=3mm}
\pgfdeclarepatternformonly[\starspread,\starsize]%
  {my fivepointed stars}%
  {\pgfpointorigin}%
  {\pgfqpoint{\starspread}{\starspread}}%
  {\pgfqpoint{\starspread}{\starspread}}%
  {%
   \pgftransformshift{\pgfqpoint{\starsize}{\starsize}}
   \pgfpathmoveto{\pgfqpointpolar{18}{\starsize}}
   \pgfpathlineto{\pgfqpointpolar{162}{\starsize}}
   \pgfpathlineto{\pgfqpointpolar{306}{\starsize}}
   \pgfpathlineto{\pgfqpointpolar{90}{\starsize}}
   \pgfpathlineto{\pgfqpointpolar{234}{\starsize}}
   \pgfpathclose%
   \pgfusepath{fill}
  }

\input{tex/macro}
\begin{document}

\title{Adaptive Local Clustering over Attributed Graphs\\ {\large Technical Report}
}

\author{\IEEEauthorblockN{Haoran Zheng}
\IEEEauthorblockA{\textit{Hong Kong Baptist University}\\
Hong Kong SAR, China \\
cshrzheng@comp.hkbu.edu.hk}
\and
\IEEEauthorblockN{Renchi Yang}
\IEEEauthorblockA{\textit{Hong Kong Baptist University}\\
Hong Kong SAR, China \\
renchi@hkbu.edu.hk}
\and
\IEEEauthorblockN{Jianliang Xu}
\IEEEauthorblockA{\textit{Hong Kong Baptist University}\\
Hong Kong SAR, China \\
xujl@hkbu.edu.hk}

}

\maketitle

\thispagestyle{plain}
\pagestyle{plain}

\begin{abstract}
Given a graph $\G$ and a seed node $v_s$, the objective of local graph clustering (LGC) is to identify a subgraph $\C_s\in \G$ (a.k.a. local cluster) surrounding $v_s$ in time roughly linear with the size of $\C_s$. This approach yields personalized clusters without needing to access the entire graph, which makes it highly suitable for numerous applications involving large graphs.
However, most existing solutions merely rely on the topological connectivity between nodes in $\G$, rendering them vulnerable to missing or noisy links that are commonly present in real-world graphs.

To address this issue, this paper resorts to leveraging the complementary nature of graph topology and node attributes to enhance local clustering quality.
To effectively exploit the attribute information, we first formulate the LGC as an estimation of the {\em bidirectional diffusion distribution} (BDD), which is specialized for capturing the multi-hop affinity between nodes in the presence of attributes.
Furthermore, we propose \algo, an efficient and effective approach for LGC that achieves superb empirical performance on multiple real datasets while maintaining strong locality.
The core components of \algo include (i) a fast and theoretically-grounded preprocessing technique for node attributes, (ii) an adaptive algorithm for diffusing any vectors over $\G$ with rigorous theoretical guarantees and expedited convergence, %
and (iii) an effective three-step scheme for BDD approximation.
Extensive experiments, comparing 17 competitors on 8 real datasets, show that \algo outperforms all competitors in terms of result quality measured against ground truth local clusters, while also being up to orders of magnitude faster.
\end{abstract}


\input{tex/introducation}

\input{tex/preliminary}
\input{tex/overview}

\input{tex/solution_secondary}

\input{tex/solution}

\input{tex/experiments}

\input{tex/relatedwork}

\section{Conclusion}
In this paper, we present \algo, an effective approach that leverages node attributes to improve the LGC quality on attributed graphs. \algo achieves remarkable performance through four major contributions: (i) a novel problem formulation based on the novel node affinity measure BDD, (ii) an adaptive RWR-based graph diffusion algorithm with faster convergence, (iii) a highly scalable preprocessing technique that enables problem reduction, and (iv) a well-thought-out three-step scheme for BDD approximation.
The superiority of \algo over 17 baselines is experimentally validated over 8 real datasets in terms of both clustering quality and practical efficiency. Regarding future work, we plan to study the local clustering on heterophilic graphs.

\balance

\bibliographystyle{IEEEtran}
\bibliography{sample-base}

\clearpage
\appendix
\input{tex/proof}

\input{tex/additional-experiments}
\input{tex/alternative-implementation}

\end{document}

%% file: tex/macro.tex
\makeatletter
\newcommand*\bigcdot{\mathpalette\bigcdot@{.5}}
\newcommand*\bigcdot@[2]{\mathbin{\vcenter{\hbox{\scalebox{#2}{$\m@th#1\bullet$}}}}}
\makeatother

\newcommand{\stitle}[1]{\vspace*{0.5em}\noindent{\bf #1.\/}}

\newcommand{\V}{\mathcal{V}\xspace}
\newcommand{\G}{\mathcal{G}\xspace}
\newcommand{\N}{\mathcal{N}\xspace}
\newcommand{\EDG}{\mathcal{E}\xspace}
\newcommand{\C}{\mathcal{C}\xspace}
\newcommand{\Y}{\mathcal{Y}\xspace}

\newcommand{\AM}{\boldsymbol{A}\xspace}
\newcommand{\DM}{\boldsymbol{D}\xspace}
\newcommand{\IM}{\boldsymbol{I}\xspace}

\newcommand{\MM}{\boldsymbol{M}\xspace}
\newcommand{\PM}{\boldsymbol{P}\xspace}

\newcommand{\YM}{\boldsymbol{Y}\xspace}
\newcommand{\XM}{\boldsymbol{X}\xspace}
\newcommand{\ZM}{\boldsymbol{Z}\xspace}
\newcommand{\LM}{\boldsymbol{L}\xspace}
\newcommand{\UM}{\boldsymbol{U}\xspace}
\newcommand{\VM}{\boldsymbol{V}\xspace}
\newcommand{\HM}{\boldsymbol{H}\xspace}

\newcommand{\NAM}{\tilde{\AM}\xspace}

\newcommand{\GM}{\boldsymbol{G}\xspace}

\newcommand{\QM}{\boldsymbol{Q}\xspace}
\newcommand{\avec}{\overrightarrow{\boldsymbol{a}}\xspace}
\newcommand{\xvec}{\overrightarrow{\boldsymbol{x}}\xspace}
\newcommand{\yvec}{\overrightarrow{\boldsymbol{y}}\xspace}
\newcommand{\yhvec}{\overrightarrow{\widehat{\boldsymbol{y}}}\xspace}
\newcommand{\zvec}{\overrightarrow{\boldsymbol{z}}\xspace}
\newcommand{\rvec}{\overrightarrow{\boldsymbol{r}}\xspace}
\newcommand{\bvec}{\overrightarrow{\boldsymbol{b}}\xspace}
\newcommand{\rrvec}{\overrightarrow{\boldsymbol{\gamma}}\xspace}
\newcommand{\pvec}{\overrightarrow{\boldsymbol{q}}\xspace}
\newcommand{\qvec}{\overrightarrow{\boldsymbol{\phi}}\xspace}

\newcommand{\fvec}{\overrightarrow{\boldsymbol{f}}\xspace}
\newcommand{\hvec}{\overrightarrow{\boldsymbol{h}}\xspace}

\newcommand{\rhovec}{\overrightarrow{\boldsymbol{\rho}}\xspace}

\newcommand{\pivec}{\overrightarrow{\boldsymbol{\pi}}\xspace}
\newcommand{\psivec}{\overrightarrow{\boldsymbol{\psi}}\xspace}

\newcommand{\evec}{{\overrightarrow{\boldsymbol{1}}}\xspace}

\newcommand{\ppr}{BDD\xspace}

\newcommand{\algo}{\texttt{LACA}\xspace}
\newcommand{\algoc}{\texttt{LACA}~(C)\xspace}
\newcommand{\algoe}{\texttt{LACA}~(E)\xspace}
\newcommand{\algoppr}{\texttt{PR-Nibble}\xspace}

\newcommand{\gdiff}{\texttt{GreedyDiffuse}\xspace}

\newcommand{\adiff}{\texttt{AdaptiveDiffuse}\xspace}

\newenvironment{customlegend}[1][]{%
    \begingroup
    \csname pgfplots@init@cleared@structures\endcsname
    \pgfplotsset{#1}%
}{%
    \csname pgfplots@createlegend\endcsname
    \endgroup
}%

\def\addlegendimage{\csname pgfplots@addlegendimage\endcsname}

\makeatletter
\newcommand\footnoteref[1]{\protected@xdef\@thefnmark{\ref{#1}}\@footnotemark}
\makeatother

\let\oldnl\nl%
\newcommand{\nonl}{\renewcommand{\nl}{\let\nl\oldnl}}%

\SetKwComment{Comment}{/* }{ */}

\makeatletter 
\g@addto@macro{\@algocf@init}{\SetKwInOut{Parameter}{Parameters}} 
\makeatother

\definecolor{myred}{HTML}{fd7f6f}
\definecolor{myred_new}{HTML}{D8D8D8}
\definecolor{myred_new2}{HTML}{D7191C}
\definecolor{myblue}{HTML}{7eb0d5}
\definecolor{mygreen}{HTML}{b2e061}
\definecolor{mypurple}{HTML}{bd7ebe}
\definecolor{myorange}{HTML}{ffb55a}
\definecolor{myyellow}{HTML}{ffee65}
\definecolor{mypurple2}{HTML}{beb9db}
\definecolor{mypink}{HTML}{fdcce5}
\definecolor{mycyan}{HTML}{8bd3c7}
\definecolor{mycyan2}{HTML}{00ffff}

\definecolor{myblue2}{HTML}{115f9a}
\definecolor{myred2}{HTML}{c23728}

\definecolor{mygreen1}{HTML}{d3ffce}

\definecolor{preprocess-color}{HTML}{cd747c}
\definecolor{query-color}{HTML}{8282bb}

\definecolor{greedy-color}{HTML}{932d37}
\definecolor{nongreedy-color}{HTML}{3d3d7f}

\definecolor{Cora-color}{HTML}{FC9871}
\definecolor{Pubmed-color}{HTML}{377483}
\definecolor{Blogcatalog-color}{HTML}{4D4D9F}
\definecolor{Flickr-color}{HTML}{4F845C}
\definecolor{Arxiv-color}{HTML}{B83945}

\definecolor{Yelp-color}{HTML}{FC9871}
\definecolor{Reddit-color}{HTML}{377483}
\definecolor{Amazon-color}{HTML}{4F845C}

\newcommand{\xmark}{\ding{55}}%

\SetKwComment{Comment}{$\triangleright$\ }{}

\newcommand*{\rowstyle}[1]{%
  \gdef\@rowstyle{#1}%
  \@rowstyle\ignorespaces%
}

\newcolumntype{=}{%
  >{\gdef\@rowstyle{}}%
}

\newcolumntype{+}{%
  >{\@rowstyle}%
}

%% file: tex/introducation.tex
\section{Introduction}
\vspace{-1ex}

Local graph clustering (LGC) seeks to identify a single cluster (a.k.a. {\em local cluster}) pertinent to a given seed node by exploring a small region around it in the input graph.
Compared to its global counterparts, LGC runs in time proportional to the size of its resulting cluster, 
regardless of the overall size of the input graph,
making it particularly suitable for analyzing large-scale graphs,
such as social networks, online shopping graphs, biological networks, and more.
In practice, LGC finds extensive applications in various domains, including community detection or recommendation on social media~\cite{karras2022global,zhang2023constrained,chen2009local}, protein grouping in bioinformatics~\cite{voevodski2009finding,liao2009isorankn}, product return or user action prediction  ~\cite{li2018tail,zhu2018local,yang2017local} on e-commerce platforms, 
and many others~\cite{fazzone2022discovering,gleich2015using,rabbany2018active,maji2011biased,mahoney2012local}.

Canonical solutions for LGC~\cite{spielman2013local,andersen2006local,kloster2014heat,yang2019efficient} are predominantly based on random walk-based graph diffusion models~\cite{masuda2017random}, where the main idea is to spread mass from the seed node to other nodes along the edges. This methodology offers high scalability and rigorous theoretical guarantees for complexity and output cluster size, but it is extremely sensitive to {\em structural heterogeneities} (e.g., high-degree nodes) in real networks, as pinpointed in~\cite{jeub2015think,wang2017capacity}.
To mitigate this issue, subsequent LGC works~\cite{wang2017capacity,fountoulakis2020p,jung2020local} formulate LGC as a combinatorial optimization problem and leverage max flow algorithms to derive the ``optimal'' local clusters. Although these approaches improve the theoretical results of previous works,
they are mostly of theoretical interest only and struggle to cope with sizable graphs
due to poor practical efficiency.
Most importantly, the above algorithms primarily focus on optimizing connectivity-based clustering quality metrics~\cite{lovasz1993random}.
Real graphs are often constructed from data riddled with noise and thus encompass substantial noisy or missing links, leading to high ground-truth {\em conductance}~\cite{lovasz1993random}, e.g., $0.765$ for {\em Flickr} and $0.649$ for {\em Yelp}.
Consequently, applying classic LGC techniques to such graphs leads to sub-optimal performance, e.g, often low precisions below $30\%$ (see Table~\ref{tbl:node-clustering}).
As a partial remedy, recent efforts~\cite{yin2017local,fu2020local,chhabra2023local,yuan2024index} incorporate {\em higher-order connectivity patterns} (e.g., motifs) into the LGC frameworks for improved clustering. However, they still rely solely on graph topology and are therefore vulnerable to missing or noisy nodes/edges. Moreover, these methods suffer from severe efficiency issues as they require enumerating the motifs in graphs in the pre-processing or on the fly.

In real life, graphs are often endowed with rich nodal attributes, e.g., user profiles in social networks and paper abstracts in citation graphs. These are termed {\em attributed graphs}.
Recent studies~\cite{huang2024cross,liao2018attributed,bothorel2015clustering} have corroborated that node attributes provide information that can effectively complement the graph topology for better performance in various tasks. 
Inspired by this, a straightforward idea is to exploit the attribute information to enhance LGC performance whilst retaining the {\em locality}.
Very recently, several attempts~\cite{freitas2018local,yudong2022local,pmlr-v202-yang23d} have been made to extend classic LGC techniques to attributed graphs by simply re-weighting each edge by the attribute similarity of its endpoints. However, 
this strategy merely accounts for connected nodes and still fails to deal with missing and noisy connections.

To tackle the foregoing issues, we propose \algo 
, a novel solution that seamlessly integrates node attributes into graph structures for effective LGC computation, while offering strong theoretical guarantees on locality and high empirical efficiency.
Specifically, \algo formulates the LGC task as a seeded random walk diffusion over graphs based on the novel notion of {\em bidirectional diffusion distribution} (BDD), an attribute-aware affinity measure dedicated to node pairs in attributed graphs.
The main idea is to model the affinity between any node pair $(v_s,v_t)$ as the expected normalized (exponential) cosine similarity (hereafter SNAS) between the attributes of ending node pairs of random walks originating from $v_s$ and $v_t$, which essentially evaluates their {\em meeting probability} from the perspectives of both of them through random walks and attribute-based transitions.
Accordingly, given a seed node $v_s$, the goal of LGC is to find a local cluster containing nodes with the highest BDD values w.r.t. $v_s$.

The exact computation of BDD incurs a signiﬁcant cost of up to $O(n^3)$ as there are $n\times n$ possible ending node pairs from random walks starting at the seed node and any of the $n$ nodes. This is {\em non-local} and infeasible for large graphs.
In response, we develop an adaptive framework for an approximate solution with a runtime linear to the output size and independent of the size of the graph.
First, we 
obtain a fast and theoretically grounded decomposition of the SNAS matrix into 
low-dimensional feature vectors through randomized techniques~\cite{halko2011finding,yu2016orthogonal}.
This step
enables the decoupling of the BDD computation and transforms the problem into the diffusion of vectors over input graphs.
On top of that, we devise a new algorithm \adiff, which diffuses non-negative vectors on graphs using efficient matrix operations in an adaptive fashion.
It not only alleviates the intensive memory access patterns in previous traversal/sampling-based diffusion approaches~\cite{andersen2006local,wang2017fora,yang2022efficient} but also overcomes their limitations of sensitivity to high-degree nodes and slow convergence,
without compromising the theoretical assurance in approximation and asymptotic performance.
Lastly, 
a carefully designed three-step scheme is employed to construct the approximate BDD.
We further conduct non-trivial theoretical analyses to reveal the approximation and complexity bounds of \algo, as well as its connection to {\em graph neural networks} (GNNs)~\cite{hamilton2017inductive}.

Our empirical studies, which evaluate \algo against 17 competitors on 8 real attributed graph datasets with ground-truth local clusters, demonstrate that \algo is consistently superior or comparable with the state-of-the-art baseline approaches in terms of result quality at a fraction of their cost. In particular, on the largest {\em Amazon2M} dataset with 61.9 million edges, \algo is able to recover the target local clusters with an average of $1.8\%$ improvement in precision and $152\times$ speedup compared to the best competitor.

%% file: tex/preliminary.tex
\section{Problem Formulation}
\subsection{Notations and Terminology}

\begin{table}[!t]
\centering
\renewcommand{\arraystretch}{1.0}
\begin{footnotesize}
\caption{Frequently used notations.} \label{tbl:notations}
\resizebox{\columnwidth}{!}{%
\begin{tabular}{|p{0.54in}|p{2.6in}|}
    \hline
    {\bf Notation}  &  {\bf Description}\\
    \hline
    $\V,\EDG,\XM$   & The node set, edge set, and node attribute matrix $\XM$ of attributed graph $\G$, respectively.\\ \hline
    $n,m,d$ & The numbers of nodes, edges, and distinct attributes, respectively.\\ \hline
    $\N(v_i), d(v_i)$ & The set of neighbors and degree of node $v_i$, respectively.\\ \hline
    $\AM,\DM,\PM$   & The adjacency, degree, and transition matrices of the graph $\G$, respectively.\\ \hline
    $\mathsf{vol}(\C)$   & The volume of a set $\C$ of nodes, i.e., $\sum_{v_i\in \C}{d(v_i)}$.\\ \hline
    $\mathsf{supp}(\xvec)$ & The support of vector $\xvec$, i.e., $\{i: \xvec_i\neq 0\}$. \\ \hline
    $\alpha$  & The restart factor in RWR, $\alpha\in (0,1)$.\\ \hline
    $s(v_i,v_j)$    & The SNAS defined by Eq.~\eqref{eq:SNAS}.\\ \hline
    $\pi(v_x,v_y)$    & The RWR score of $v_y$ w.r.t. $v_x$ (See Eq.~\eqref{eq:pi-matrix}).\\ \hline
    $\rhovec_t,\rhovec^{\prime}_t$ & The exact and approximate BDD values of $v_t$ w.r.t. $v_s$ respectively (See definition in Eq.~\eqref{eq:csim}).\\ \hline
    $\ZM, \zvec^{(i)}$  & The TNAM and its $i$-th row vector (See Eq.~\eqref{eq:s-zz}).\\ \hline
    $\epsilon, \sigma$  & The diffusion threshold and balancing parameter in Algo.~\ref{alg:iter-fwd}.\\ \hline
    $k$  & The dimension of TNAM vectors $\zvec^{(i)}\ \forall{v_i\in \V}$.\\ \hline
\end{tabular}%
}
\end{footnotesize}
\vspace{-3ex}
\end{table}

Let $\G=(\V,\EDG)$ be a connected, undirected, and unweighted graph (a.k.a. network), where $\V=\{v_1,v_2,\ldots,v_n\}$ is a set of $n$ nodes and $\EDG\in \V\times\V$ is a set of $m$ edges. For each edge $(v_i,v_j)\in \EDG$, we say $v_i$ and $v_j$ are neighbors to each other. We use $\N(v_i)$ to denote the set of neighbors of $v_i$ and $d(v_i)=|\N(v_i)|$ as its degree. 
Let $\AM$ be the adjacency matrix of $\G$ where $\AM_{i,j}=1$ if $(v_i, v_j)\in \EDG$, otherwise $\AM_{i,j}=0$. 
The diagonal matrix $\DM$ is used to symbolize the degree matrix of $\G$ where $\DM_{i,i}=d(v_i)$. 
The {\em transition matrix} of $\G$ is defined by $\DM^{-1}\AM$, where $\PM_{i,j}=\frac{1}{d(v_i)}$ if $(v_i,v_j)\in \EDG$, otherwise $\PM_{i,j}=0$. 
Accordingly, $\PM^{\ell}_{i,j}$ signifies the probability of a length-$\ell$ random walk originating from node $v_i$ ending at node $v_j$.

A graph is referred to as an {\em attributed graph} if each node $v_i$ in $\V$ is endowed with a $d$-dimensional attribute vector $\xvec^{(i)}$, which is the $i$-th row vector in the node attribute matrix $\XM$ of $\G$.
We assume $\xvec^{(i)}$ is $L_2$-normalized, i.e., $\|\xvec^{(i)}\|_2=1$.
$\XM_{i,j}$ and $\xvec^{(i)}_j$ represent the $(i,j)$-th element in matrix $\XM$ and $j$-th entry in vector $\xvec^{(i)}$, respectively. 
The {\em support} of vector $\xvec^{(i)}$ is defined as $\mathsf{supp}(\xvec^{(i)})=\{j: \xvec^{(i)}_j\neq 0\}$, which comprises the indices of non-zero entries in $\xvec^{(i)}$.
The {\em volume} of a set $\C$ of nodes is defined as $\mathsf{vol}(\C)=\sum_{v_i\in \C}{d(v_i)}$. 
Given a length-$n$ vector $\xvec$, its {\em volume} $\mathsf{vol}(\xvec)$ is defined as $\sum_{i\in \mathsf{supp}(\xvec)}{d(v_i)}$.
Table~\ref{tbl:notations} lists notations frequently used throughout this paper.

A {\em local cluster} $\C_s$ w.r.t. a seed node $v_s$ in graph $\G$ is defined as a subset of $\V$ containing $v_s$. 
Intuitively, in an attributed graph $\G$, a good local cluster should be internally cohesive and well-separated from the remainder of $\G$ in terms of both topological connections and attribute similarity.
Given the seed node $v_s$, the general goal of LGC over attributed graph $\G$ is to identify such a local cluster $\C_s$ with the runtime cost roughly linear to its volume $\mathsf{vol}(\C_s)$.

\subsection{Symmetric Normalized Attribute Similarity}\label{sec:SNAS}
We quantify the the similarity of two nodes $v_i$ and $v_j$ in $\G$ in terms of attributes as follows: 
\begin{footnotesize}
\begin{equation}\label{eq:SNAS}
s(v_i,v_j) = \frac{f(\xvec^{(i)},\xvec^{(j)})}{\sqrt{\sum_{v_\ell\in \V}{f(\xvec^{(i)},\xvec^{(\ell)})}} \sqrt{\sum_{\xvec^{(\ell)}\in \V}{f(\xvec^{(j)},\xvec^{(\ell)})}}},
\end{equation}
\end{footnotesize}
where $f(\cdot,\cdot)$ can be any metric function defined over two vectors. The denominator in Eq.~\eqref{eq:SNAS} 
is to ensure the $s(v_i,v_j)$ values w.r.t. any node $v_i\in \V$ to be symmetric and normalized to a comparable range ($0\le s(v_i,v_j)\le 1$), which facilitates the design of the BDD in subsequent section.
Particularly, $s(v_i,v_j)$ is referred to as the SNAS of nodes $v_i$ and $v_j$. 

We adopt two classic metric functions for $f(\cdot,\cdot)$, i.e., {\em cosine similarity} and {\em exponent cosine similarity}~\cite{li2022graph}. When $f(\cdot,\cdot)$ is the cosine similarity, $f(\xvec^{(i)},\xvec^{(j)})=\xvec^{(i)}\cdot\xvec^{(j)}$ since $\|\xvec^{(i)}\|_2=1$. Then, the SNAS $s(v_i,v_j)$ can be computed via
\begin{footnotesize}
\begin{equation}\label{eq:SNAS-cos}
\frac{\xvec^{(i)}\cdot\xvec^{(j)}}{\sqrt{\sum_{v_\ell\in \V}{\xvec^{(i)}\cdot\xvec^{(\ell)}}} \sqrt{\sum_{v_\ell\in \V}{\xvec^{(j)}\cdot\xvec^{(\ell)}}} }.
\end{equation}
\end{footnotesize}
Analogously, when the exponent cosine similarity is adopted,
\begin{small}
\begin{equation}\label{eq:exp-cos}
f(\xvec^{(i)},\xvec^{(j)})
=\exp{\left({\xvec^{(i)}\cdot \xvec^{(j)}}/{\delta}\right)}
\end{equation}
\end{small}
and the SNAS of nodes $v_i,v_j$ is formulated as
\begin{footnotesize}
\begin{equation}\label{eq:SNAS-ecos}
\frac{\exp{\left({\xvec^{(i)}\cdot \xvec^{(j)}}/{\delta}\right)}}{\sqrt{\sum_{v_\ell \in \V}{\exp{\left({\xvec^{(i)}\cdot \xvec^{(\ell)}}/{\delta}\right)}}} \sqrt{\sum_{v_\ell \in \V}{\exp{\left({\xvec^{(j)}\cdot \xvec^{(\ell)}}/{\delta}\right)}}} },
\end{equation}
\end{footnotesize}
where $\delta$ (typically $1$ or $2$) is the sensitivity factor. Essentially, Eq. \eqref{eq:SNAS-ecos} can be deemed as a variant of the {\em softmax function}.

\begin{figure}[!t]
\centering
\includegraphics[width=\linewidth]{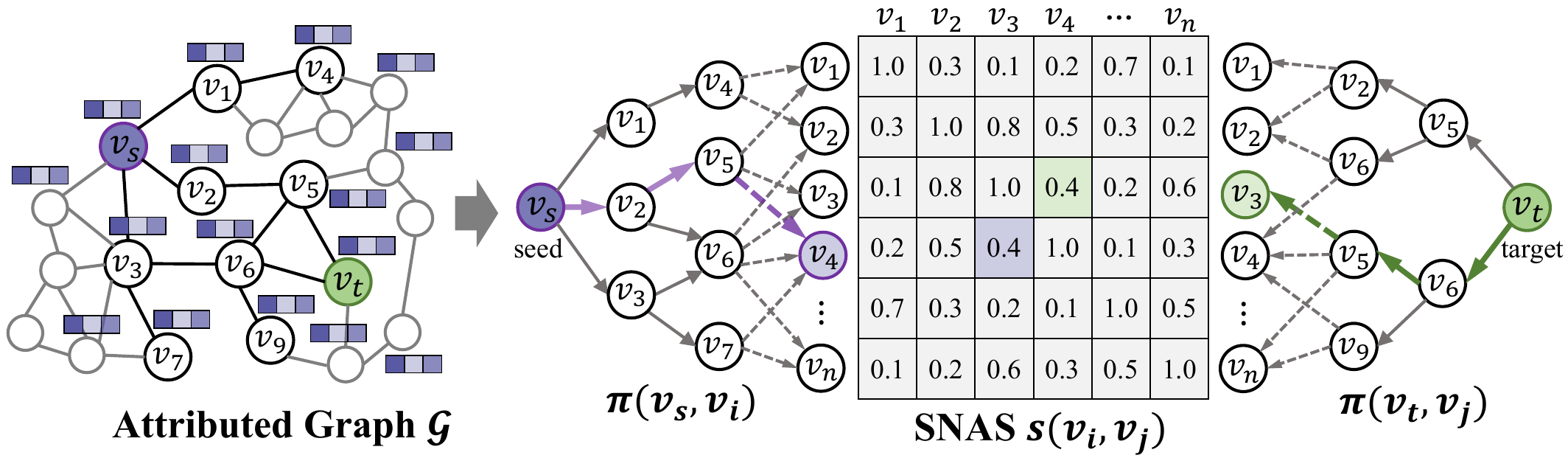}
\vspace{-3ex}
\caption{Figurative Illustration of \ppr.}\label{fig:BDD}
\vspace{-3ex}
\end{figure}

\subsection{Bidirectional Diffusion Distribution}\label{sec:BDD-desc}
We further propose the {\em bidirectional diffusion} (\ppr) to model the likelihood of two nodes in $\G$ to be in the same cluster. 
In particular, 
Unlike previous works that are based on biased graph proximity from the seed's view, 
\ppr integrates the strength of topological connections and the attribute similarity of the node pair in a coherent framework from the perspectives of both the seed and target.
More precisely, given a seed node $v_s\in \V$, for any target node $v_t\in\V$, the \ppr $\rhovec_t$
of node pair $(v_s,v_t)$ is defined by
\begin{small}
\begin{equation}\label{eq:csim}
 \rhovec_t = \sum_{v_i,v_j\in \V}{\pi(v_s,v_i)\cdot s(v_i,v_j) \cdot \pi(v_t,v_j)},
\end{equation}
\end{small}
where $s(v_i,v_j)$ is the SNAS of intermediate nodes $v_i,v_j$ and 
\begin{small}
\begin{equation}\label{eq:pi-matrix}
\textstyle \pi(v_x,v_y) = \sum_{\ell=0}^{\infty}{(1-\alpha)\alpha^\ell\cdot \PM^\ell_{x,y}}
\end{equation}
\end{small}
stands for the RWR ({\em random walk with restart}~\cite{tong2006fast,jeh2003scaling}) score of node $v_y$ w.r.t. node $v_x$.
At each step, an RWR over $\G$ either stops at the current node with $1-\alpha$ probability or navigates to one of its neighbors uniformly at random with probability $\alpha$.
In essence, $\pi(v_s,v_i)$ (resp. $\pi(v_t,v_j)$) is the probability that an RWR originating from $v_s$ (resp. $v_t$) terminates at node $v_i$ (resp. $v_j$). Put differently, $\pi(v_s,v_i)\ \forall{v_i\in \V}$ (resp. $\pi(v_t,v_j)\ \forall{v_j\in \V}$) describe the distribution of mass ($1.0$ in total) disseminated from $v_s$ (resp. $v_t$) to all the nodes in $\G$ via random walks, respectively. Particularly, given a vector $\avec\in\mathbb{R}^n$, we refer to the below process as an {\em RWR-based graph diffusion}:
\begin{small}
\begin{equation}\label{eq:RWR-diffusion}
\sum_{v_x\in \V}{\avec_x \cdot \pi(v_x,v_y)}\ \forall{v_y\in \V},
\end{equation}
\end{small}
where $\avec_x\cdot {\pi(v_x,v_y)}$ can be interpreted as the amount of mass in $\avec$ spread from $v_x$ to $v_y$ via random walks on $\G$.

Let nodes $(v_i,v_j)$ be the {\em ending node pair} of two bidirectional random walks with restart from seed node $v_s$ and target node $v_t$. The \ppr $\rhovec_t$ of $(v_s,v_t)$ in Eq. \eqref{eq:csim} is therefore the overall SNAS of all such ending pairs, as illustrated in Fig.~\ref{fig:BDD}.
Intuitively, if two nodes $v_s$ and $v_t$ are located in the same cluster, their proximal nodes (i.e., nodes with high RWR scores) are more likely to have high SNAS (i.e., high attribute homogeneity), resulting in a high \ppr value $\rhovec_t$. Particularly, the injection of the SNAS reduces the likelihood of transiting to undesired nodes caused by noisy links, while increasing the affinity to desired nodes with high attribute similarities but low connectivity due to missing links.

\stitle{Remark} When the input graph $\G$ is {\em non-attributed}, we set the SNAS $s(v_i,v_j)=1$ if $v_i=v_j$ and 0 otherwise $\forall{v_i,v_j\in \V}$. The \ppr $\rho(v_q,v_t)$ is then a variant of the CoSimRank~\cite{rothe2014cosimrank} metric, which measures the likelihood of random walks from two nodes meeting each other over the graph.

\subsection{Problem Statement}
Based thereon, the LGC task for seed node $v_s$ over attributed graph $\G$ is framed as 
an {\em approximation} of the \ppr vector $\rhovec$ (denoted as $\rhovec^{\prime}$), followed by a simple extraction of the nodes with $|\C_s|$ 
{\em largest} values in $\rhovec^{\prime}$ as the predicted local cluster $\C_s$. 
In turn, our major focus of the LGC problem lies in the computation of $\rhovec^{\prime}$. 
More formally, given a diffusion threshold $\epsilon$, we aim to estimate a BDD vector $\rhovec^{\prime}$ such that both its output volume $\mathsf{vol}(\rhovec^{\prime})$ (i.e., the size of the explored region on $\G$) and entailed computation cost are bounded by $\textstyle O\big({1}/{\epsilon}\big)$, i.e., achieving the locality.

Notice that the direct computation of BDD $\rhovec$ in Eq.~\eqref{eq:csim} requires the RWR scores $\pi(v_s,v_i)$ of all intermediate nodes $v_i\in \V$ w.r.t. the seed $v_s$, the RWR scores $\pi(v_t,v_j)$ of all intermediate nodes $v_j\in \V$ w.r.t. all possible target nodes $v_t\in \V$, and the SNAS values $s(v_i,v_j)$ of all possible node pairs $(v_i,v_j)\in \V\times \V$. 
The latter two ingredients involve up to $O(n^2)$ node pairs, rendering the {\em local} estimation of $\rhovec$ particularly challenging. 
Additionally, it remains unclear how to provide approximation accuracy assurance for $\rhovec^{\prime}$.

%% file: tex/overview.tex
\section{Solution Overview}\label{sec:high-level-idea}

\begin{figure}[!t]
    \centering
    \includegraphics[width=\columnwidth]{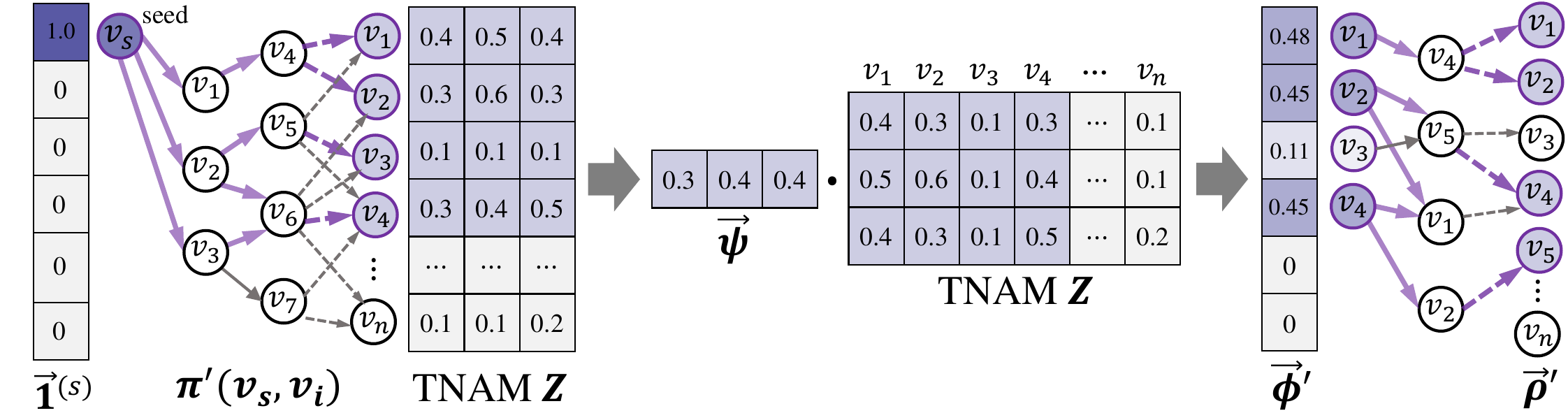}
    \vspace{-3ex}
    \caption{Basic Idea of \algo.}
    \label{fig:idea}
    \vspace{-3ex}
\end{figure}

To address the preceding technical challenges, we streamline the approximation of BDD via a three-step framework through our careful analyses as follows.

\subsection{Problem Transformation}

Firstly, by the definition of BDD in Eq.~\eqref{eq:csim} and the symmetric property of RWR scores~\cite{lofgren2015bidirectional} (i.e., $\pi(v_t,v_j)\cdot d(v_t)=\pi(v_j,v_t)\cdot d(v_j)$), we can rewrite the \ppr value $\rhovec_t$ of any target node $v_t\in \V$ w.r.t. the seed node $v_s$ as
\begin{small}
\begin{align}
\rhovec_t = \frac{1}{d(v_t)}\sum_{v_i\in \V}{ \qvec_i\cdot {\pi(v_i,v_t)}}, \label{eq:rho-t}
\end{align}
\end{small}
where $\qvec$ is called the RWR-SNAS vector w.r.t. $v_s$ and
\begin{small}
\begin{equation}\label{eq:sigma-vec}
\qvec_i
= \sum_{v_j\in \V}{{\pi(v_s,v_j)}\cdot s(v_j,v_i)\cdot d(v_i)}.
\end{equation}
\end{small}
Intuitively, if the RWR-SNAS vector $\qvec$ is at hand, Eq.~\eqref{eq:rho-t} implies that an {\em approximate} \ppr vector $\rhovec^{\prime}$ can be obtained 
via the RWR-based graph diffusion (see Eq.~\eqref{eq:RWR-diffusion}) of $\qvec$ over the graph $\G$ with a diffusion threshold.

\begin{figure}[!t]
    \centering
    \includegraphics[width=\columnwidth]{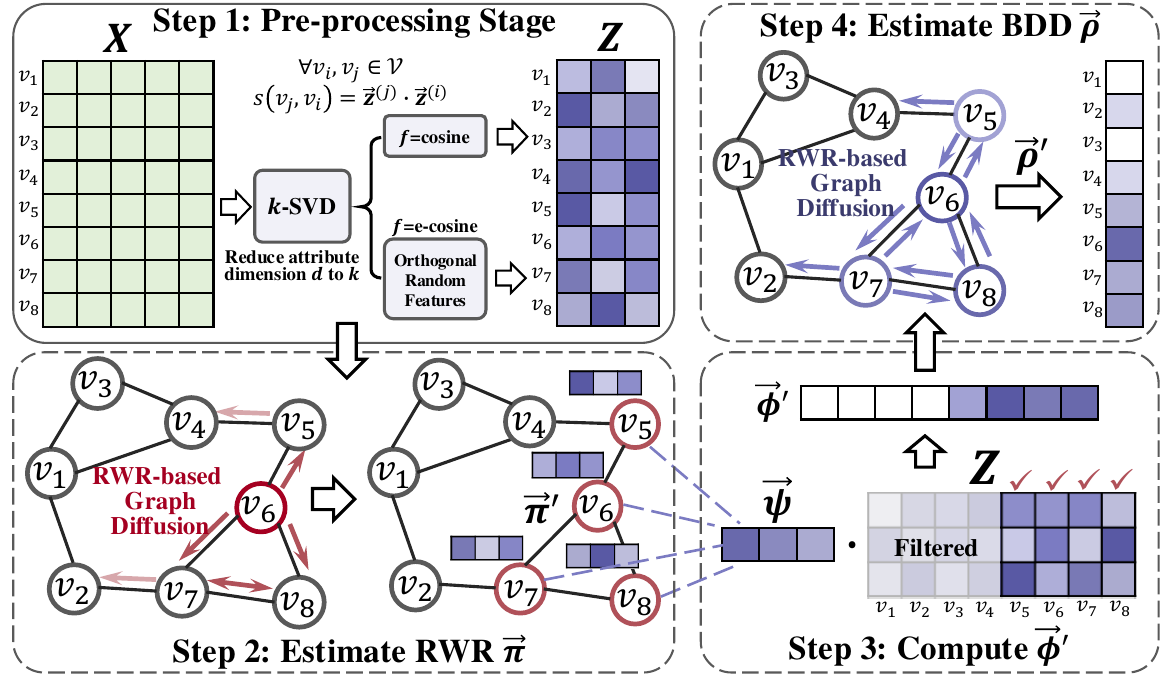}
    \vspace{-3ex}
    \caption{Overview of \algo.}
    \label{fig:overview}
    \vspace{-3ex}
\end{figure}

However, the direct and exact computation of the RWR-SNAS vector $\qvec$ in Eq.~\eqref{eq:sigma-vec} is still immensely expensive as it requires calculating the SNAS $s(v_j,v_i)$ for all possible node pairs (up to $O(n^2)$) in $\G$.
Towards this end, we propose to decompose each SNAS $s(v_j,v_i)$ as 
the product of two length-$k$ ($k\ll d$ and is a constant) vectors, i.e.,
\begin{equation}\label{eq:s-zz}
s(v_j,v_i) = \zvec^{(j)}\cdot \zvec^{(i)},
\end{equation}
where $\ZM\in \mathbb{R}^{n\times k}$ is a transformed node attribute matrix $\XM$ (hereafter TNAM).
In doing so, the RWR-SNAS vector $\qvec$ in Eq.~\eqref{eq:sigma-vec} can be reformulated as
\begin{equation}\label{eq:sigma-vec-2}
\qvec_i = \pivec \cdot \ZM \cdot \zvec^{(i)}\cdot d(v_i),
\end{equation}
where $\pivec$ denotes the RWR vector w.r.t. seed node $v_s$, i.e., $\pivec_i=\pi(v_s,v_i)\ \forall{v_i\in \V}$. 
Given an estimation $\pivec^{\prime}$ of $\pivec$, the term $\pivec \cdot \ZM $ in Eq.~\eqref{eq:sigma-vec-2} can be approximated by
\begin{small}
\begin{equation}\label{eq:pi-Z}
 \psivec = \sum_{i\in \mathsf{supp}(\pivec^{\prime})}{\pivec^{\prime}_i\cdot \zvec^{(i)}}\ \in \mathbb{R}^k,
\end{equation}
\end{small}
and accordingly, we can estimate $\qvec$ by
\begin{equation}\label{eq:sigma-vec-3}
\qvec^{\prime}_i =  \psivec \cdot \zvec^{(i)}\cdot d(v_i)\ \forall{v_i\in \{v_i| i\in \mathsf{supp}(\pivec^{\prime})\}}.
\end{equation}
Note that $\psivec$ is shared by the computations of all possible $\qvec^{\prime}_i$. If we can calculate an approximate RWR vector $\pivec^{\prime}$ with support size (i.e., the number of non-zero entries) $\mathsf{supp}(\pivec^{\prime})=O(1/\epsilon)$ in $O(1/\epsilon)$ time, the construction times and support sizes of $\psivec$ and $\qvec^{\prime}$ are also bounded by $O(1/\epsilon)$.

As illustrated in Fig.~\ref{fig:idea}, our above idea transforms the computation of the BDD for each target node $v_t\in \V$ in Fig.~\ref{fig:BDD} into (i) aggregation of TNAM vectors of nodes into vector $\psivec$ based on their RWR scores in $\pivec^{\prime}$, (ii) construction of the RWR-SNAS vector $\qvec^{\prime}$ by Eq.~\eqref{eq:sigma-vec-3}, and (iii) RWR-based graph diffusion of $\qvec^{\prime}$ over $\G$ to get $\rhovec^{\prime}$.

\begin{figure}[!t]
\centering
\includegraphics[width=\linewidth]{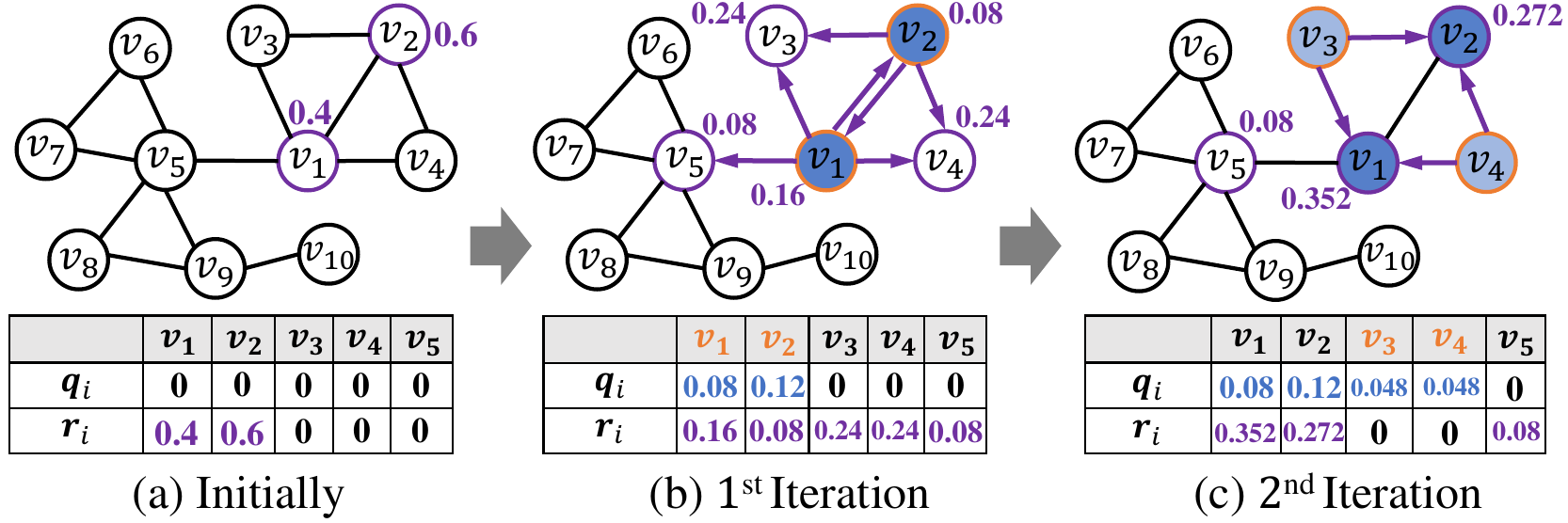}
\vspace{-2ex}
\caption{\gdiff with $\alpha=0.8$ and $\epsilon=0.1$.}\label{fig:gdiff-example}
\vspace{-1ex}
\end{figure}

\subsection{A Three-Step Framework}
The BDD approximation thus involves four subtasks, i.e., the computations of the TNAM $\ZM$, approximate RWR vector $\pivec^{\prime}$, RWR-SNAS vector $\qvec^{\prime}$, and approximate \ppr vector $\rhovec^{\prime}$.

Since $\qvec^{\prime}$ can be properly computed using Eq.~\eqref{eq:sigma-vec-3}, our main tasks are to construct $\ZM$, $\pivec^{\prime}$, and $\rhovec^{\prime}$. 
Let $\evec^{(s)}$ be a vector with $1$ at $s$-th entry and $0$ everywhere else. The exact RWR score $\pivec_t$ of any node $v_t\in \V$ can be represented as
\begin{small}
\begin{equation*}
\pivec_t = \sum_{v_i\in \V}{\evec^{(s)}_i \cdot \pi(v_i,v_t)},
\end{equation*}
\end{small}
which can also be regarded as an RWR-based graph diffusion. This inspires us to design a unified graph diffusion algorithm that obtains estimations $\pivec^{\prime}$ and $\rhovec^{\prime}$ by diffusing $\evec^{(s)}$ and $\qvec^{\prime}$ over $\G$ based on RWR, respectively, while fulfilling the volume and runtime bounds (i.e., $O(1/\epsilon)$), and high practical efficiency.
As for TNAM $\ZM$, we propose to generate it through a preprocessing of the input node attribute matrix $\XM$ as $\ZM$ can be reused in the LGC task of any seed node $v_s\in \V$.

As summarized in Fig.~\ref{fig:overview}, our proposed \algo includes a preprocessing algorithm that converts $\XM$ into a $k$-dimensional TNAM $\ZM$ thereby enabling the problem transformation in Eq.~\eqref{eq:sigma-vec-2} (Algo.~\ref{alg:Z}), and a three-step scheme for the online approximation of BDD vector.
Therein, Step 1 identifies a small set of nodes around the seed node by estimating their RWR $\pivec^{\prime}$ using an RWR-based graph diffusion algorithm (Algo.~\ref{alg:fwd} or~\ref{alg:iter-fwd}) with $\evec^{(s)}$ as input, while Step 2 aggregates their TNAM vectors as $\psivec$ (Eq.~\eqref{eq:pi-Z}) to build the RWR-SNAS vector $\qvec^{\prime}$ (Eq.~\eqref{eq:sigma-vec-2}). In the final step (Step 3), \algo conducts another RWR-based graph diffusion with $\qvec^{\prime}$ over $\G$ to derive the approximate BDD vector $\rhovec^{\prime}$ (Algo.~\ref{alg:main}).

In succeeding sections, we first elaborate on the algorithmic design of our generalized RWR-based graph diffusion approaches (Algo.~\ref{alg:fwd} and~\ref{alg:iter-fwd}) for the estimations of RWR vector $\pivec$ and BDD vector $\rhovec$ in Section~\ref{sec:adiff}. After that, Section~\ref{sec:complete} describes the technical details of Algo.~\ref{alg:Z} for constructing TNAM $\ZM$ and the complete \algo algorithm (Algo.~\ref{alg:main}), followed by theoretical analyses in terms of accuracy approximation, volume, and runtime.

%% file: tex/solution_secondary.tex
\begin{algorithm}[!t]
\caption{\gdiff}\label{alg:fwd}
\KwIn{Transition matrix $\PM$, restart factor $\alpha$, diffusion threshold $\epsilon$, initial vector $\fvec$}
\KwOut{Diffused vector $\pvec$}
$\rvec \gets \fvec;\ \pvec \gets \mathbf{0}$\;
\While{true}{
    Compute sparse vector $\rrvec$\Comment*[r]{Eq.~\eqref{eq:gamma_vec}}
    \lIf{$\rrvec$ is $\mathbf{0}$}{\textbf{break}}
    $\rvec \gets \rvec-\rrvec$\;
    Update $\pvec$ and $\rrvec$\Comment*[r]{Eq.~\eqref{eq:greedy}}
    $\rvec \gets \rvec + \rrvec$\;
}
\Return{$\pvec$}\;
\end{algorithm}

\section{RWR-based Graph Diffusion Algorithms}\label{sec:adiff}
We unify the computations of $\pivec^{\prime}$ and $\rhovec^{\prime}$ as diffusing an input vector $\fvec$ along edges over $\G$ to get $\pvec$ satisfying
\begin{small}
\begin{equation}\label{eq:f-q-eps}
\forall{v_t\in \V},\ 0\le \sum_{v_i\in \V}{\fvec_i\cdot {\pi(v_i,v_t)}-\pvec_t \le \epsilon\cdot d(v_t)},
\end{equation}
\end{small}
which is $\pivec^{\prime}$ and $\rhovec^{\prime}$ when $\fvec$ is set to $\evec^{(s)}$ and $\qvec^{\prime}$, respectively. 
Instead of using deterministic graph traversals or random walk samples for diffusion as in prior algorithms~\cite{andersen2006local,wang2017fora}, we first develop \gdiff by leveraging matrix operations and a greedy strategy for cache-friendly memory access patterns and higher efficiency.
Further, we upgrade \gdiff to \adiff for faster termination without compromising the theoretical guarantees through an adaptive scheme.

\subsection{The \gdiff Approach}\label{sec:gdiff}

Algo.~\ref{alg:fwd} presents the pseudo-code of \gdiff for ``diffusing'' any initial non-negative vector $\fvec$. 
Algo.~\ref{alg:fwd} is greedy in the sense that the diffusion operations are solely conducted for nodes whose residues are beyond a certain threshold so as to minimize the total amount of operations needed to satisfy the desired accuracy guarantee.

Given the transition matrix $\PM$ of the input graph $\G$ and the diffusion threshold, Algo. \ref{alg:fwd} begins by initializing a {\em residual} vector $\rvec$ as $\fvec$ and a {\em reserve} vector $\pvec$ as $\mathbf{0}$ at Line 1.
Afterward, it starts an iterative process for diffusing and converting the residuals in $\rvec$, which continuously transfers the residuals into the reserve vector $\pvec$ (Lines 2-7). Specifically, in each iteration, we first identify the residuals in $\rvec$ whose corresponding $\rvec_i/d(v_i)$ values are equal to or beyond $\epsilon\cdot \|\fvec\|_1$ and move them to a temporary vector $\rrvec$ for subsequent diffusion. More precisely, we obtain a sparse vector $\rrvec$ at Line~3 as follows:
\begin{small}
\begin{equation}\label{eq:gamma_vec}
\rrvec_i = \begin{cases}
\rvec_i & \text{if } (\rvec\DM^{-1})_i = \frac{\rvec_i}{d(v_i)} \ge \epsilon, \\
0 & \text{otherwise}.
\end{cases}
\end{equation}
\end{small}

Next, we update residual vector $\rvec$ as $\rvec-\rrvec$ such that $\rvec$ contain the residuals below the threshold and then convert $(1-\alpha)$ portion of residuals in $\rrvec$ into reserve vector $\pvec$ (Lines 5-6). $\forall{v_i}\in \V$, its remaining $\alpha$ fraction of residual in $\rrvec$ is later evenly scattered to its out-neighbors. That is, each node $v_j\in \V$ receives a total of
$\sum_{v_i\in \N(v_j)}{\alpha\cdot \frac{\rrvec_i}{d(v_i)}}$
residual from its incoming neighbors $\N(v_j)$, which can be written as a sparse matrix-vector multiplication as follows (Line 6):
\begin{equation}\label{eq:greedy}
\begin{gathered}
\pvec \gets \pvec + (1-\alpha)\rrvec,\ \rrvec \gets \alpha\rrvec\PM
\end{gathered}
\end{equation}
These residuals in $\rrvec$ will be added back to $\rvec$ for the next round of conversion and diffusion (Line 7).

\begin{theorem}\label{lem:LP}
Given initial vector $\fvec$, restart factor $\alpha$, and diffusion threshold $\epsilon$, Algo.~\ref{alg:fwd} outputs a diffused vector $\pvec$ satisfying Eq.~\eqref{eq:f-q-eps} using $\footnotesize O\left(\max\left\{|\mathsf{supp}(\fvec)|,\frac{\|\fvec\|_1}{(1-\alpha)\epsilon}\right\}\right)$ time.
\begin{proof}
All missing proofs can be found in Appendix~\ref{sec:proof}.
\end{proof}
\end{theorem}

Algo. \ref{alg:fwd} repeats the above procedure until the resulting $\rrvec$ turns to be a zero vector (Line 4), i.e., all non-converted residuals in $\rvec$ fall below the desired threshold in Eq.~\eqref{eq:gamma_vec}. Eventually, Algo. \ref{alg:fwd} returns $\pvec$ as the diffused vector of $\fvec$. Theorem~\ref{lem:LP}
establishes the approximation accuracy guarantees of Algo. \ref{alg:fwd} and indicates that \gdiff runs in time proportional to the size of $\fvec$ and $\frac{1}{(1-\alpha)\epsilon}$, but independent of the size of the input graph $\G$.
This indicates that \gdiff enables the local estimation of RWR $\pivec$ and $\rhovec$ if $\evec^{(s)}$ and $\qvec^{\prime}$ are given as input, respectively.
\stitle{A Running Example} Consider the example $\G$ in Fig.~\ref{fig:gdiff-example}. The input vector $\fvec$ is a length-10 vector in which the first and second entries are $0.4$ and $0.6$. We conduct \gdiff over $\G$ with restart factor $\alpha=0.8$ and diffusion threshold $\epsilon=0.1$. Initially, the residuals of $v_1$ and $v_2$ are $0.4$ and $0.6$ as in $\fvec$, while the reserves of all nodes are $0$ as in Fig.~\ref{fig:gdiff-example}(a). Since $\frac{\rvec_1}{d(v_1)}=0.4/4\ge\epsilon$ and $\frac{\rvec_2}{d(v_2)}=0.6/3\ge \epsilon$, \gdiff converts the $1-\alpha=0.2$ portion of their residuals into their reserves and distributes the rest to their neighbors evenly. Specifically, nodes $v_2$-$v_5$ receive a residual of $\frac{0.4\alpha}{d(v_1)}=0.08$ from $v_1$, respectively, while each of nodes $v_1$, $v_3$, and $v_4$ receives $\frac{0.6\alpha}{d(v_2)}=0.16$ residual. Notably, both $v_3$ and $v_4$ have a total residual of $0.24$, which satisfy $\frac{0.24}{d(v_3)}=\frac{0.24}{d(v_4)}=0.12\ge \epsilon$. Thus, in the second iteration, \gdiff merely performs diffusion operations on $v_3$ and $v_4$. A residual of $0.24(1-\alpha)=0.048$ will be converted into their reserves, and a residual of $\frac{0.24\alpha}{d(v_3)}=\frac{0.24\alpha}{d(v_4)}=0.096$ will be transferred to $v_1$ and $v_2$ from $v_3$ and $v_4$, respectively. The residuals at $v_1$, $v_2$, and $v_5$ are updated to $0.352$, $0.272$, and $0.08$, respectively, leading to $\frac{0.352}{d(v_1)}=0.088$, $\frac{0.272}{d(v_2)}=0.0907$, $\frac{0.08}{d(v_5)}=0.016$, all of which are less than $\epsilon=0.1$. \gdiff then terminates and returns the reserve values as the result. 

\input{tex/figs/residual}

\subsection{An Empirical Study of \gdiff}
Although \gdiff enjoys favorable theoretical properties in Theorem~\ref{lem:LP}, it suffers from slow convergence on real graphs due to its aggressive strategy in Eq.~\eqref{eq:greedy}. 
To exemplify, we evaluate the residual sum $\|\rvec\|_1$ at the end of each iteration in Algo.~\ref{alg:fwd} when adopting greedy (Lines 5-7) and {\em non-greedy} operations (Eq.~\eqref{eq:non-greedy}) on {\em PubMed} and {\em ArXiv} datasets (Table~\ref{tbl:exp-data}).
\begin{equation}\label{eq:non-greedy}
\begin{gathered}
\pvec \gets \pvec + (1-\alpha){\rvec},\ {\rvec} \gets \alpha{\rvec}\PM
\end{gathered}
\end{equation}
Distinct from the greedy way in Eq.~\eqref{eq:greedy}, non-greedy operations in Eq.~\eqref{eq:non-greedy} directly convert and diffuse the residuals of {\em all} nodes in one shot in each iteration.
From Fig.~\ref{fig:residual}, we can observe that Algo.~\ref{alg:fwd} using the greedy strategy needs $2\times$ more iterations to terminate and near $4\times$ more iterations to attain the same residual sum compared to its non-greedy variant on both datasets, leading to inferior empirical efficiency.
The reason is that \gdiff always attempts to sift out a small moiety of low-degree nodes
(Eq.~\eqref{eq:gamma_vec}) for residual conversion and diffusion in each iteration (Line 5), making it sensitive to high-degree nodes and leaving the bulk of residual untouched.
Such a way tends to trigger relentless residual accumulation and propagation among a minority of nodes, causing numerous iterations but fewer non-zero entries in $\pvec$.
As reported in Table~\ref{tbl:degree-compare-1}, on real datasets on \textit{ArXiv} and \textit{Yelp}, the average node degrees of local clusters output by \gdiff are notably lower than the average node degrees of the entire graphs and those by the non-greedy strategy.

In contrast, non-greedy operations transform $1-\alpha=20\%$ of residuals into reserves in each iteration, making $\|\rvec\|_1$ decrease rapidly after a few iterations, e.g., $\|\rvec\|_1=0.107$ after only $10$ iterations. 
Meanwhile, the residual is evenly distributed across more nodes, each with a small value, enabling early termination.  
However, due to its brute-force nature, the non-greedy strategy entails up to $O(m)$ cost in $\alpha{\rvec}\PM$ of each iteration in the worst case, especially on dense graphs.

\begin{algorithm}[!t]
\caption{\adiff}\label{alg:iter-fwd}
\KwIn{$\PM$, $\alpha$, $\sigma$, $\epsilon$, $\fvec$}
\KwOut{Diffused vector $\pvec$}
$\rvec \gets \fvec;\ \pvec \gets \mathbf{0};\ C_{tot}\gets 0$\;
\While{true}{
    Line 3 is the same as Line 3 in Algo. \ref{alg:fwd}\;
    \If{$\frac{|\mathsf{supp}(\rrvec)|}{|\mathsf{supp}(\rvec)|}> \sigma$ and $C_{tot} + \mathsf{vol}(\rvec)< \frac{\|\fvec\|_1}{(1-\alpha)\epsilon}$}{
        $C_{tot}\gets C_{tot} + \mathsf{vol}(\rvec)$\;
        Update $\pvec$ and $\rvec$\Comment*[r]{Eq.~\eqref{eq:non-greedy}}
    }\Else{
    {\nonl{Lines 8-11 are the same as Lines 4-7 in Algo.~\ref{alg:fwd}}\;
    \setcounter{AlgoLine}{11}}
    }
}
\Return{$\pvec$}\;
\end{algorithm}

\subsection{The \adiff Approach}\label{sec:adiff}

Inspired by the preceding analysis, we propose combining greedy and non-greedy operations in an adaptive way to overcome the limitations of both, ensuring fast termination and high locality.

The pseudo-code of this \adiff method is provided in Algo.~\ref{alg:iter-fwd}, which additionally requires inputting a parameter $\sigma\in [0,1]$ and a variable $C_{tot}$ tracking the total cost incurred by non-greedy diffusion operations. In contrast to Algo.~\ref{alg:fwd}, in each iteration, after calculating $\rrvec$ by Eq.~\eqref{eq:gamma_vec} at Line 3, Algo.~\ref{alg:iter-fwd} adaptively selects the greedy strategy (Lines 8-11) or non-greedy way (Lines 5-6) based on the following conditions. The rationale is to first deplete the residuals through non-greedy diffusion as much as possible (for faster convergence) and then disseminate the rest carefully with greedy operations (for rigorous guarantees). To be specific, \adiff conducts non-greedy diffusion operations when the fraction of nodes with residues above the threshold (Eq.~\eqref{eq:gamma_vec}), i.e., $\textstyle {|\mathsf{supp}(\rrvec)|}/{|\mathsf{supp}(\rvec)|}$, outstrips $\sigma$, and in the meantime, the total cost $C_{tot}$ after conducting such non-greedy operations, i.e., $C_{tot} + \mathsf{vol}(\rvec)$, is still less than the total cost using \gdiff, i.e., $\textstyle \frac{\|\fvec\|_1}{(1-\alpha)\epsilon}$. 
Notice that the smaller $\sigma$ is, the more non-greedy operations will be conducted. When $\sigma=0$, \adiff prioritize executing Lines 5-6 over Lines 8-11.
Once the non-greedy (Lines 5-6) strategy is chosen, $C_{tot}$ is increased by the volume of $\rvec$, i.e., the amount of work needed in computing $\alpha{\rvec}\PM$.
In turn, the same theoretical properties as \gdiff can be proved in the following lemma for \adiff:
\begin{theorem}\label{lem:acc-adaptive}
Algo.~\ref{alg:iter-fwd} outputs a vector $\pvec$ such that Eq.~\eqref{eq:f-q-eps} holds $\forall{v_t\in \V}$ using 
$\footnotesize O\left(\max\left\{\left|\mathsf{supp}(\fvec)\right|,\frac{\|\fvec\|_1}{(1-\alpha)\epsilon}\right\}\right)$ time.
\end{theorem}

\begin{table}[!t]
\centering
\caption{Average node degrees of local clusters ($\epsilon=10^{-7}$).}
\label{tbl:degree-compare-1}
\begin{small}
\addtolength{\tabcolsep}{-0.25em}
\begin{tabular}{l|c | c | c}
\hline
\multirow{1}{*}{\bf Dataset} & \multicolumn{1}{c|}{{\bf {Global avg. degree}}} & \multicolumn{1}{c|}{{\bf {Greedy}}} & \multicolumn{1}{c}{{\bf {Non-greedy}}} \\ \cline{2-4}
\hline
\textit{PubMed}  & 13.77 & 12.49 & 14.39 \\
\textit{Yelp} & 20.47 & 17.83 & 22.65 \\
\hline
\end{tabular}
\end{small}
\vspace{-3ex}
\end{table}

In addition, Lemma~\ref{lem:q-size-adaptive} states that the support size and the volume of the vector $\pvec$ returned by Algo.~\ref{alg:iter-fwd} is bounded, solely dependent on $\|\fvec\|_1$, $\alpha$, and $\epsilon$.
\begin{lemma}\label{lem:q-size-adaptive}
Let $\fvec$ and $\pvec$ be the input and output of Algo.~\ref{alg:iter-fwd}, respectively. Then, $|\mathsf{supp}(\pvec)|\le \mathsf{vol}(\pvec) \le \frac{\beta\|\fvec\|_1}{(1-\alpha)\epsilon}$, where $1\le \beta\le 2$. In particular, when $\sigma\ge 1$, $\beta=1$.
\end{lemma}

%% file: tex/figs/residual.tex
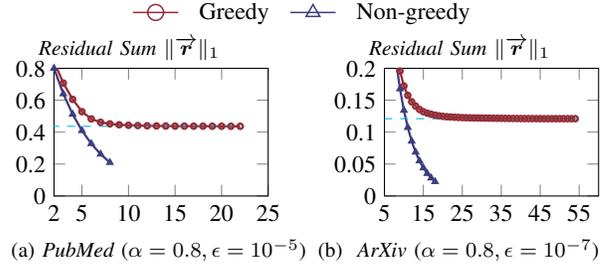
\begin{figure}[!t]
\centering
\begin{small}
\begin{tikzpicture}
    \begin{customlegend}[legend columns=2,
        legend entries={Greedy, Non-greedy},
        legend style={at={(0.45,1.35)},anchor=north,draw=none,font=\small,column sep=0.2cm}]
    \addlegendimage{line width=0.2mm,mark size=3pt,color=greedy-color,mark=o}
    \addlegendimage{line width=0.2mm,mark size=3pt,color=nongreedy-color,mark=triangle}
    \end{customlegend}
\end{tikzpicture}
\\[-\lineskip]
\vspace{-4mm}
\subfloat[{\em PubMed} ($\alpha=0.8, \epsilon=10^{-5}$)]{
\begin{tikzpicture}[scale=1,every mark/.append style={mark size=1pt}]
    \begin{axis}[
        height=\columnwidth/2.7,
        width=\columnwidth/2.0,
        ylabel={\it Residual Sum $\|\rvec\|_1$},
        xmin=2, xmax=25,
        ymin=0, ymax=0.8,
        xtick={2,5,10,15,20,25},
        xticklabel style = {font=\small},
        yticklabel style = {font=\small},
        xticklabels={2,5,10,15,20,25},
        ytick={0,0.2,0.4,0.6,0.8},
        yticklabels={0,0.2,0.4,0.6,0.8},
        every axis y label/.style={font=\footnotesize,at={(current axis.north west)},right=10mm,above=0mm},
        legend style={fill=none,font=\footnotesize,at={(0.02,0.99)},anchor=north west,draw=none},
    ]

        \addplot[line width=0.3mm, color=greedy-color,mark=o]  %
        plot coordinates {
(	1,	1	)
(	2,	0.841852209	)
(	3,	0.707469358	)
(	4,	0.605748576	)
(	5,	0.528580299	)
(	6,	0.483183928	)
(	7,	0.461900459	)
(	8,	0.451947432	)
(	9,	0.446183203	)
(	10,	0.443152421	)
(	11,	0.441279224	)
(	12,	0.439821007	)
(	13,	0.438999498	)
(	14,	0.438293929	)
(	15,	0.437847657	)
(	16,	0.437552854	)
(	17,	0.437442592	)
(	18,	0.437143131	)
(	19,	0.437002083	)
(	20,	0.436898672	)
(	21,	0.436818015	)
(	22,	0.436808632	)
    };

        \addplot[line width=0.3mm, color=nongreedy-color,mark=triangle]  %
        plot coordinates {
(	1,	1	)
(	2,	0.8	)
(	3,	0.64	)
(	4,	0.512	)
(	5,	0.4096	)
(	6,	0.32768	)
(	7,	0.262144	)
(	8,	0.2097152	)
    };

\addplot[cyan,dashed,domain=0:22, samples=2] {0.436808632};
    \end{axis}
\end{tikzpicture}\hspace{4mm}\label{fig:alpha-email}%
}%
\subfloat[ {\em ArXiv} ($\alpha=0.8, \epsilon=10^{-7}$)]{
\begin{tikzpicture}[scale=1,every mark/.append style={mark size=1pt}]
    \begin{axis}[
        height=\columnwidth/2.7,
        width=\columnwidth/2.0,
        ylabel={\it Residual Sum $\|\rvec\|_1$},
        xmin=5, xmax=60,
        ymin=0.00, ymax=0.2,
        xtick={5,15,25,35,45,55},
        xticklabel style = {font=\small},
        yticklabel style = {font=\small},
        xticklabels={5,15,25,35,45,55},
        ytick={0, 0.05, 0.1, 0.15, 0.2},
        yticklabels={0, 0.05, 0.1, 0.15, 0.2},
        yticklabel style={/pgf/number format/fixed},
        scaled y ticks=false,
        every axis y label/.style={font=\footnotesize,at={(current axis.north west)},right=10mm,above=0mm},
        legend style={fill=none,font=\footnotesize,at={(0.02,0.99)},anchor=north west,draw=none},
    ]

        \addplot[line width=0.3mm, color=greedy-color,mark=o]  %
        plot coordinates {
(	1,	1	)
(	2,	0.804914286	)
(	3,	0.64880889	)
(	4,	0.525592776	)
(	5,	0.426888164	)
(	6,	0.348694724	)
(	7,	0.28562072	)
(	8,	0.234605538	)
(	9,	0.195793875	)
(	10,	0.172300548	)
(	11,	0.157608224	)
(	12,	0.147563016	)
(	13,	0.14088269	)
(	14,	0.135568283	)
(	15,	0.131993203	)
(	16,	0.129706793	)
(	17,	0.128026415	)
(	18,	0.126686867	)
(	19,	0.125567909	)
(	20,	0.124742945	)
(	21,	0.124153466	)
(	22,	0.12370902	)
(	23,	0.123339853	)
(	24,	0.123040003	)
(	25,	0.122795751	)
(	26,	0.122595786	)
(	27,	0.122439449	)
(	28,	0.122308411	)
(	29,	0.122179745	)
(	30,	0.122068874	)
(	31,	0.121975409	)
(	32,	0.121883891	)
(	33,	0.12178478	)
(	34,	0.121709325	)
(	35,	0.121627024	)
(	36,	0.12155611	)
(	37,	0.121476635	)
(	38,	0.121415865	)
(	39,	0.121357948	)
(	40,	0.121312445	)
(	41,	0.121267761	)
(	42,	0.12124067	)
(	43,	0.121212758	)
(	44,	0.121189839	)
(	45,	0.121168568	)
(	46,	0.12115315	)
(	47,	0.12114395	)
(	48,	0.12113594	)
(	49,	0.121126204	)
(	50,	0.121121745	)
(	51,	0.121118481	)
(	52,	0.121115279	)
(	53,	0.121113612	)
(	54,	0.121113033	)
    };

        \addplot[line width=0.3mm, color=nongreedy-color,mark=triangle]  %
        plot coordinates {
(	1,	1	)
(	2,	0.8	)
(	3,	0.64	)
(	4,	0.512	)
(	5,	0.4096	)
(	6,	0.32768	)
(	7,	0.262144	)
(	8,	0.2097152	)
(	9,	0.16777216	)
(	10,	0.134217728	)
(	11,	0.107374182	)
(	12,	0.085899346	)
(	13,	0.068719477	)
(	14,	0.054975581	)
(	15,	0.043980465	)
(	16,	0.035184372	)
(	17,	0.028147498	)
(	18,	0.022517998	)
    };

\addplot[cyan,dashed,domain=5:55, samples=2] {0.121113033};
    \end{axis}
\end{tikzpicture}\hspace{0mm}\label{fig:alpha-fb}%
}%
\end{small}
 \vspace{-1ex}
\caption{Greedy v.s. Non-greedy.} \label{fig:residual}
\vspace{-3ex}
\end{figure}

%% file: tex/solution.tex
\section{The \algo Approach}\label{sec:complete}
This section presents our local algorithm \algo for the estimation of the BDD vector for LGC.
We first elucidate the algorithmic details of Algo.~\ref{alg:Z} for constructing TNAM $\ZM$.
The complete algorithmic details of \algo (Algo.~\ref{alg:main}) and related analyses are provided in Section \ref{sec:main-algo}. 
Lastly, we conduct an in-depth theoretical analysis to unveil the connection between \algo and GNNs~\cite{hamilton2017inductive}.

\subsection{Construction of TNAM $\ZM$}\label{sec:construct-z}

\begin{algorithm}[!t]
\caption{\texttt{TNAM Construction}}\label{alg:Z}
\KwIn{Attribute matrix $\XM$, function $f(\cdot,\cdot)$, and dimension $k$}
\KwOut{The TNAM $\ZM$}
$\UM, \boldsymbol{\Lambda}, \VM \gets k\textrm{-}\mathsf{SVD}(\XM)$\;
\Switch{$f(\cdot,\cdot)$}{
\Case{cosine similarity function}{
    $\YM\gets \UM{\boldsymbol{\Lambda}}$\;
}\Case{exponential cosine similarity function}{
    Sample a Gaussian matrix $\GM \sim \mathcal{N}(0,1)^{k\times k}$\;
    $\QM\gets \mathsf{QRDecomposition}(\GM)$\;
    Sample diagonal matrix $\boldsymbol{\Sigma}_{ii} \sim \chi(k)$ $\forall{i} \in \{1, \ldots, k\}$\;
    Compute $\YM$\Comment*[r]{Eq.~\eqref{eq:Y-ecos}}
}
}
Compute ${\yvec}^{\ast}$\Comment*[r]{Eq.~\eqref{eq:y-mean-z}}
\lFor{$v_i\in \V$}{Compute $\zvec^{(i)}$;\Comment*[f]{Eq.~\eqref{eq:y-mean-z}}\DontPrintSemicolon}
\Return{$\ZM$}\;
\end{algorithm}

\stitle{{Basic Idea}}
To realize the idea of transforming $s(v_i,v_j)$ into $\zvec^{(i)}\cdot \zvec^{(j)}$ (Eq.~\eqref{eq:s-zz}), the key is to find length-$k$ vectors $\yvec^{(i)}\ \forall{v_i\in \V}$ (i.e., an $n\times k$ matrix $\YM$) such that $f(v_i,v_j)=\yvec^{(i)}\cdot \yvec^{(j)}$. Accordingly, Eq.~\eqref{eq:SNAS} can be rewritten as
\begin{small}
\begin{equation*}
\textstyle s(v_i,v_j) = \frac{\yvec^{(i)}\cdot \yvec^{(j)}}{\sqrt{\yvec^{(i)} \cdot {\yvec}^{\ast}} \cdot \sqrt{\yvec^{(j)} \cdot {\yvec}^{\ast}} } = \zvec^{(i)}\cdot \zvec^{(j)},\ \text{where}
\end{equation*}
\end{small}
\begin{small}
\begin{equation}\label{eq:y-mean-z}
\textstyle {\yvec}^{\ast}=\sum_{v_\ell\in \V}{\yvec^{(\ell)}}\ \text{and}\ \zvec^{(i)}={\yvec^{(i)}}/{\sqrt{\yvec^{(i)} \cdot {\yvec}^{\ast}}}.
\end{equation}
\end{small}
Recall that the SNAS metrics in Section~\ref{sec:SNAS} are defined upon the dot product $\xvec^{(i)}\cdot \xvec^{(j)}\ \forall{v_i,v_j\in \V}$. Let $\UM\in \mathbb{R}^{n\times k}$ and diagonal matrix $\boldsymbol{\Lambda}\in \mathbb{R}^{k\times k}$ consist of the top-$k$ left singular vectors and top-$k$ singular values of $\XM$, respectively. Lemma~\ref{lem:SVD} connotes that $\UM\boldsymbol{\Lambda}$ can be used as the $k$-dimensional approximation of $\XM$ for the construction of $\YM$.
\begin{lemma}\label{lem:SVD}
Let $\lambda_{k+1}$ be the $(k+1)$-th largest singular value of $\XM$. Then, $\textstyle \left\|\left(\UM{\boldsymbol{\Lambda}}\right)\cdot \left(\UM{\boldsymbol{\Lambda}}\right)^{\top}-\XM\XM^{\top}\right\|_2 \le {\lambda_{k+1}}^2$.
\end{lemma}

\stitle{{Details}}
In Algo.~\ref{alg:Z}, we describe the pseudo-code for constructing vectors $\yvec^{(i)}$ and $\zvec^{(i)}$ for each node $v_i\in \V$ based on the input node attribute matrix $\XM$, the metric functions $f(\cdot,\cdot)$ in Section \ref{sec:SNAS}, and a small integer $k\ll d$ (typically $32$) to cope with the high-dimension $d$ of $\XM$.
That is, %
Algo.~\ref{alg:Z} first applies
a {\em $k$-truncated singular value decomposition} ($k$-SVD)~\cite{halko2011finding} over $\XM$ to obtain its top-$k$ left and right singular vectors $\UM$, $\VM$, and the diagonal singular value matrix $\boldsymbol{\Lambda}$ (Line 1).
$\UM{\boldsymbol{\Lambda}}$ then substitutes $\XM$ for subsequent generation of vectors $\yvec^{(i)}\ \forall{v_i\in \V}$.
When $f(\cdot,\cdot)$ is the cosine similarity function, it is straightforward to get $\YM=\UM{\boldsymbol{\Lambda}}$ (Lines 3-4).

However, when $f(\cdot,\cdot)$ is the exponential cosine similarity function (Eq.~\eqref{eq:exp-cos}), constructing $\yvec^{(i)}$ exactly involves the materialization of $f(v_i,v_j)$ for all node pairs in $\V\times\V$ and a matrix factorization, which is prohibitive for large graphs. 
As a workaround, we capitalize on the {\em orthogonal random features}~\cite{yu2016orthogonal} to create estimators $\yvec^{(i)}\ \forall{v_i\in \V}$ such that $\yvec^{(i)}\cdot \yvec^{(j)}\approx f(v_i,v_j)\ \forall{v_i,v_j\in \V}$.

More concretely, Algo.~\ref{alg:Z} first randomly generates a $k\times k$ random Gaussian matrix $\GM$ with every entry sampled from the standard normal distribution independently at Line 6, followed by a QR decomposition of $\GM$ at Line 7. This step produces a uniformly distributed random orthogonal matrix $\QM \in \mathbb{R}^{k\times k}$~\cite{muirhead2009aspects}. Algo.~\ref{alg:Z} further builds a $k\times k$ diagonal matrix $\boldsymbol{\Sigma}$ with diagonal entries sampled i.i.d. from the $\chi$-distribution with $k$ degrees of freedom (Line 8), enforcing the norms of the rows of $\boldsymbol{\Sigma}\QM$ and $\GM$ identically distributed. Based thereon, we construct matrix $\YM$ at Line 9 as follows:
\begin{small}
\begin{equation}\label{eq:Y-ecos}
\textstyle \YM\gets \sqrt{\frac{2\exp\left({1}/{\delta}\right)}{k}}\cdot sin(\widehat{\YM}) \mathbin\Vert cos(\widehat{\YM}),
\end{equation}
\end{small}
where $\textstyle \widehat{\YM}\gets \frac{1}{\delta}\UM{\boldsymbol{\Lambda}}\boldsymbol{\Sigma}\QM$ and $\mathbin\Vert$ stands for a horizontal concatenation of two matrices. Theorem~\ref{lem:yy-exp-cos} indicates that $\yvec^{(i)}\cdot \yvec^{(j)}$ is an {\em unbiased} estimator of $f(v_i,v_j)$ for any node pair $(v_i,v_j)\in \V\times \V$.
\begin{theorem}\label{lem:yy-exp-cos}
$\mathbb{E}\left[\yvec^{(i)}\cdot \yvec^{(j)}\right] = f(v_i,v_j)$ in Eq.~\eqref{eq:exp-cos}.
\end{theorem}

After computing vector $\yvec^{(i)}$ for each node $v_i\in \V$, Algo.~\ref{alg:Z} first computes the sum of these vectors, i.e., ${\yvec}^{\ast}$, and finally constructs $\zvec^{(i)}$ for each node $v_i\in \V$ by Eq.~\eqref{eq:y-mean-z} (Lines 10-11). The total processing cost entailed by Algo.~\ref{alg:Z} is linear to the size of the input node attribute matrix $\XM$, as proved in the following lemma:
\begin{lemma}\label{lem:Z-cost}
The runtime cost of Algo.~\ref{alg:Z} is $O(nd)$.
\end{lemma}

\subsection{Complete Algorithm and Analysis}\label{sec:main-algo}

\begin{algorithm}[!t]
\caption{\algo}\label{alg:main}
\KwIn{$\G=(\V,\EDG)$, TNAM $\ZM$, seed node $v_s$, restart factor $\alpha$, parameter $\sigma$, diffusion threshold $\epsilon$}
\KwOut{Approximate \ppr vector $\rhovec^{\prime}$}
\tcc{Step 1: Estimate RWR vector $\pivec^{\prime}$}
Create a unit vector $\evec^{(s)} \in \mathbb{R}^n$\;
$\pivec^{\prime}\gets \adiff{}(\PM, \alpha, \sigma, \epsilon, \evec^{(s)})$\;
\tcc{Step 2: Compute RWR-SNAS vector $\qvec^{\prime}$}
Compute $\psivec$\Comment*[r]{Eq.~\eqref{eq:pi-Z}}
\lFor{$i\in \mathsf{supp}(\pivec^{\prime})$}{Compute $\qvec^{\prime}_i$;\Comment*[f]{Eq.~\eqref{eq:sigma-vec-3}}\DontPrintSemicolon}
\tcc{Step 3: Estimate BDD vector $\rhovec^{\prime}$}
$\rhovec^{\prime} \gets \adiff{}(\PM, \alpha, \sigma, {\epsilon\cdot \|\qvec^{\prime}\|_1}, \qvec^{\prime})$\;
\lFor{$i\in \mathsf{supp}(\rhovec^{\prime})$}{
$\rhovec^{\prime}_i \gets \frac{\rhovec^{\prime}_i}{d(v_i)}$
}
\Return{$\rhovec^{\prime}$}\;
\end{algorithm}

In Algo.~\ref{alg:main}, we present the complete pseudo-code of \algo, which takes as input the attributed graph $\G$, the TNAM $\ZM$ obtained in the preprocessing stage, seed node $v_s$, diffusion threshold $\epsilon$, and parameters $\alpha$ and $\sigma$. In the first place, Algo.~\ref{alg:main} invokes \adiff{} (Algo.~\ref{alg:iter-fwd}) with a unit vector $\evec^{(s)}$ as input, which has value $1$ at entry $s$ and $0$ everywhere else (Lines 1-2). 
By Lemma~\ref{lem:q-size-adaptive}, the support size $|\mathsf{supp}(\pivec^{\prime})|$ of the returned RWR vector $\pivec^{\prime}$ is bounded by $O\left(\frac{1}{(1-\alpha)\epsilon}\right)$. Next, $\pivec^{\prime}$ is used for producing vector $\psivec \gets \pivec^{\prime} \cdot \ZM$ at Line 3 and subsequently the RWR-SNAS vector $\qvec^{\prime}$ at Line 4. In particular, in lieu of computing $\qvec^{\prime}_i$ for each node $v_i\in \V$ by Eq.~\eqref{eq:sigma-vec-3}, \algo merely accounts for nodes with non-zero entries in vector $\pivec^{\prime}$, i.e., $i\in \mathsf{supp}(\rhovec^{\prime})$, whereby the number of non-zero entries in $\qvec^{\prime}$ can be guaranteed to be bounded by $O\left(\frac{1}{(1-\alpha)\epsilon}\right)$. After that, \algo starts to diffuse the RWR-SNAS vector $\qvec^{\prime}$ over graph $\G$ using the \adiff with diffusion threshold $\epsilon\cdot \|\qvec^{\prime}\|_1$, and parameters $\alpha$ and $\sigma$ (Line 5). Let $\rhovec^{\prime}$ be the output of the above diffusion process. 
\algo then gives $\rhovec^{\prime}$ a final touch by dividing each non-zero entry $\rhovec^{\prime}_i$ in $\rhovec^{\prime}$ by $\frac{1}{d(v_i)}$ (Line 6) and returns $\rhovec^{\prime}$ as the approximate BDD vector at Line 7.
On the basis of Theorem~\ref{lem:acc-adaptive}, we can establish the accuracy guarantee of \algo as follows: 
\begin{theorem}\label{lem:main-acc}
When the TNAM $\ZM$ and SNAS $s(v_i,v_j)\ \forall{v_i,v_j\in \V}$ satisfy Eq.~\eqref{eq:s-zz}, $\rhovec^{\prime}$ output by Algo.~\ref{alg:main} ensures $\forall{v_t\in \V}$
\begin{footnotesize}
\begin{equation*}
 0 \le \rhovec_t-\rhovec^{\prime}_t \le  \left(1+\sum_{v_i\in \V}{d(v_i)\cdot \max_{v_j\in \V}{s(v_i,v_j)}}\right)\cdot \epsilon.
\end{equation*}
\end{footnotesize}
\end{theorem}

\stitle{{Volume and Complexity Analysis}}
Recall that $\rhovec^{\prime}$ is obtained by calling \adiff with $\fvec=\qvec^{\prime}$ and diffusion threshold $\epsilon\cdot \|\qvec^{\prime}\|_1$. Both the support size $|\mathsf{supp}(\rhovec^{\prime})|$ and volume $\mathsf{vol}(\rhovec^{\prime})$ of $\rhovec^{\prime}$ are therefore bounded by $\textstyle O\left(\frac{1}{(1-\alpha)\epsilon}\right)$ using Lemma~\ref{lem:q-size-adaptive}.

Next, we analyze the time complexity of \algo. First, Line 2 invokes Algo. \ref{alg:fwd} with a one-hot vector $\evec^{(s)}$, i.e., $\|\evec^{(s)}\|_1=1$, entailing $\textstyle O\left(\frac{1}{(1-\alpha)\epsilon}\right)$ time as per Theorem~\ref{lem:acc-adaptive}.
The cost of Lines 3 and 4 is dependent on the number of non-zero elements in vector $\pivec^{\prime}$, i.e., $|\mathsf{supp}(\pivec^{\prime})|$, and the dimension $k$ of $\ZM$, which is $\textstyle O\left(\frac{k}{(1-\alpha)\epsilon}\right)$ time by Lemma~\ref{lem:q-size-adaptive}.
Analogously, we can derive that the computational complexities of Lines 6-7 are $\textstyle O\left(\max\left\{\left|\mathsf{supp}(\qvec^{\prime})\right|,\frac{\|\qvec^{\prime}\|_1}{(1-\alpha)\epsilon\cdot \|\qvec^{\prime}\|_1}\right\}\right)=O\left(\frac{1}{(1-\alpha)\epsilon}\right)$. Overall, the time complexity of \algo is $\footnotesize O\left(\frac{k}{(1-\alpha)\epsilon}\right)$, which equals $O\left({1}/{\epsilon}\right)$ when $\alpha$ and $k$ are regarded as constants and is linear to the volume of its output $\rhovec^{\prime}$.

\subsection{Theoretical Connection to GNNs}\label{sec:connect-GNN}
Recent studies \cite{ma2021unified,zhu2021interpreting} demystify that learning node representations $\HM$ via existing canonical GNN architectures can be characterized by a graph smoothing process in Definition~\ref{def:lap-smooth}.
\begin{definition}[Graph Signal Denoising\textnormal{~\cite{ma2021unified}}]\label{def:lap-smooth}
Let $\LM$ be the normalized Laplacian matrix of $\G$ and $\HM^{\circ}\in \mathbb{R}^{n\times k}$ be a feature matrix. The graph signal denoising is to optimize $\HM$:
\begin{equation}\label{eq:GNN-obj}
\arg\min_{\HM}{(1-\alpha)\|\HM-\HM^{\circ}\|^2_F+\alpha\cdot trace(\HM^{\top}\LM\HM)},
\end{equation}
where $\|\cdot\|_F$ stands for the matrix Frobenius norm.
\end{definition}
The fitting term $\|\HM-\HM^{\circ}\|^2_F$ in Eq.~\eqref{eq:GNN-obj} seeks to make the final node representations $\HM$ close to the initial feature matrix $\HM^{\circ}$, while the graph Laplacian regularization term $trace(\HM^{\top}\LM\HM)$ forces learned representations of two adjacent nodes over $\G$ to be similar. The hyperparameter $\alpha\in [0,1]$ controls the smoothness of $\HM$ through graph regularization. 
\begin{lemma}\label{lem:GNN-sol}
The closed-form solution to Eq.~\eqref{eq:GNN-obj} is $\HM = \sum_{\ell=0}^{\infty}(1-\alpha)\alpha^\ell \NAM^{\ell}\HM^{\circ}$, where $\NAM=\DM^{-\frac{1}{2}}\AM\DM^{-\frac{1}{2}}$.
\end{lemma}
By applying the gradient descent to solve Eq. \eqref{eq:GNN-obj}, Lemma~\ref{lem:GNN-sol} states that the final representation $\hvec^{(i)}$ of any node $v_i\in \V$ can be formulated as $\hvec^{(i)} = \sum_{\ell=0}^{\infty}{(1-\alpha)\alpha^\ell \NAM^{\ell}}{\hvec^{\circ}}^{(i)}$, where the normalized adjacency matrix $\NAM$ can also be replaced by the transition matrix $\PM$ in popular GNN models~\cite{bojchevski2020scaling}.

If we let TNAM $\ZM$ be the initial feature matrix $\HM^{\circ}$ input to GNN models, the eventual smoothed node representations (a.k.a. embeddings) are $\HM=\sum_{\ell=0}^{\infty}{(1-\alpha)\alpha^\ell \PM^\ell}\ZM$,
When Eq.~\eqref{eq:s-zz} holds, combining Eq.~\eqref{eq:csim} and Eq.~\eqref{eq:pi-matrix} leads to $\forall{v_t\in \V}$, $\rhovec_t = \hvec^{(s)}\cdot \hvec^{(t)}$,
implying that $\rhovec^{\prime}$ output by \algo essentially approximates $\hvec^{(s)}\cdot\HM^{\top}$. In this view, our LGC task that extracts a local cluster $\C_s$ from $\G$ based on BDD values is equivalent to identifying the $K$-NN ($K=|\C_s|$) of $\hvec^{(s)}$ among $n$ GNN-like embeddings $\{\hvec^{(t)}|v_i\in \V\}$. 
Distinctly, our \algo approach fulfills this goal without explicitly materializing the GNN-like embeddings $\HM$ and incurring the $\tilde{O}(n)$ cost by the $K$-NN search, but undergoes a local exploration of $\G$ in time linear to the volume of $\C_s$, regardless of $n$ and $m$.

%% file: tex/experiments.tex
\section{Experiments} \label{sec:exp}
This section experimentally evaluates our proposed \algo against 17 alternative solutions to LGC on 8 real datasets, in terms of both local clustering quality and efficiency. All experiments are conducted on a Linux machine powered by Intel Xeon(R) Gold 6330 2.00GHz CPUs and 2TB memory. Due to space limits, additional experimental results regarding the parameter analysis, ablation study, scalability tests, and the LGC quality of \algo on non-attributed graphs are deferred to Appendix~\ref{sec:add-exp}. For reproducibility, the source code, datasets, and detailed parameter settings are available at \url{https://github.com/HaoranZ99/laca}.

\begin{table}[!t]
\centering
\renewcommand{\arraystretch}{1.0}
\begin{footnotesize}
\caption{Statistics of Datasets.}\label{tbl:exp-data}
\vspace{-1mm}
\resizebox{\columnwidth}{!}{%
\begin{tabular}{l|r|r|r|c|c}
	\hline
	{\bf Dataset} & \multicolumn{1}{c|}{$\boldsymbol n$ } & \multicolumn{1}{c|}{$\boldsymbol m$ } & \multicolumn{1}{c|}{$m/n$ } & \multicolumn{1}{c|}{$\boldsymbol d$} & {\bf $\overline{|\Y_s|}$} \\
	\hline
    {\em Cora}~\cite{sen2008collective} & 2,708 & 5,429 & 2.01 & 1,433 & 488 \\
    {\em PubMed}~\cite{sen2008collective} & 19,717 & 44,338 & 2.25 & 500 & 7,026 \\
    {\em BlogCL}~\cite{tang2009relational} & 5,196 & 343,486 & 66.11 & 8,189 & 869 \\
    {\em Flickr}~\cite{huang2017label} & 7,575 & 479,476 & 63.30 & 12,047 & 846 \\
    {\em ArXiv}~\cite{hu2020open} & 169,343 & 1,166,243 & 6.89 & 128 & 12,828 \\
    {\em Yelp}~\cite{zeng2020graphsaint} & 716,847 & 7,335,833 & 10.23 & 300 & 476,555 \\
    {\em Reddit}~\cite{zeng2020graphsaint} & 232,965 & 11,606,919 & 49.82 & 602 & 9,418 \\
    {\em Amazon2M}~\cite{chiang2019cluster} & 2,449,029 & 61,859,140 & 25.26 & 100 & 260,129 \\
    \hline
\end{tabular}%
}
\end{footnotesize}
\vspace{-3ex}
\end{table}

\subsection{Experimental Setup}
\stitle{{Datasets}}
Table \ref{tbl:exp-data} lists the statistics of the datasets used in the experiments. The numbers of nodes, edges, and distinct attributes of the graph data are denoted as $n$, $m$, and $d$, respectively. $\overline{|\Y_s|}$ stands for the average size of the ground-truth local clusters of all nodes in the graph. \textit{Cora}, \textit{PubMed}~\cite{sen2008collective}, and \textit{ArXiv}~\cite{hu2020open} are citation networks, where nodes and edges represent publications and citation links among them, respectively. The attributes of each node are bag-of-words embeddings of the corresponding publication. The ground-truth local cluster $\Y_s$ of each publication contains the publications in its same subject areas.
\textit{BlogCL}~\cite{tang2009relational} and \textit{Flickr}~\cite{huang2017label} are social networks extracted from the BlogCatalog and Flickr websites, respectively.
$\Y_s$ of each user $v_s\in \V$ includes users who are in the same topic categories or interest groups.
\textit{Yelp} and \textit{Reddit} datasets are collected from in \cite{zeng2020graphsaint}. {\em Yelp} contains friendships between Yelp users, those who have been to the same types of business constitute local clusters. {\em Reddit} connects online posts if the same user comments on both and the communities of posts are used as local clusters.
\textit{Amazon2M}~\cite{chiang2019cluster} is a co-purchasing network of Amazon products wherein each node corresponds to a product and each edge represents that two products are purchased together. 
The ground-truth local clusters are generated based on the categories of products.

\begin{table}[t]
\centering
\renewcommand{\arraystretch}{1.3}
\begin{footnotesize}
\caption{Evaluated methods.}\label{tbl:exp-algo}
\vspace{-1mm}
\resizebox{\columnwidth}{!}{%
\begin{tabular}{l|p{1cm}|c|c}
	\hline
	{\bf Method} & \multicolumn{1}{c|}{\bf Category} & \multicolumn{1}{c|}{\bf Preprocessing Cost} & \multicolumn{1}{c}{\bf Online Cost} \\
	\hline
\texttt{PR-Nibble}~\cite{andersen2006local}	  & \multirow{6}{1cm}{\centering Local Graph Clustering}  & - & \multirow{2}{2cm}{\centering $\textstyle \tilde{O}\left(\frac{1}{\epsilon}\right)$}     \\\cline{1-1}\cline{3-3}
\texttt{APR-Nibble}	 &   & \centering $O(md)$ &      \\\cline{1-1}\cline{3-4}
\texttt{HK-Relax}~\cite{kloster2014heat}	 &   & \multirow{3}{1cm}{\centering -} & $\textstyle \tilde{O}\left(\frac{\log\left(1 / \epsilon\right)}{\epsilon}\right)$  \\\cline{1-1}\cline{4-4}
\texttt{CRD}~\cite{wang2017capacity}	  &   &   &  $\textstyle O\left(\frac{1}{\epsilon}\right)$  \\\cline{1-1}\cline{4-4}
\texttt{$p$-Norm FD}~\cite{fountoulakis2020p}  &   &  &  \multirow{2}{2cm}{\centering $\textstyle O\left(\frac{\underset{{v_i\in \V}}{\max}{d(v_i)^2}}{\epsilon}\right)$}     \\\cline{1-1}\cline{3-3}
\texttt{WFD}~\cite{pmlr-v202-yang23d}  &   & \centering $O(md)$ &      \\\cline{1-1}
	\hline
\texttt{Jaccard}~\cite{liben2003link}	 & \multirow{4}{1cm}{\centering Link Similarity}  & \multirow{4}{1cm}{\centering -} & \multirow{4}{1cm}{\centering $\tilde{O}(n)$} \\\cline{1-1}
\texttt{Adamic-Adar}~\cite{liben2003link}	 &   &  & \\\cline{1-1}
\texttt{Common-Nbrs}~\cite{liben2003link}	 &   &   & \\\cline{1-1}
\texttt{SimRank}~\cite{jeh2002simrank}	 &   &   & \\\cline{1-1}
	\hline
\texttt{SimAttr}~\cite{yin2010unified,rahutomo2012semantic} & \multirow{2}{1cm}{\centering Attribute Similarity} & - & $\tilde{O}(nd)$ \\\cline{1-1}\cline{3-4}
\texttt{AttriRank}~\cite{hsu2017unsupervised} &  & $O(nd^2+m)$ & $\tilde{O}(n)$ \\\cline{1-1}
	\hline
\texttt{Node2Vec}~\cite{grover2016node2vec} & \multirow{4}{1cm}{\centering Node Embedding} & $O(n)$ & \multirow{4}{2cm}{\centering $\tilde{O}(n)$} \\\cline{1-1}\cline{3-3}
\texttt{SAGE}~\cite{hamilton2017inductive} &  & $O(nd^2)$ & \\\cline{1-1}\cline{3-3}
\texttt{PANE}~\cite{yang2020scaling,yang2023pane} &  & $\textstyle O((m+n)\cdot d)$ & \\\cline{1-1}\cline{3-3}
\texttt{CFANE}~\cite{pan2021unsupervised} &  & $O((m+n)\cdot d)$ & \\\cline{1-1}
	\hline
\algo & \multirow{1}{1cm}{\centering Ours} & $O(nd)$ & $\textstyle \tilde{O}\left(\frac{1}{\epsilon}\right)$ \\ \cline{1-1}
\hline
\end{tabular}%
}
\end{footnotesize}
\vspace{-4ex}
\end{table}

\stitle{{Competitors}}
We dub our \algo algorithms using metric functions cosine similarity and exponential cosine similarity as \algoc and \algoe, respectively.
We experimentally compare \algoc and \algoe against 17 methods adopted for LGC, which can be categorized into four groups:
\begin{enumerate}[leftmargin=*]
\item LGC-based methods: \texttt{PR-Nibble}~\cite{andersen2006local}, \texttt{APR-Nibble}, \texttt{HK-Relax}~\cite{kloster2014heat}, \texttt{CRD}~\cite{wang2017capacity}, \texttt{$p$-Norm FD}~\cite{fountoulakis2020p}, and \texttt{WFD}~\cite{pmlr-v202-yang23d};
\item Link Similarity-based methods: \texttt{Jaccard}~\cite{liben2003link}, \texttt{Adamic-Adar}~\cite{liben2003link}, \texttt{Common-Nbrs}~\cite{liben2003link}, and \texttt{SimRank}~\cite{jeh2002simrank};
\item Attribute Similarity-based methods: \texttt{SimAttr (C)}~\cite{yin2010unified}, \texttt{SimAttr (E)}~\cite{rahutomo2012semantic}, and \texttt{AttriRank}~\cite{hsu2017unsupervised};
\item Network Embedding-based methods: \texttt{Node2Vec}~\cite{grover2016node2vec}, \texttt{SAGE}~\cite{hamilton2017inductive}, \texttt{PANE}~\cite{yang2020scaling,yang2023pane}, and \texttt{CFANE}~\cite{pan2021unsupervised}.
\end{enumerate}

\input{tex/figs/precision}

Amid LGC-based approaches, \texttt{PR-Nibble}~\cite{andersen2006local}, \texttt{APR-Nibble}, and \texttt{HK-Relax}~\cite{kloster2014heat} are based on random walk graph diffusion, while \texttt{CRD}~\cite{wang2017capacity}, \texttt{$p$-Norm FD}~\cite{fountoulakis2020p}, and \texttt{WFD}~\cite{pmlr-v202-yang23d} leverages maximum flow algorithms, all of which are {\em local} algorithms.
\texttt{APR-Nibble} is a variant of \texttt{PR-Nibble} wherein edges are weighted by the Gaussian kernel of their endpoints' attribute vectors, similar to \texttt{WFD}~\cite{pmlr-v202-yang23d}. Groups 2)-4) comprise all {\em global} methods. Link similarity-based and attribute similarity-based methods calculate the link-based and attribute-based similarities between the seed node and {\em all} nodes, respectively. The local clusters are then generated by sorting all nodes according to the similarity scores.
The fourth category of methods first encodes all nodes in the input graph into low-dimensional embedding vectors and then obtains the local clusters for given seed nodes through the $K$-NN, spectral clustering (SC), or DBSCAN over the embedding vectors. Particularly, \texttt{SAGE}~\cite{hamilton2017inductive}, \texttt{PANE}~\cite{yang2023pane}, and \texttt{CFANE}~\cite{pan2021unsupervised} incorporate both topology and attribute semantics into the embeddings, whereas \texttt{Node2Vec}~\cite{grover2016node2vec} disregard nodal attributes.
Table~\ref{tbl:exp-algo} summarizes the preprocessing cost and average complexity of these algorithms for generating local clusters.

\stitle{{Implementations and Parameter Settings}} 
For \texttt{PR-Nibble}, \texttt{CRD}, and \texttt{APR-Nibble}, we use their implementations provided by~\cite{fountoulakis2018short}. We also employ the well-known NetworkX \cite{hagberg2008exploring} package for the computation of link similarities of nodes, including \texttt{Jaccard}, \texttt{Adamic-Adar}, \texttt{Common-Nbrs}, and \texttt{SimRank}.
As for other competitors, we obtain their source codes from the respective authors. All competitors are implemented in Python, except \texttt{$p$-Norm FD} and \texttt{WFD}, which have been implemented in Julia.
For a fair comparison, we run grid searches for parameters and report the results corresponding to the best precision for LGC-based methods, \algoc, and \algoe.
The parameters in global methods
are set as suggested in their respective papers.
On each dataset, we randomly select 500 seed nodes $\mathcal{S}$ from the graph for LGC tasks. All evaluation results reported in the experiments are averaged over seed nodes in $\mathcal{S}$.

\subsection{Quality Evaluation}\label{sec:quality}
\subsubsection{\bf {Precision}}
In this set of experiments, we empirically evaluate the average precisions of the local clusters returned by \algoc, \algoe, and the 17 competitors in four categories, respectively, based on the ground-truth local clusters. More concretely, for each seed node $v_s$ in $\mathcal{S}$, we run all the evaluated methods such that the predicted local cluster $\C_s$ satisfies $|\C_s|=|\Y_s|$ and calculate the precision as ${|\C_s\cap \Y_s|}/{|\C_s|}$. Table~\ref{tbl:node-clustering} reports the average precision scores achieved by all the evaluated approaches on 8 datasets. 
The best results among all methods are highlighted in bold, and the best performance by the competitors is underlined.
{We exclude any method with a preprocessing time exceeding 3 days or an average running time for LGC over 2 hours.}

\input{tex/figs/recall_new}

\input{tex/figs/running_time}

Overall, our methods \algoc and \algoe achieve the top-$2$ best average ranks in terms of precision on all datasets.
Specifically, on small datasets {\em Cora}, {\em PubMed}, {\em BlogCL}, and {\em Flickr}, \algoc and \algoe consistently outperform all the competitors, often by a large margin. For instance, on {\em Cora} and {\em Flickr}, compared to the state-of-the-art baseline \texttt{CFANE} and \texttt{PANE}, \algoc is able to take a lead by $6.1\%$ and $11.5\%$, respectively.
Similar observations can be made on the medium-sized graph {\em ArXiv} with over one million edges and the largest dataset {\em Amazon2M} with 61.9 million edges, where both \algoc or \algoe outperforms the best competitor by a margin of up to $2.6\%$ and $1.8\%$, respectively.
Over other large graphs {\em Yelp} and {\em Reddit}, \algoc and \algoe also obtain comparable or superior prediction precision. Note that on {\em Yelp} \algo is slightly inferior to \texttt{SimAttr (C)} and \texttt{SimAttr (E)}, with merely a $0.4\%$ decline in precision, but significantly dominates other competitors by a substantial margin of at least $4.6\%$. The reason is that the ground-truth clusters in the {\em Yelp} dataset are more relevant to node attributes than graph structures. \algo is the only LGC solution that can handle such graphs without downgrading the precision, exhibiting the effectiveness of our algorithmic designs in combining attribute and structure information.
This also underlines a limitation of \algo on graph datasets with high-quality attributes but substantial poor/corrupted structures, e.g., heterophilic graphs.
In addition, we can observe that \algoc attains comparable or superior performance to \algoe in most datasets. The only exception is on {\em Amazon2M}, where \algoe obtains an improvement of $5.9\%$ over \algoc.

\subsubsection{\bf {Recall when varying $\epsilon$}}
This set of experiments studies the effectiveness of \algo and other LGC-based methods in recovering the ground-truth local clusters when the sizes of their predicted ones (i.e., runtime budget) are varied. 
For a fair comparison, we only compare \algoc and \algoe with graph diffusion-based baselines \texttt{PR-Nibble}, \texttt{APR-Nibble}, and \texttt{HK-Relax} as the sizes of their output local clusters can be controlled by diffusion threshold $\epsilon$ and are bounded by $O(1/\epsilon)$. We additionally include an ablated version of \algo, dubbed as \algo (w/o SNAS), for comparison, which disables attribute information in \algo.
For all evaluated algorithms, we vary the size of the output local clusters by decreasing $\epsilon$ from $1.0$ to $10^{-8}$. For each seed node $v_s\in \mathcal{S}$, given the predicted local cluster $C_s$, we compute the recall by $|\C_s\cap \Y_s|/|\Y_s|$. Intuitively, the smaller $\epsilon$ is, the higher the recall should be. Fig.~\ref{fig:recall} depicts the average recall scores by all six methods when varying $\epsilon$ on six datasets. The $x$-axis and $y$-axis represent $\epsilon$ and the average recall, respectively.

From Fig.~\ref{fig:recall}, we can make the following observations. In most cases, both \algoc and \algoe consistently outperform other methods under the same diffusion thresholds $\epsilon$. In particular, on all datasets, the superiority of \algoc and \algoe is pronounced when $\epsilon\ge 10^{-3}$, indicating that \algo is effective and favorable when the budget for the sizes of local clusters and costs is low (e.g., only a small portion of the graph is allowed to be explored).
When $\epsilon\le 10^{-6}$, we can make qualitatively analogous observations on all datasets except {\em ArXiv}, where \algoc, \algoe, \algo (w/o SNAS), and \texttt{HK-Relax} perform comparably.
Notice that on {\em BlogCL}, {\em Flickr}, and {\em ArXiv}, \texttt{Hk-Relax} outmatches \algoc and \algoe when $\epsilon$ is roughly $10^{-4}$ or $10^{-5}$. The reason is that \texttt{HK-Relax} produces $\C_s$ with larger sizes 
than that by \algo under the same $\epsilon$ owing to its higher running time (Table~\ref{tbl:exp-algo}).
Furthermore, it can be seen that \algo (w/o SNAS) obtains similar results to \algoc and \algoe when $\epsilon$ is large and inferior ones when $\epsilon$ is small, which are still superior to those by other methods on most datasets. 
{\em This phenomenon implies that (i) our BDD (even without attributes) is more effective than existing graph diffusion metrics ({personalized PageRank}~\cite{jeh2003scaling} and {heat kernel PageRank}~\cite{chung2007heat}) in exploiting topological features for LGC, and (ii) the SNAS (attribute information) in \algo is crucial
for identifying the nodes in $\C_s$ that are far from the seed.}

Further, in Appendix~\ref{sec:add-exp}, we evaluate the average external connectivity (i.e., conductance) and attribute variance (a.k.a. WCSS) of nodes in local clusters output by all methods. 
We also showcase that our \algo's framework remains effective on non-attributed graphs in Appendix~\ref{sec:add-exp}.

\subsection{Efficiency Evaluation}
For ease of comparison, on each dataset, we assess the empirical efficiency of \algoc and \algoe only against the methods that yield the top-$4$ best results among all competitors in terms of precision (Table~\ref{tbl:node-clustering}). 
In Fig.~\ref{fig:time}, we show the running times (measured in wall-clock time) required by the preprocessing phase and the online phase (i.e., the procedure generating a local cluster for a single seed node) of each evaluated approach. The $y$-axis represents the running time in seconds on a log scale. 

The first observation we can make from Fig.~\ref{fig:time} is 
that 
on small datasets {\em Cora}, {\em PubMed}, {\em BlogCL}, and {\em Flickr}, \algoc, \algoe and the-state-of the-art solutions (\texttt{CFANE} or \texttt{PANE}) are highly fast in the online stage, all of which take less than $0.1$ seconds to finish a LGC task on average. However, \texttt{CFANE} and \texttt{PANE} obtain such a high online efficiency at the cost of up to 19.4 and 2.5 minutes for constructing node embeddings of all nodes in the preprocessing, whereas \algoc and \algoe require at most 2.8 seconds. 

On larger graphs {\em ArXiv}, {\em Reddit}, and {\em Amazon2M} comprising millions or tens of millions of edges,
\texttt{CFANE} fails to report the results within 3 days and LGC-based methods \texttt{HK-Relax}, \texttt{$p$-Norm FD} and \texttt{WFD} are the state-of-the-art solutions. On such datasets, \algoc and \algoe are able to gain a significant speedup of $196\times$, $209\times$, and $152\times$, respectively, in the online stage on average. Note that the total preprocessing costs of \algoc and \algoe are still insignificant, which often take a few seconds, even less than the average cost for a single LGC task. For instance, on the largest dataset {\em Amazon2M}, with 62 million edges, \algoe attains a precision of $0.521$ using an average of $36.7$ seconds for Algo.~\ref{alg:main} (online stage) and $13.4$ seconds for TNAM construction (preprocessing stage). In comparison, the best competitor \texttt{WFD} consumes $92.9$ minutes to get a precision of $0.503$. 
The empirical observations are consistent with the theoretical evidence that \texttt{HK-Relax}, \texttt{$p$-Norm FD} and \texttt{WFD} have worse asymptotic complexities than \algo as in Table~\ref{tbl:exp-algo}.
On {\em Yelp}, where the best methods \texttt{SimAttr (C)} and \texttt{SimAttr (E)} are simply based on attributes, \algoc and \algoe have comparable prediction precision using slightly higher running time.
In summary, our \algo methods can achieve significantly higher efficiency for LGC on most attributed graphs with various volumes and meanwhile yield state-of-the-art result quality.

\subsection{ Real-world Example} \label{sec:real-example}

\begin{figure}[h]
    \centering
    \subfloat[Run \algo on ``Jian Pei"\label{fig:revision-academic-1}]{
        \includegraphics[width=0.47\columnwidth]{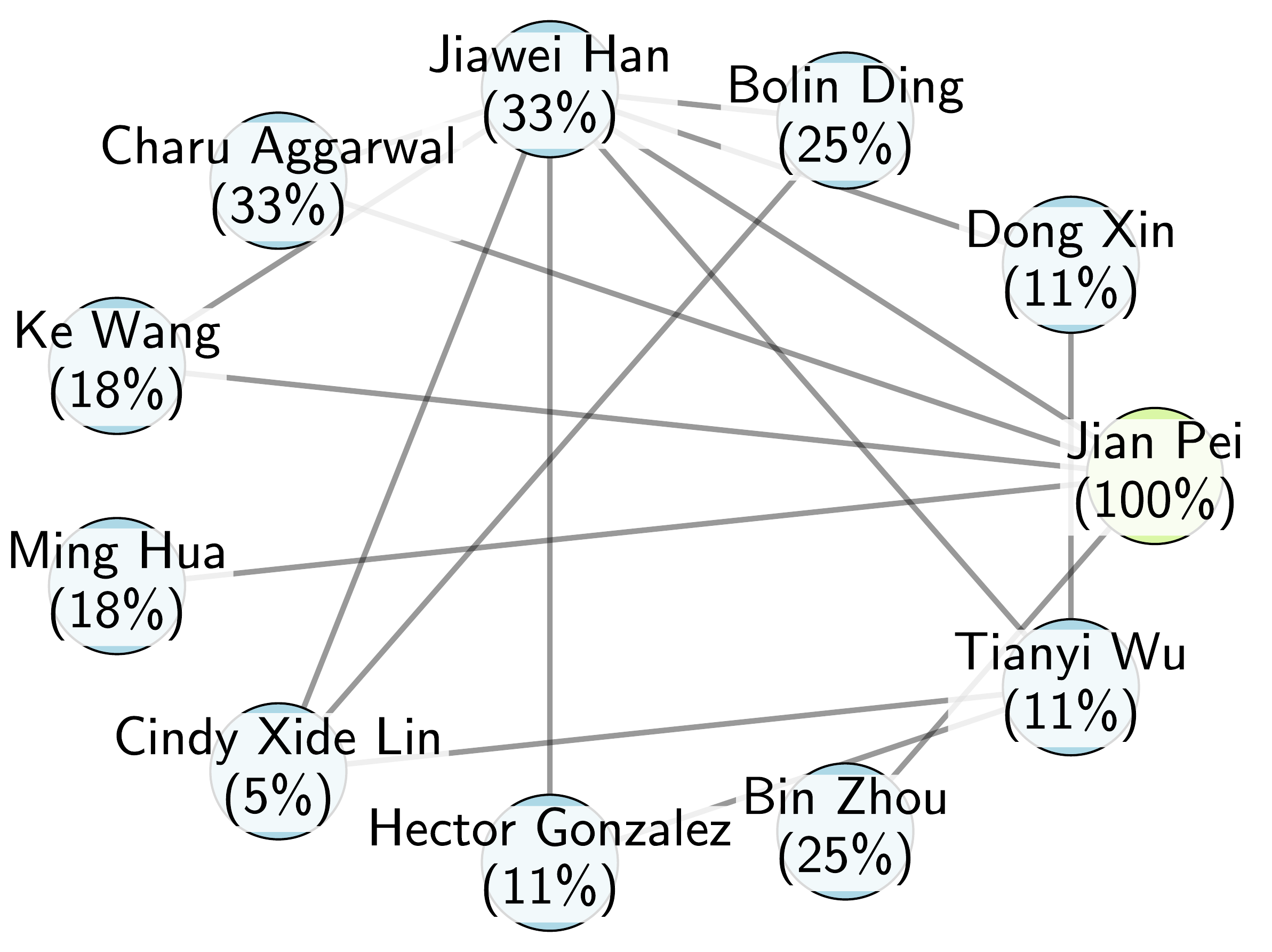}
    }
    \hfill
    \subfloat[Run \texttt{PR-Nibble} on ``Jian Pei"\label{fig:revision-academic-1-1}]{
        \includegraphics[width=0.47\columnwidth]{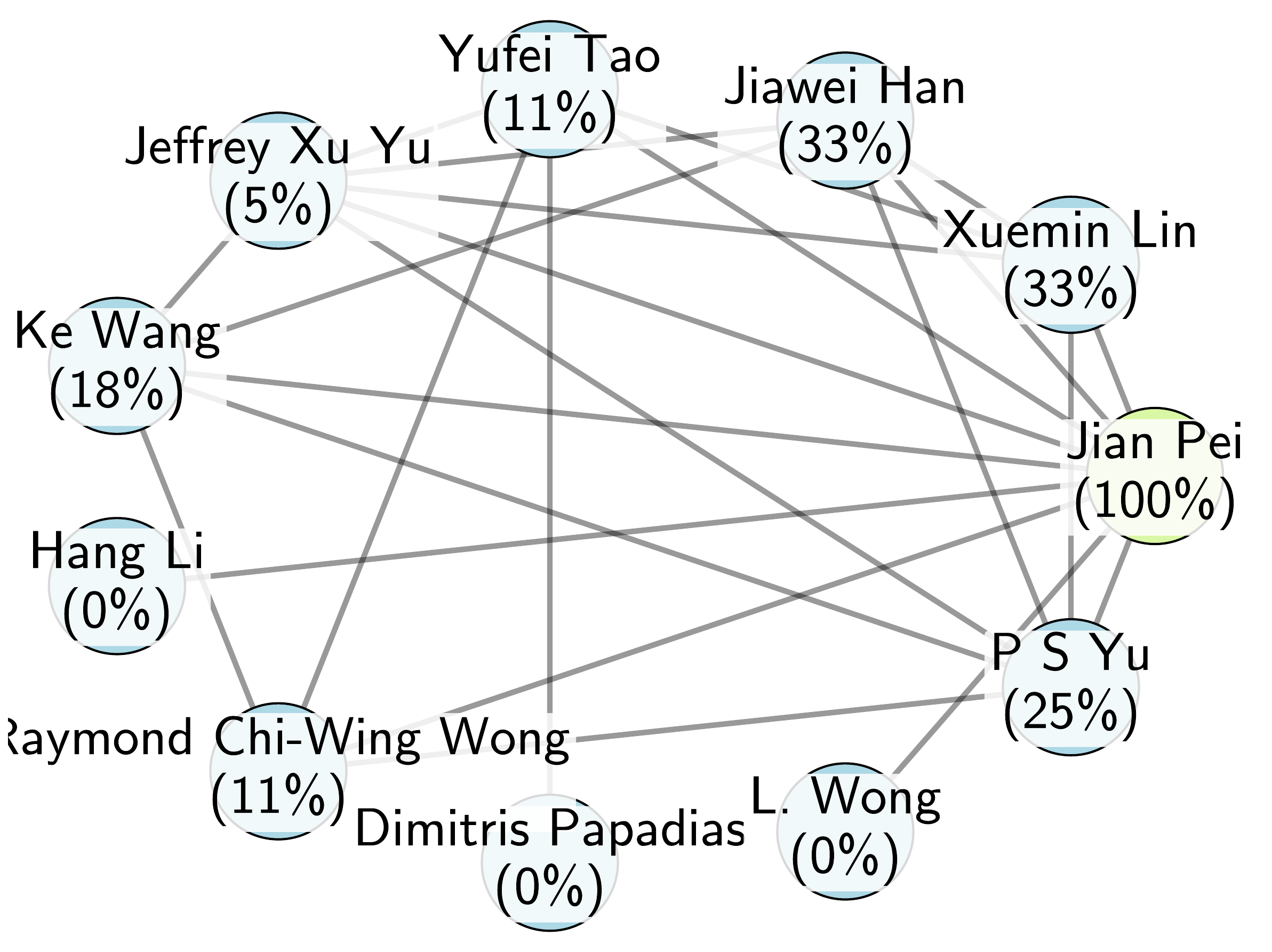}
    }
    \vspace{-1ex}
    \caption{A real-world scenario on an academic network.}
    \label{fig:academic-combined-1}
    \vspace{-4ex}
\end{figure}

Our proposed adaptive local clustering method can be applied to real-world scenarios, including the following three representative applications: game club recommendation~\cite{zhang2023constrained}, social network analysis (e.g., Twitter community detection)~\cite{fang2016effective, fang2018effective, yang2019efficient}, and academic collaboration networks~\cite{fang2020effective}. A specific validation is conducted using the academic collaboration graph from the AMiner Coauthor dataset~\cite{zhang2015panther}, which contains 1.7M scholars with co-authorships and keyword-based research interests.
Applying \algo starting from the seed scholar ``Jian Pei", we identified 10 scholars with both strong co-authorship ties and aligned research interests, as shown in Fig.~\ref{fig:revision-academic-1}.
This group includes direct co-authors such as ``Jiawei Han" (similarity: 33\%) and ``Charu Aggarwal" (33\%) as well as a subgroup centered around ``Jiawei Han" (e.g., ``Bolin Ding" (25\%)).
In contrast, \texttt{PR-Nibble}—an LGC baseline—selected three scholars (e.g., ``Hang Li," ``Dimitris Papadias") with 0\% similarity despite direct co-authorships (Fig.~\ref{fig:revision-academic-1-1}).
This stark difference (3/10 zero-similarity nodes in \texttt{PR-Nibble} vs. 0/10 in ours) demonstrates how ignoring attributes, as in LGC methods, risks recommending collaborators with mismatched expertise, even when strong structural ties exist.

%% file: tex/figs/precision.tex
\begin{table*}[!t]
\centering
\renewcommand{\arraystretch}{1.0}
\caption{The average precision evaluated with ground-truth. Best is \textbf{bolded} and best baseline \underline{underlined}.}
\begin{footnotesize}
\addtolength{\tabcolsep}{-0.25em}
\resizebox{\textwidth}{!}{
\begin{tabular}{c c|c | c | c| c| c | c| c| c| c}
\hline
& \multirow{1}{*}{\bf Method} & \multicolumn{1}{c|}{{\bf {\em Cora}}~\cite{sen2008collective}} & \multicolumn{1}{c|}{{\bf {\em PubMed}}~\cite{sen2008collective}}  & \multicolumn{1}{c|}{{\bf {\em BlogCL}}~\cite{tang2009relational}}  & \multicolumn{1}{c|}{{\bf {\em Flickr}}~\cite{huang2017label}} & \multicolumn{1}{c|}{{\bf {\em ArXiv}}~\cite{zeng2020graphsaint}} & \multicolumn{1}{c|}{{\bf {\em Yelp}}~\cite{hu2020open}} & \multicolumn{1}{c|}{{\bf {\em Reddit}}~\cite{hu2020open}} & \multicolumn{1}{c|}{{\bf {\em {Amazon2M}}}~\cite{chiang2019cluster}} & \multicolumn{1}{c}{\bf Rank}  \\ \cline{3-11}
\hline
\multirow{6}{1cm}{\centering Local Graph Clustering} & \texttt{PR-Nibble}~\cite{andersen2006local}	&	0.413	&	0.481	&	0.263	&	0.198	&	0.299	&	0.214	&	0.651	&	0.364	& \color{black}9.25 \\
& \texttt{APR-Nibble} 	&	0.396	&	0.479	&	0.252	&	0.173	&	0.282	&	0.093	&	0.435	&	0.129	& \color{black}13.13 \\
& \texttt{HK-Relax}~\cite{kloster2014heat} 	&	0.477	&	0.476	&	0.284	&	0.219	&	\underline{0.351}	&	0.214	&	0.751	&	0.116	& \color{black}8.63 \\
& \texttt{CRD}~\cite{wang2017capacity} 	&	0.149	&	0.112	&	0.236	&	0.192	&	0.166	&	0.098	&	0.641	&	0.072	& \color{black}16.38 \\
& \texttt{$p$-Norm FD}~\cite{fountoulakis2020p} 	&	0.263	&	0.131	&	0.159	&	0.132	&	0.225	&	0.034	&	\underline{0.806}	&	0.441	& \color{black}16.5 \\
& \texttt{WFD}~\cite{pmlr-v202-yang23d}	&	0.298	&	0.17	&	0.181	&	0.133	&	0.253	&	0.043	&	0.589	&	\underline{0.503}	& \color{black}15.63 \\ \hline
\multirow{4}{1cm}{\centering Link Similarity} & \texttt{Jaccard}~\cite{liben2003link} 	&	0.231	&	0.358	&	0.282	&	0.209	&	0.116	&	0.662	&	0.343	&	0.142	& \color{black}13.13 \\
& \texttt{Adamic-Adar}~\cite{liben2003link} 	&	0.231	&	0.358	&	0.265	&	0.163	&	0.117	&	0.662	&	0.33	&	0.142	& \color{black}15 \\
& \texttt{Common-Nbrs}~\cite{liben2003link} 	&	0.231	&	0.358	&	0.259	&	0.156	&	0.115	&	0.662	&	0.318	&	0.142	& \color{black}16 \\
& \texttt{SimRank}~\cite{jeh2002simrank} 	&	0.436	&	0.519	&	0.273	&	0.186	&	-	&	-	&	-	&	-	& \color{black}9.25 \\ \hline
\multirow{3}{1cm}{\centering Attribute Similarity} & \texttt{SimAttr (C)}~\cite{yin2010unified} 	&	0.288	&	0.469	&	0.306	&	0.182	&	0.154	&	\underline{\textbf{0.758}}	&	0.035	&	0.194	& \color{black}10.88 \\
& \texttt{SimAttr (E)}~\cite{rahutomo2012semantic}	&	0.288	&	0.469	&	0.306	&	0.182	&	0.154	&	\underline{\textbf{0.758}}	&	0.035	&	0.194	& \color{black}10.88 \\
& \texttt{AttriRank}~\cite{hsu2017unsupervised} 	&	0.181	&	0.363	&	0.184	&	0.124	&	0.076	&	0.666	&	0.047	&	0.122	& \color{black}18.25 \\ \hline
\multirow{12}{1cm}{\centering Network Embedding} & \texttt{Node2Vec} ($K$-NN) 	&	0.181	&	0.363	&	0.167	&	0.111	&	0.074	&	0.665	&	-	&	-	& \color{black}20.5\\
& \color{black}\texttt{Node2Vec} (SC) 	& \color{black}0.41	& \color{black}0.488	&\color{black} 0.263	& \color{black}0.182	& \color{black}-	& \color{black}-	& \color{black}-	&\color{black} - & \color{black}12.25\\
& \color{black}\texttt{Node2Vec} (DBSCAN) 	&\color{black} 0.419	&\color{black} 0.488	& \color{black}0.263	& \color{black}0.182	&\color{black} 0.316	&\color{black} 0.679	&\color{black} -	& \color{black}- & \color{black}9.5\\
& \texttt{SAGE} ($K$-NN) &	0.423	&	0.434	&	0.209	&	0.143	&	-	&	-	&	-	&	-	& \color{black}18\\
& \color{black}\texttt{SAGE} (SC) 	& \color{black}0.403	&\color{black} 0.399	& \color{black}0.226	& \color{black}0.155	& \color{black}-	& \color{black}-	& \color{black}-	& \color{black}- & \color{black}18.5\\
& \color{black}\texttt{SAGE} (DBSCAN) 	& \color{black}0.326	& \color{black}0.359	& \color{black}0.226	& \color{black}0.155	& \color{black}-	& \color{black}-	& \color{black}-	& \color{black}- & \color{black}20\\
& \texttt{CFANE} ($K$-NN)    	&	\underline{0.495}	&	\underline{0.531}	&	\underline{0.505}	&	0.2	&	-	&	-	&	-	&	-	&  \underline{\color{black}4}\\
& \color{black}\texttt{CFANE} (SC)     	& \color{black}0.494	& \color{black}\underline{0.531}	& \color{black}\underline{0.505}	& \color{black}0.198	&\color{black} -	&\color{black} -	&\color{black} -	&\color{black} - & \color{black}4.5\\
& \color{black}\texttt{CFANE} (DBSCAN)    	&\color{black} 0.383	& \color{black}0.44	&\color{black} 0.36	& \color{black}0.164	&\color{black} -	&\color{black} -	&\color{black} -	&\color{black} - & \color{black}14.25\\
& \texttt{PANE} ($K$-NN)  	&	0.445	&	0.497	&	0.456	&	\underline{0.332}	&	0.147	&	0.708	&	0.263	&	0.197	& \color{black}6.88\\
& \color{black}\texttt{PANE} (SC)   		&\color{black} 0.445	&\color{black} 0.497	& \color{black}0.456	& \color{black}\underline{0.332}	& \color{black}-	& \color{black}-	&\color{black} -	&\color{black} - & \color{black}5\\
& \color{black}\texttt{PANE} (DBSCAN)  	& \color{black}0.422	&\color{black} 0.477	&\color{black} 0.267	& \color{black}0.264	& \color{black}0.145	&\color{black} 0.635	&\color{black} 0.232	&\color{black} 0.177 &  \color{black}10.88\\ \hline
\multirow{2}{1cm}{\centering Ours} & \algoc       	&	\textbf{0.556}	&	{0.552}	&	\textbf{0.51}	&	\textbf{0.447}	&	\textbf{0.377}	&	{0.754}	&	\textbf{0.808}	&	{0.465}	& \textbf{1.63} \\
& \algoe        	&	{0.552}	&	\textbf{0.555}	&	0.493	&	{0.39}	&	\textbf{0.377}	&	0.739	&	\textbf{0.808}	&	\textbf{0.521}	& \color{black}2 \\ \hline
\end{tabular}
}
\end{footnotesize}
\label{tbl:node-clustering}
\vspace{-2ex}
\end{table*}

%% file: tex/figs/recall_new.tex
 \begin{figure}[!t]
\centering
\begin{small}
\begin{tikzpicture}
    \begin{customlegend}[legend columns=3,
        legend entries={\algoc, \algoe, \algo (w/o SNAS), \algoppr, \texttt{HK-Relax}, \texttt{APR-Nibble}},
        legend style={at={(0.45,1.35)},anchor=north,draw=none,font=\footnotesize,column sep=0.15cm}]
    \addlegendimage{line width=0.3mm,mark size=3pt,mark=o, color=Cora-color}
    \addlegendimage{line width=0.3mm,mark size=3pt,mark=star, color=Pubmed-color}
    \addlegendimage{line width=0.3mm,mark size=3pt,mark=diamond, color=Arxiv-color}
    \addlegendimage{line width=0.3mm,mark size=3pt,mark=pentagon, color=Blogcatalog-color}
    \addlegendimage{line width=0.3mm,mark size=3pt,mark=square, color=Flickr-color}
    \addlegendimage{line width=0.3mm,mark size=4pt,mark=+, color=Reddit-color}
    \end{customlegend}
\end{tikzpicture}
\\[-\lineskip]
\vspace{-4mm}
\subfloat[{\em Cora}]{
\begin{tikzpicture}[scale=1,every mark/.append style={mark size=2pt}]
    \begin{axis}[
        height=\columnwidth/2.6,
        width=\columnwidth/2.0,
         ylabel={\it Recall},
        xmin=1e-8, xmax=1,
        ymin=0, ymax=0.6,
        xtick={1e-10, 1e-8, 1e-6, 1e-4, 1e-2, 1},
        ytick={0,0.1,0.2,0.3,0.4,0.5,0.6},
        xticklabel style = {font=\tiny},
        yticklabel style = {font=\footnotesize},
        xticklabels={$10^{-10}$, $10^{-8}$, $10^{-6}$, $10^{-4}$, $10^{-2}$, $10^{0}$},
        yticklabels={0,0.1,0.2,0.3,0.4,0.5,0.6},
        xmode=log,
        every axis y label/.style={font=\footnotesize,at={(current axis.north west)},right=2mm,above=0mm},
        legend style={fill=none,font=\small,at={(0.02,0.99)},anchor=north west,draw=none},
    ]
    \addplot[line width=0.3mm, mark=o, color=Cora-color]  %
        plot coordinates {
(1e-8,	0.547	)
(1e-7,	0.549	)
(1e-6,	0.556	)
(1e-5,	0.551	)
(1e-4,	0.517	)
(1e-3,	0.318	)
(1e-2,	0.204	)
(1e-1,	0.181	)
(1,	0.189	)
    };

    \addplot[line width=0.3mm, mark=star, color=Pubmed-color]  %
        plot coordinates {
(1e-8,	0.543	)
(1e-7,	0.546	)
(1e-6,	0.552	)
(1e-5,	0.537	)
(1e-4,	0.505	)
(1e-3,	0.317	)
(1e-2,	0.204	)
(1e-1,	0.191	)
(1,	0.189	)
    };

    \addplot[line width=0.3mm, mark=diamond, color=Arxiv-color]  %
        plot coordinates {
(1e-8,	0.486	)
(1e-7,	0.486	)
(1e-6,	0.486	)
(1e-5,	0.488	)
(1e-4,	0.477	)
(1e-3,	0.299	)
(1e-2,	0.203	)
(1e-1,	0.191	)
(1,	0.189	)
    };

    \addplot[line width=0.3mm, mark=pentagon, color=Blogcatalog-color]  %
        plot coordinates {
(1e-8,	0.413	)
(1e-7,	0.399	)
(1e-6,	0.393	)
(1e-5,	0.384	)
(1e-4,	0.274	)
(1e-3,	0.061	)
(1e-2,	0.014	)
(1e-1,	0.003	)
(1,	0.003	)
    };

    \addplot[line width=0.3mm, mark=+, color=Reddit-color]  %
        plot coordinates {
(1e-8,	0.396	)
(1e-7,	0.389	)
(1e-6,	0.390	)
(1e-5,	0.336	)
(1e-4,	0.209	)
(1e-3,	0.061	)
(1e-2,	0.016	)
(1e-1,	0.005	)
(1,	0.003	)
    };

    \addplot[line width=0.3mm, mark=square, color=Flickr-color]  %
        plot coordinates {
(1e-8,	0.477	)
(1e-7,	0.467	)
(1e-6,	0.458	)
(1e-5,	0.439	)
(1e-4,	0.425	)
(1e-3,	0.352	)
(1e-2,	0.127	)
(1e-1,	0.030	)
(1,	0.010	)
    };
    
    \end{axis}
\end{tikzpicture}\hspace{4mm}\label{fig:vary-dim}%
}%
\subfloat[{\em PubMed}]{
\begin{tikzpicture}[scale=1,every mark/.append style={mark size=2pt}]
    \begin{axis}[
        height=\columnwidth/2.6,
        width=\columnwidth/2.0,
         ylabel={\it Recall},
        xmin=1e-8, xmax=1,
        ymin=0, ymax=0.6,
        xtick={1e-10, 1e-8, 1e-6, 1e-4, 1e-2, 1},
        ytick={0,0.1,0.2,0.3,0.4,0.5,0.6,0.7},
        xticklabel style = {font=\tiny},
        yticklabel style = {font=\footnotesize},
        xticklabels={$10^{-10}$, $10^{-8}$, $10^{-6}$, $10^{-4}$, $10^{-2}$, $10^{0}$},
        yticklabels={0,0.1,0.2,0.3,0.4,0.5,0.6,0.7},
        xmode=log,
        every axis y label/.style={font=\footnotesize,at={(current axis.north west)},right=2mm,above=0mm},
        legend style={fill=none,font=\small,at={(0.02,0.99)},anchor=north west,draw=none},
    ]
    \addplot[line width=0.3mm, mark=o, color=Cora-color]  %
        plot coordinates {
(1e-8,	0.552	)
(1e-7,	0.552	)
(1e-6,	0.552	)
(1e-5,	0.548	)
(1e-4,	0.430	)
(1e-3,	0.360	)
(1e-2,	0.356	)
(1e-1,	0.356	)
(1,	0.356	)
    };

    \addplot[line width=0.3mm, mark=star, color=Pubmed-color]  %
        plot coordinates {
(1e-8,	0.555	)
(1e-7,	0.555	)
(1e-6,	0.554	)
(1e-5,	0.544	)
(1e-4,	0.356	)
(1e-3,	0.356	)
(1e-2,	0.356	)
(1e-1,	0.356	)
(1,	0.356	)
    };

    \addplot[line width=0.3mm, mark=diamond, color=Arxiv-color]  %
        plot coordinates {
(1e-8,	0.537	)
(1e-7,	0.537	)
(1e-6,	0.537	)
(1e-5,	0.532	)
(1e-4,	0.473	)
(1e-3,	0.368	)
(1e-2,	0.357	)
(1e-1,	0.356	)
(1,	0.356	)
    };

    \addplot[line width=0.3mm, mark=pentagon, color=Blogcatalog-color]  %
        plot coordinates {
(1e-8,	0.340	)
(1e-7,	0.396	)
(1e-6,	0.481	)
(1e-5,	0.178	)
(1e-4,	0.028	)
(1e-3,	0.004	)
(1e-2,	0.001	)
(1e-1,	0.000	)
(1,	0.000	)
    };

    \addplot[line width=0.3mm, mark=+, color=Reddit-color]  %
        plot coordinates {
(1e-8,	0.360	)
(1e-7,	0.423	)
(1e-6,	0.479	)
(1e-5,	0.176	)
(1e-4,	0.028	)
(1e-3,	0.004	)
(1e-2,	0.001	)
(1e-1,	0.000	)
(1,	0.000	)
    };

    \addplot[line width=0.3mm, mark=square, color=Flickr-color]  %
        plot coordinates {
(1e-8,	0.475	)
(1e-7,	0.476	)
(1e-6,	0.476	)
(1e-5,	0.476	)
(1e-4,	0.395	)
(1e-3,	0.059	)
(1e-2,	0.009	)
(1e-1,	0.002	)
(1,	0.001	)
    };
    
    \end{axis}
\end{tikzpicture}\hspace{0mm}\label{fig:vary-dim}%
}%
\vspace{-3ex}
\subfloat[{\em BlogCL}]{
\begin{tikzpicture}[scale=1,every mark/.append style={mark size=2pt}]
    \begin{axis}[
        height=\columnwidth/2.6,
        width=\columnwidth/2.0,
         ylabel={\it Recall},
        xmin=1e-8, xmax=1,
        ymin=0, ymax=0.5,
        xtick={1e-10, 1e-8, 1e-6, 1e-4, 1e-2, 1},
        ytick={0,0.1,0.2,0.3,0.4,0.5,0.6,0.7},
        xticklabel style = {font=\tiny},
        yticklabel style = {font=\footnotesize},
        xticklabels={$10^{-10}$, $10^{-8}$, $10^{-6}$, $10^{-4}$, $10^{-2}$, $10^{0}$},
        yticklabels={0,0.1,0.2,0.3,0.4,0.5,0.6,0.7},
        xmode=log,
        every axis y label/.style={font=\footnotesize,at={(current axis.north west)},right=2mm,above=0mm},
        legend style={fill=none,font=\small,at={(0.02,0.99)},anchor=north west,draw=none},
    ]
    \addplot[line width=0.3mm, mark=o, color=Cora-color]  %
        plot coordinates {
(1e-8,	0.461	)
(1e-7,	0.467	)
(1e-6,	0.487	)
(1e-5,	0.194	)
(1e-4,	0.173	)
(1e-3,	0.168	)
(1e-2,	0.168	)
(1e-1,	0.167	)
(1,	0.167	)
    };

    \addplot[line width=0.3mm, mark=star, color=Pubmed-color]  %
        plot coordinates {
(1e-8,	0.448	)
(1e-7,	0.454	)
(1e-6,	0.469	)
(1e-5,	0.194	)
(1e-4,	0.172	)
(1e-3,	0.168	)
(1e-2,	0.168	)
(1e-1,	0.167	)
(1,	0.167	)
    };

    \addplot[line width=0.3mm, mark=diamond, color=Arxiv-color]  %
        plot coordinates {
(1e-8,	0.285	)
(1e-7,	0.287	)
(1e-6,	0.305	)
(1e-5,	0.203	)
(1e-4,	0.176	)
(1e-3,	0.168	)
(1e-2,	0.168	)
(1e-1,	0.167	)
(1,	0.167	)
    };

    \addplot[line width=0.3mm, mark=pentagon, color=Blogcatalog-color]  %
        plot coordinates {
(1e-8,	0.243	)
(1e-7,	0.244	)
(1e-6,	0.263	)
(1e-5,	0.062	)
(1e-4,	0.011	)
(1e-3,	0.001	)
(1e-2,	0.001	)
(1e-1,	0.001	)
(1,	0.001	)
    };
    
    \addplot[line width=0.3mm, mark=+, color=Reddit-color]  %
        plot coordinates {
(1e-8,	0.228	)
(1e-7,	0.236	)
(1e-6,	0.252	)
(1e-5,	0.235	)
(1e-4,	0.058	)
(1e-3,	0.013	)
(1e-2,	0.001	)
(1e-1,	0.001	)
(1,	0.001	)
    };

    \addplot[line width=0.3mm, mark=square, color=Flickr-color]  %
        plot coordinates {
(1e-8,	0.273	)
(1e-7,	0.273	)
(1e-6,	0.273	)
(1e-5,	0.273	)
(1e-4,	0.284	)
(1e-3,	0.054	)
(1e-2,	0.008	)
(1e-1,	0.001	)
(1,	0.001	)
    };
    
    \end{axis}
\end{tikzpicture}\hspace{4mm}\label{fig:vary-dim}%
}%
\subfloat[{\em Flickr}]{
\begin{tikzpicture}[scale=1,every mark/.append style={mark size=2pt}]
    \begin{axis}[
        height=\columnwidth/2.6,
        width=\columnwidth/2.0,
         ylabel={\it Recall},
        xmin=1e-8, xmax=1,
        ymin=0, ymax=0.5,
        xtick={1e-10, 1e-8, 1e-6, 1e-4, 1e-2, 1},
        ytick={0,0.1,0.2,0.3,0.4,0.5,0.6,0.7},
        xticklabel style = {font=\tiny},
        yticklabel style = {font=\footnotesize},
        xticklabels={$10^{-10}$, $10^{-8}$, $10^{-6}$, $10^{-4}$, $10^{-2}$, $10^{0}$},
        yticklabels={0,0.1,0.2,0.3,0.4,0.5,0.6,0.7},
        xmode=log,
        every axis y label/.style={font=\footnotesize,at={(current axis.north west)},right=2mm,above=0mm},
        legend style={fill=none,font=\small,at={(0.02,0.99)},anchor=north west,draw=none},
    ]
    \addplot[line width=0.3mm, mark=o, color=Cora-color]  %
        plot coordinates {
(1e-8,	0.357	)
(1e-7,	0.369	)
(1e-6,	0.432	)
(1e-5,	0.139	)
(1e-4,	0.114	)
(1e-3,	0.111	)
(1e-2,	0.111	)
(1e-1,	0.110	)
(1,	0.110	)
    };

    \addplot[line width=0.3mm, mark=star, color=Pubmed-color]  %
        plot coordinates {
(1e-8,	0.316	)
(1e-7,	0.337	)
(1e-6,	0.364	)
(1e-5,	0.133	)
(1e-4,	0.112	)
(1e-3,	0.111	)
(1e-2,	0.111	)
(1e-1,	0.110	)
(1,	0.110	)
    };

    \addplot[line width=0.3mm, mark=diamond, color=Arxiv-color]  %
        plot coordinates {
(1e-8,	0.173	)
(1e-7,	0.177	)
(1e-6,	0.203	)
(1e-5,	0.140	)
(1e-4,	0.112	)
(1e-3,	0.111	)
(1e-2,	0.111	)
(1e-1,	0.110	)
(1,	0.110	)
    };

    \addplot[line width=0.3mm, mark=pentagon, color=Blogcatalog-color]  %
        plot coordinates {
(1e-10,	0.134	)
(1e-9,	0.134	)
(1e-8,	0.135	)
(1e-7,	0.141	)
(1e-6,	0.198	)
(1e-5,	0.017	)
(1e-4,	0.003	)
(1e-3,	0.001	)
(1e-2,	0.001	)
(1e-1,	0.001	)
(1,	0.001	)
    };
    
    \addplot[line width=0.3mm, mark=+, color=Reddit-color]  %
        plot coordinates {
(1e-8,	0.278	)
(1e-7,	0.282	)
(1e-6,	0.163	)
(1e-5,	0.038	)
(1e-4,	0.008	)
(1e-3,	0.002	)
(1e-2,	0.001	)
(1e-1,	0.000	)
(1,	0.000	)
    };

    \addplot[line width=0.3mm, mark=square, color=Flickr-color]  %
        plot coordinates {
(1e-8,	0.219	)
(1e-7,	0.219	)
(1e-6,	0.219	)
(1e-5,	0.219	)
(1e-4,	0.208	)
(1e-3,	0.020	)
(1e-2,	0.003	)
(1e-1,	0.001	)
(1,	0.001	)
    };
    
    \end{axis}
\end{tikzpicture}\hspace{0mm}\label{fig:vary-dim}%
}%
\vspace{-3ex}
\subfloat[{\em ArXiv}]{
\begin{tikzpicture}[scale=1,every mark/.append style={mark size=2pt}]
    \begin{axis}[
        height=\columnwidth/2.6,
        width=\columnwidth/2.0,
         ylabel={\it Recall},
        xmin=1e-8, xmax=1,
        ymin=0, ymax=0.4,
        xtick={1e-10, 1e-8, 1e-6, 1e-4, 1e-2, 1},
        ytick={0,0.1,0.2,0.3,0.4,0.5,0.6,0.7},
        xticklabel style = {font=\tiny},
        yticklabel style = {font=\footnotesize},
        xticklabels={$10^{-10}$, $10^{-8}$, $10^{-6}$, $10^{-4}$, $10^{-2}$, $10^{0}$},
        yticklabels={0,0.1,0.2,0.3,0.4,0.5,0.6,0.7},
        xmode=log,
        every axis y label/.style={font=\footnotesize,at={(current axis.north west)},right=2mm,above=0mm},
        legend style={fill=none,font=\small,at={(0.02,0.99)},anchor=north west,draw=none},
    ]
    \addplot[line width=0.3mm, mark=o, color=Cora-color]  %
        plot coordinates {
(1e-8,	0.330	)
(1e-7,	0.342	)
(1e-6,	0.369	)
(1e-5,	0.186	)
(1e-4,	0.098	)
(1e-3,	0.080	)
(1e-2,	0.077	)
(1e-1,	0.076	)
(1,	0.076	)
    };

    \addplot[line width=0.3mm, mark=star, color=Pubmed-color]  %
        plot coordinates {
(1e-8,	0.329	)
(1e-7,	0.335	)
(1e-6,	0.369	)
(1e-5,	0.185	)
(1e-4,	0.098	)
(1e-3,	0.080	)
(1e-2,	0.077	)
(1e-1,	0.076	)
(1,	0.076	)
    };

    \addplot[line width=0.3mm, mark=diamond, color=Arxiv-color]  %
        plot coordinates {
(1e-8,	0.323	)
(1e-7,	0.333	)
(1e-6,	0.336	)
(1e-5,	0.172	)
(1e-4,	0.097	)
(1e-3,	0.081	)
(1e-2,	0.077	)
(1e-1,	0.076	)
(1,	0.076	)
    };

    \addplot[line width=0.3mm, mark=pentagon, color=Blogcatalog-color]  %
        plot coordinates {
(1e-10,	0.290	)
(1e-9,	0.292	)
(1e-8,	0.299	)
(1e-7,	0.298	)
(1e-6,	0.172	)
(1e-5,	0.041	)
(1e-4,	0.008	)
(1e-3,	0.002	)
(1e-2,	0.001	)
(1e-1,	0.000	)
(1,	0.000	)
    };

    \addplot[line width=0.3mm, mark=+, color=Reddit-color]  %
        plot coordinates {
(1e-8,	0.278	)
(1e-7,	0.282	)
(1e-6,	0.163	)
(1e-5,	0.038	)
(1e-4,	0.008	)
(1e-3,	0.002	)
(1e-2,	0.001	)
(1e-1,	0.000	)
(1,	0.000	)
    };

    \addplot[line width=0.3mm, mark=square, color=Flickr-color]  %
        plot coordinates {
(1e-8,	0.351	)
(1e-7,	0.346	)
(1e-6,	0.342	)
(1e-5,	0.258	)
(1e-4,	0.040	)
(1e-3,	0.011	)
(1e-2,	0.004	)
(1e-1,	0.001	)
(1,	0.000	)
    };

    \end{axis}
\end{tikzpicture}\hspace{4mm}\label{fig:vary-dim}%
}%
\subfloat[{\em Yelp}]{
\begin{tikzpicture}[scale=1,every mark/.append style={mark size=2pt}]
    \begin{axis}[
        height=\columnwidth/2.6,
        width=\columnwidth/2.0,
         ylabel={\it Recall},
        xmin=1e-8, xmax=1,
        ymin=0, ymax=0.76,
        xtick={1e-10, 1e-8, 1e-6, 1e-4, 1e-2, 1},
        ytick={0,0.1,0.2,0.3,0.4,0.5,0.6,0.7},
        xticklabel style = {font=\tiny},
        yticklabel style = {font=\footnotesize},
        xticklabels={$10^{-10}$, $10^{-8}$, $10^{-6}$, $10^{-4}$, $10^{-2}$, $10^{0}$},
        yticklabels={0,0.1,0.2,0.3,0.4,0.5,0.6,0.7},
        xmode=log,
        every axis y label/.style={font=\footnotesize,at={(current axis.north west)},right=2mm,above=0mm},
        legend style={fill=none,font=\small,at={(0.02,0.99)},anchor=north west,draw=none},
    ]
    \addplot[line width=0.3mm, mark=o, color=Cora-color]  %
        plot coordinates {
(1e-8,	0.742	)
(1e-7,	0.662	)
(1e-6,	0.662	)
(1e-5,	0.662	)
(1e-4,	0.662	)
(1e-3,	0.662	)
(1e-2,	0.662	)
(1e-1,	0.662	)
(1,	0.662	)
    };

    \addplot[line width=0.3mm, mark=star, color=Pubmed-color]  %
        plot coordinates {
(1e-8,	0.725	)
(1e-7,	0.681	)
(1e-6,	0.663	)
(1e-5,	0.662	)
(1e-4,	0.662	)
(1e-3,	0.662	)
(1e-2,	0.662	)
(1e-1,	0.662	)
(1,	0.662	)
    };

    \addplot[line width=0.3mm, mark=diamond, color=Arxiv-color]  %
        plot coordinates {
(1e-8,	0.686	)
(1e-7,	0.687	)
(1e-6,	0.673	)
(1e-5,	0.662	)
(1e-4,	0.662	)
(1e-3,	0.662	)
(1e-2,	0.662	)
(1e-1,	0.662	)
(1,	0.662	)
    };

    \addplot[line width=0.3mm, mark=pentagon, color=Blogcatalog-color]  %
        plot coordinates {
(1e-10,	0.124	)
(1e-9,	0.128	)
(1e-8,	0.214	)
(1e-7,	0.068	)
(1e-6,	0.004	)
(1e-5,	0.000	)
(1e-4,	0.000	)
(1e-3,	0.000	)
(1e-2,	0.000	)
(1e-1,	0.000	)
(1,	0.000	)
    };

    \addplot[line width=0.3mm, mark=+, color=Reddit-color]  %
        plot coordinates {
(1e-8,	0.093	)
(1e-7,	0.041	)
(1e-6,	0.004	)
(1e-5,	0.000	)
(1e-4,	0.000	)
(1e-3,	0.000	)
(1e-2,	0.000	)
(1e-1,	0.000	)
(1,	0.000	)
    };
    
    \addplot[line width=0.3mm, mark=square, color=Flickr-color]  %
        plot coordinates {
(1e-8,	0.214	)
(1e-7,	0.139	)
(1e-6,	0.025	)
(1e-5,	0.002	)
(1e-4,	0.000	)
(1e-3,	0.000	)
(1e-2,	0.000	)
(1e-1,	0.000	)
(1,	0.000	)
    };

    \end{axis}
\end{tikzpicture}\hspace{0mm}\label{fig:vary-dim}%
}%
\end{small}
 \vspace{-1ex}
\caption{Recall when varying $\epsilon$.} \label{fig:recall}
\vspace{-4ex}
\end{figure}

%% file: tex/figs/running_time.tex
\pgfplotsset{ every non boxed x axis/.append style={x axis line style=-} }
\pgfplotsset{ every non boxed y axis/.append style={y axis line style=-} }
\begin{figure*}[!t]
\centering
\begin{small}
\begin{tikzpicture}
    \begin{customlegend}[
        legend entries={Online Stage, Preprocessing Stage},
        legend columns=2,
        area legend,
        legend style={at={(0.45,1.15)},anchor=north,draw=none,font=\small,column sep=0.15cm}]
        \addlegendimage{preaction={fill,query-color}}
        \addlegendimage{preaction={fill,preprocess-color}, pattern={crosshatch dots}}
    \end{customlegend}
\end{tikzpicture}
\\[-\lineskip]
\vspace{-4mm}

\subfloat[{\em Cora}]{
\begin{tikzpicture}[scale=1]
\begin{axis}[
    ybar,
    ybar=0pt,
    height=\columnwidth/2.6,
    width=\columnwidth/1.8,
    axis lines=left,
    ylabel={\it running time (sec)},
    xtick=data,
    bar width=0.18cm,
    enlarge x limits=true,
    xticklabel=\empty,
    xticklabel style = {font=\tiny},
    x tick label style={rotate=25,anchor=east},
    symbolic x coords={{\bf \algoc}, \algoe, \underline{\textsf{CFANE}}, \textsf{HK-Relax}, \textsf{PANE}, \textsf{SimRank}},
    ymin=0.001,
    ymax=100,
    ytick={0.001,0.01,0.1,1,10,100},
    ymode=log,
    log origin y=infty,
    log basis y={10},
    every axis y label/.style={at={(current axis.north west)},right=10mm,above=0mm},
    legend style={at={(0.02,0.98)},anchor=north west,cells={anchor=west},font=\tiny}
    ]
\addplot[preaction={fill,query-color}] coordinates {
    ({\bf \algoc},0.084)
    (\algoe,0.084)
    (\underline{\textsf{CFANE}},0.004)
    (\textsf{HK-Relax},0.473)
    (\textsf{PANE},0.003)
    (\textsf{SimRank},0.898)

};

\addplot[preaction={fill,preprocess-color}, pattern={crosshatch dots}] coordinates {
    ({\bf \algoc},0.01134)
    (\algoe,0.0458)
    (\underline{\textsf{CFANE}},82.013)
    (\textsf{PANE},0.616)
    
};
\end{axis}
\end{tikzpicture}\vspace{-2mm}\hspace{4mm}\label{fig:time-cora}%
}%
\subfloat[{\em PubMed}]{
\begin{tikzpicture}[scale=1]
\begin{axis}[
    ybar,
    ybar=0pt,
    height=\columnwidth/2.6,
    width=\columnwidth/1.8,
    axis lines=left,
    ylabel={\it running time (sec)},
    xtick=data,
    bar width=0.18cm,
    enlarge x limits=true,
    xticklabel=\empty,
    x tick label style={rotate=25,anchor=east},
    symbolic x coords={\algoc, {\bf \algoe}, \underline{\textsf{CFANE}}, \textsf{SimRank}, \textsf{PANE}, \textsf{PR-Nibble}},
    ymin=0.01,
    ymax=1000,
    ytick={0.01,0.1,1,10,100,1000},
    ymode=log,
    log origin y=infty,
    log basis y={10},
    every axis y label/.style={at={(current axis.north west)},right=10mm,above=0mm},
    legend style={at={(0.02,0.98)},anchor=north west,cells={anchor=west},font=\tiny},
    ]
\addplot[preaction={fill,query-color}] coordinates {
    (\algoc,0.181)
    ({\bf \algoe},0.577)
    (\underline{\textsf{CFANE}},0.025)
    (\textsf{SimRank},110.8)
    (\textsf{PANE},0.021)
    (\textsf{PR-Nibble},0.194)
};

\addplot[preaction={fill,preprocess-color}, pattern={crosshatch dots}] coordinates {
    (\algoc,0.03934)
    ({\bf \algoe},2.74094)
    (\underline{\textsf{CFANE}},415.047)
    (\textsf{PANE},3.69)
};
\end{axis}
\end{tikzpicture}\vspace{-2mm}\hspace{4mm}\label{fig:time-pubmed}%
}%
\subfloat[{\em BlogCL}]{
\begin{tikzpicture}[scale=1]
\begin{axis}[
    ybar,
    ybar=0pt,
    height=\columnwidth/2.6,
    width=\columnwidth/1.8,
    axis lines=left,
    ylabel={\it running time (sec)},
    xtick=data,
    bar width=0.18cm,
    enlarge x limits=true,
    xticklabel=\empty,
    x tick label style={rotate=25,anchor=east},
    symbolic x coords={{\bf \algoc}, \algoe, \underline{\textsf{CFANE}}, \textsf{PANE}, \textsf{SimAttr}, \textsf{HK-Relax} },
    ymin=0.001,
    ymax=1000,
    ytick={0.001,0.01,0.1,1,10,100,1000},
    ymode=log,
    log origin y=infty,
    log basis y={10},
    every axis y label/.style={at={(current axis.north west)},right=10mm,above=0mm},
    legend style={at={(0.02,0.98)},anchor=north west,cells={anchor=west},font=\tiny},
    ]
\addplot[preaction={fill,query-color}] coordinates {
    ({\bf \algoc},0.037)
    (\algoe,0.031)
    (\underline{\textsf{CFANE}},0.006)
    (\textsf{PANE},0.006)
    (\textsf{SimAttr},0.132)
     (\textsf{HK-Relax},0.715)
};

\addplot[preaction={fill,preprocess-color}, pattern={crosshatch dots}] coordinates {
    ({\bf \algoc},0.0194)
    (\algoe,0.066566)
    (\underline{\textsf{CFANE}},711.863)
    (\textsf{PANE},70.616)
};
\end{axis}
\end{tikzpicture}\vspace{-2mm}\hspace{4mm}\label{fig:time-blogcatalog}%
}%
\subfloat[{\em Flickr}]{
\begin{tikzpicture}[scale=1]
\begin{axis}[
    ybar,
    ybar=0pt,
    height=\columnwidth/2.6,
    width=\columnwidth/1.8,
    axis lines=left,
    ylabel={\it running time (sec)},
    xtick=data,
    bar width=0.18cm,
    enlarge x limits=true,
    xticklabel=\empty,
    x tick label style={rotate=25,anchor=east},
    symbolic x coords={{\bf \algoc}, \algoe, \underline{\textsf{PANE}}, \textsf{HK-Relax}, \textsf{Jaccard}, \textsf{CFANE}},
    ymin=0.01,
    ymax=2001,
    ytick={0.01,0.1,1,10,100,1000},
    ymode=log,
    log origin y=infty,
    log basis y={10},
    every axis y label/.style={at={(current axis.north west)},right=10mm,above=0mm},
    legend style={at={(0.02,0.98)},anchor=north west,cells={anchor=west},font=\tiny},
    ]
\addplot[preaction={fill,query-color}] coordinates {
    ({\bf \algoc},0.044)
    (\algoe,0.033)
    (\underline{\textsf{PANE}},0.018)
    (\textsf{HK-Relax},15.135)
    (\textsf{Jaccard},0.4053705735)
    (\textsf{CFANE},0.021)
};

\addplot[preaction={fill,preprocess-color}, pattern={crosshatch dots}] coordinates {
    ({\bf \algoc},0.049595)
    (\algoe,0.0823)
    (\underline{\textsf{PANE}},150.241)
    (\textsf{CFANE},1164.876)
};
\end{axis}
\end{tikzpicture}\vspace{-2mm}\hspace{2mm}\label{fig:time-flickr}%
}%

\vspace{-3ex}
\subfloat[{\em ArXiv}]{
\begin{tikzpicture}[scale=1]
\begin{axis}[
    ybar,
    ybar=0pt,
    height=\columnwidth/2.6,
    width=\columnwidth/1.8,
    axis lines=left,
    ylabel={\it running time (sec)},
    xtick=data,
    bar width=0.18cm,
    enlarge x limits=true,
    xticklabel=\empty,
    x tick label style={rotate=25,anchor=east},
    symbolic x coords={{\bf \algoc}, {\bf \algoe}, \underline{\textsf{HK-Relax}}, \textsf{PR-Nibble}, \textsf{APR-Nibble}, \textsf{WFD}},
    ymin=0.1,
    ymax=1000,
    ytick={0.01,0.1,1,10,100,1000},
    ymode=log,
    log origin y=infty,
    log basis y={10},
    every axis y label/.style={at={(current axis.north west)},right=10mm,above=0mm},
    legend style={at={(0.02,0.98)},anchor=north west,cells={anchor=west},font=\tiny},
    ]
\addplot[preaction={fill,query-color}] coordinates {
    ({\bf \algoc},0.737)
    ({\bf \algoe},0.699)
    (\underline{\textsf{HK-Relax}},140.686)
    (\textsf{PR-Nibble},5.942)
    (\textsf{APR-Nibble},0.824)
    (\textsf{WFD},5.310996502)
};

\addplot[preaction={fill,preprocess-color}, pattern={crosshatch dots}] coordinates {
    ({\bf \algoc},0.7927)
    ({\bf \algoe},1.45309)
    (\textsf{APR-Nibble},305.706)
};

\end{axis}
\end{tikzpicture}\vspace{-2mm}\hspace{4mm}\label{fig:time-arxiv}%
}%
\subfloat[{\em Yelp}]{
\begin{tikzpicture}[scale=1]
\begin{axis}[
    ybar,
    ybar=0pt,
    height=\columnwidth/2.6,
    width=\columnwidth/1.8,
    axis lines=left,
    ylabel={\it running time (sec)},
    xtick=data,
    bar width=0.18cm,
    enlarge x limits=true,
    xticklabel=\empty,
    x tick label style={rotate=25,anchor=east},
    symbolic x coords={\algoc, \algoe, \underline{{\bf \textsf{SimAttr}}}, \textsf{PANE}, \textsf{AttrRank}, \textsf{Node2Vec} },
    ymin=0.1,
    ymax=100000,
    ytick={0.1,1,10,100,1000,10000,100000},
    ymode=log,
    log origin y=infty,
    log basis y={10},
    every axis y label/.style={at={(current axis.north west)},right=10mm,above=0mm},
    legend style={at={(0.02,0.98)},anchor=north west,cells={anchor=west},font=\tiny},
    ]
\addplot[preaction={fill,query-color}] coordinates {
    (\algoc,4.949)
    (\algoe,15.263)
    (\underline{{\bf \textsf{SimAttr}}},2.557)
    (\textsf{PANE},0.829)
    (\textsf{AttrRank},0.09064185095)
    (\textsf{Node2Vec},0.7514391818)
};

\addplot[preaction={fill,preprocess-color}, pattern={crosshatch dots}] coordinates {
    (\algoc,3.246347)
    (\algoe,2.27097)
    (\textsf{PANE},680.275)
    (\textsf{AttrRank},531.1376717)
    (\textsf{Node2Vec},70848.43828)
};
\end{axis}
\end{tikzpicture}\vspace{-2mm}\hspace{4mm}\label{fig:time-yelp}%
}%
\subfloat[{\em Reddit}]{
\begin{tikzpicture}[scale=1]
\begin{axis}[
    ybar,
    ybar=0pt,
    height=\columnwidth/2.6,
    width=\columnwidth/1.8,
    axis lines=left,
    ylabel={\it running time (sec)},
    xtick=data,
    bar width=0.18cm,
    enlarge x limits=true,
    xticklabel=\empty,
    x tick label style={rotate=25,anchor=east},
    symbolic x coords={{\bf \algoc}, {\bf \algoe}, \underline{\textsf{$p$-Norm FD}}, \textsf{HK-Relax}, \textsf{PR-Nibble}, \textsf{CRD}},
    ymin=0.1,
    ymax=1000,
    ytick={0.1,1,10,100,1000},
    ymode=log,
    log origin y=infty,
    log basis y={10},
    every axis y label/.style={at={(current axis.north west)},right=10mm,above=0mm},
    legend style={at={(0.02,0.98)},anchor=north west,cells={anchor=west},font=\tiny},
    ]
\addplot[preaction={fill,query-color}] coordinates {
    ({\bf \algoc},0.399)
    ({\bf \algoe},0.299)
    (\underline{\textsf{$p$-Norm FD}},72.95444407)
    (\textsf{HK-Relax},127.334)
    (\textsf{PR-Nibble},4.004)
    (\textsf{CRD},411.0324511)
};

\addplot[preaction={fill,preprocess-color}, pattern={crosshatch dots}] coordinates {
    ({\bf \algoc},1.10873)
    ({\bf \algoe},0.2599)
};
\end{axis}
\end{tikzpicture}\vspace{-2mm}\hspace{4mm}\label{fig:time-reddit}%
}%
\subfloat[{\em Amazon2M}]{
\begin{tikzpicture}[scale=1]
\begin{axis}[
    ybar,
    ybar=0pt,
    height=\columnwidth/2.6,
    width=\columnwidth/1.8,
    axis lines=left,
    ylabel={\it running time (sec)},
    xtick=data,
    bar width=0.18cm,
    enlarge x limits=true,
    xticklabel=\empty,
    x tick label style={rotate=25,anchor=east},
    symbolic x coords={\algoc, {\bf \algoe}, \underline{\textsf{WFD}}, \textsf{$p$-Norm FD},\textsf{PR-Nibble}, \textsf{PANE}},
    ymin=1,
    ymax=10000,
    ytick={1,10,100,1000,10000},
    ymode=log,
    log origin y=infty,
    log basis y={10},
    every axis y label/.style={at={(current axis.north west)},right=10mm,above=0mm},
    legend style={at={(0.02,0.98)},anchor=north west,cells={anchor=west},font=\tiny},
    ]
\addplot[preaction={fill,query-color}] coordinates {
    (\algoc,36.883)
    ({\bf \algoe},36.679)
    (\underline{\textsf{WFD}},5574.513)
    (\textsf{$p$-Norm FD},3042.659)
    (\textsf{PR-Nibble},42.407)
    (\textsf{PANE},2.541)
};

\addplot[preaction={fill,preprocess-color}, pattern={crosshatch dots}] coordinates {
    (\algoc,6.9274)
    ({\bf \algoe},13.42497)
    (\textsf{PANE},1571.433)
};

\end{axis}
\end{tikzpicture}\vspace{-2mm}\hspace{2mm}\label{fig:time-amazon2m}%
}%
\vspace{-1mm}
\end{small}
\caption{Running times. Best method and competitor (in terms of precision) are bolded and \underline{underlined}, respectively.} \label{fig:time}
\vspace{-4mm}
\end{figure*}

%% file: tex/relatedwork.tex
\section{Related Work}
\subsection{Local Graph Clustering}
Local graph clustering (LGC) aims to find a high-quality local cluster without traversing the whole graph. The common characteristic of such methods is to optimize the conductance of the local cluster so that the target cluster is internally tightly connected while loosely connected to nodes outside the target cluster. Literature on LGC methods commonly distinguishes between two main types: random walk-based and flow-based methods (see \cite{baltsou2022local} for a systematic survey).
Among random walk-based methods~\cite{spielman2013local, andersen2006local,andersen2016almost, kloster2014heat, yang2019efficient, zhang2023constrained, yin2017local, yudong2022local, chhabra2023local, yuan2024index}, \texttt{Nibble} \cite{spielman2013local}, \texttt{PR-Nibble} \cite{andersen2006local} and subsequent \cite{andersen2016almost, zhang2023constrained, yin2017local, yudong2022local, chhabra2023local, yuan2024index} are proposed to perform local clustering via approximate personalized PageRank. Similarly, \cite{kloster2014heat, yang2019efficient} discuss the use of approximate heat kernel, including an LGC approach \texttt{HK-Relax} \cite{kloster2014heat}.
The flow-based methods include \cite{wang2017capacity, jung2020local, fountoulakis2020p, pmlr-v202-yang23d, de2023local}. 
\texttt{CRD} \cite{wang2017capacity}
converts capacity diffusion into a maximum-flow problem. \texttt{$p$-Norm FD} \cite{fountoulakis2020p} incorporates spectral and maximum-flow-based methods by choosing different $p$ values.

These methods usually provide theoretical guarantees on running time and approximation quality. However, these guarantees may not be useful in practical efficacy.
Considering this, recent works \cite{freitas2018local, yudong2022local, pmlr-v202-yang23d, de2023local} utilized additional graph resources besides connectivity
More specifically, \texttt{WFD} \cite{pmlr-v202-yang23d} adjusts edge weights based on attribute similarity using the Gaussian kernel, followed by local diffusion through a flow-based approach. \cite{de2023local} extracts noisy node labels from additional graph resources for diffusion on reweighed graphs. Unlike these methods requiring costly preprocessing as reported in Section~\ref{sec:exp}, ours seamlessly integrate node attribute-derived affinities into the diffusion process, speeding up preprocessing and improving attribute utilization.

\subsection{Community Search}
Community search aims to search densely connected communities for a user-specified query~\cite{hlx2019community,hlx2017community}. Similar to local clustering, community search is query-dependent and does not need to compute the graph globally. Researchers have devised different community models to define densely connected communities, including the most popular ones: $k$-core \cite{sozio2010community, cui2014local, barbieri2015efficient} and $k$-truss \cite{huang2014querying, huang2015approximate, liu2020vac}.
To incorporate attribute cohesiveness apart from structure cohesiveness, in recent years, a series of works \cite{fang2016effective, fang2017effective, huang2017attribute, zhu2020when, liu2020vac} on community search over attributed graphs have been developed. These methods usually consider keywords as the attributes of the nodes. Among these, \texttt{ACQ} \cite{fang2016effective, fang2017effective} and \texttt{ATC} \cite{huang2017attribute} apply a core-based and a truss-based approach, respectively, both of which aimed at maximizing the number of keywords matching the query in the resulting community. Our approach differs as we do not use the intersections of keywords as the metric of attribute similarity. Although \texttt{VAC} \cite{liu2020vac} relaxes restrictions on using keywords as attributes, it still diverges from ours in both the objective and the method for calculating attribute similarity, especially in its repetitive computation to derive the minimum attribute similarity in the best $k$-truss.
Another key difference is that community search literature imposes rigid topological constraints on the resulting community, which can lead to inferior outcomes. Moreover, these methods usually require offline index construction to enhance query processing speed, resulting in significant time and space overheads \cite{yudong2022local}. In contrast, our approach follows the LGC methods and designs an efficient attribute similarity computation method, eliminating the need for indices.
Therefore, the problem setting of our work is orthogonal to that of community search.

%% file: tex/proof.tex
\subsection{Theoretical Proofs}\label{sec:proof}
\subsubsection{Proof of Theorem \ref{lem:LP}}
\begin{proof}
We denote $\rrvec^{\ell}$ as the vector $\rrvec$ obtained at Line 3 and by $\bvec^{\ell}$ the remaining residual at Line 5 in $\ell$-th iteration. Then, according to Line 6, $\pvec$ can be represented by
\begin{equation}\label{eq:qvec}
\pvec = (1-\alpha)\sum_{\ell}^{\infty}{\rrvec^{\ell}}.
\end{equation}
Next, we consider $\{\rrvec^{1},\rrvec^{2},\ldots,\rrvec^{\ell},\rrvec^{\ell+1},\ldots,\rrvec^{\infty}\}$. By Lines 3 and 7-8, we have
\begin{align*}
&\rrvec^{1} = \fvec-\bvec^{1},\ \rrvec^{1} = \bvec^{1}+\alpha\rrvec^{1}\PM-\bvec^{2}\\
&\cdots \\
&\rrvec^{\ell} = \bvec^{\ell-1}+\alpha\rrvec^{\ell-1}\PM-\bvec^{\ell},\ \rrvec^{\ell+1} = \bvec^{\ell}+\alpha\rrvec^{\ell}\PM-\bvec^{\ell+1}\\
&\cdots.
\end{align*}
Then, Eq. \eqref{eq:qvec} can be rewritten as
\begin{small}
\begin{align}
\frac{\pvec}{1-\alpha} &= \fvec + \alpha\sum_{\ell=1}^{\infty}{\rrvec^{\ell}}\PM - \bvec^{\infty}\notag \\
& =\sum_{\ell=0}^{\infty}{\alpha^t\fvec\PM^\ell} -  \sum_{\ell=0}^{\infty}{\alpha^t\bvec^{\ell}\PM^\ell} \label{eq:q-bound}.
\end{align}
\end{small}
Recall that $\bvec^{\ell}$ always satisfies $\forall{v_i\in \V}$, ${\bvec^{\ell}_i}/{d(v_i)}<\epsilon$ (see Lines 3 and 5).
Let $\bvec$ be a length-$n$ vector where each $i$-th entry is $\epsilon\cdot d(v_i)$. Together with Eq. \eqref{eq:q-bound} and the matrix definition of $\pi(v_i,v_j)$ defined in Eq. \eqref{eq:pi-matrix}, we can bound each entry $\pvec_t$ by
\begin{align*}
\pvec_t &\ge (1-\alpha)\sum_{\ell=0}^{\infty}{\alpha^t(\fvec\PM^\ell)_t} -  (1-\alpha)\sum_{\ell=0}^{\infty}{\alpha^t(\bvec\PM^\ell)_t} \\
& = \sum_{v_i\in \V}{\fvec_i\cdot \pi(v_i,v_t)} - \sum_{v_i\in \V}{\bvec_i\cdot \pi(v_i,v_t)} \\
& = \sum_{v_i\in \V}{\fvec_i\cdot \pi(v_i,v_t)} - \sum_{v_i\in \V}{\epsilon\cdot d(v_i)\cdot  \pi(v_i,v_t)}.
\end{align*}
By the fact of $d(v_i)\cdot\pi(v_i,v_t)=d(v_t)\cdot\pi(v_t,v_i)$ (Lemma 1 in~\cite{lofgren2015bidirectional}) and $\sum_{v_i\in \V}{\pi(v_t,v_i)}=1$, the above inequality can be simplified as
\begin{align*}
\pvec_t & \ge \sum_{v_i\in \V}{\fvec_i\cdot \pi(v_i,v_t)} - \epsilon \cdot d(v_t) \cdot \sum_{v_i\in \V}{\pi(v_t,v_i)} \\
& = \sum_{v_i\in \V}{\fvec_i\cdot \pi(v_i,v_t)} - \epsilon \cdot d(v_t),
\end{align*}
which finishes the proof of Eq.~\eqref{eq:f-q-eps}.

In what follows, we analyze the time complexity of Algo.~\eqref{alg:fwd}.
First, suppose that Lines 3-7 in Algo.~\ref{alg:fwd} are executed for $L$ iterations. Now, we consider any iteration $\ell$. For ease of exposition, we denote $\rrvec^{\ell}$ as the vector $\rrvec$ obtained at Line 3 in $\ell$-th iteration and by $\rvec^{\ell}$ and $\pvec^{\ell}$ the residual and reserve vectors $\rvec$ and $\pvec$ at the beginning of $\ell$-th iteration, respectively. 
Accordingly, for each non-zero entry $\rrvec^{\ell}_i$ in $\rrvec^{\ell}$, $\rrvec^{\ell}_i\ge d(v_i)\cdot \epsilon$. 

First, we prove that at the beginning of any $\ell$-th iteration, the following equation holds:
\begin{equation}\label{eq:r-q-f}
\|\rvec^{\ell}\|_1 + \|\pvec^{\ell}\|_1 = \|\fvec\|_1.
\end{equation}
We prove this by induction. For the base case where $\ell=1$, i.e., at the beginning of the first iteration, we have $\rvec^{1}=\fvec$ and $\pvec^{1}=\mathbf{0}$, and thus, Eq.~\eqref{eq:r-q-f} holds. Next, we assume Eq.~\eqref{eq:r-q-f} holds at the beginning of $\ell$-th ($\ell>0$) iteration, i.e., $\|\rvec^{\ell}\|_1 + \|\pvec^{\ell}\|_1 = \|\fvec\|_1$.
According to Lines 5-7, we have
\begin{align*}
\pvec^{\ell+1} = \pvec^{\ell} + (1-\alpha)\cdot \rrvec^{\ell}\\
\rvec^{\ell+1} = \rvec^{\ell} -\rrvec + \alpha \cdot \rrvec^{\ell} \PM.
\end{align*}
As such,
\begin{align}
& \|\rvec^{\ell+1}\|_1 + \|\pvec^{\ell+1}\|_1 \notag\\
& = \|\pvec^{\ell}\|_1 + \|\rvec\|_1 + \alpha\cdot\|\rrvec^{\ell} \PM- \rrvec^{\ell}\|_1 \label{eq:r-q-f-ell}.
\end{align}
Note that
\begin{small}
\begin{align*}
\|\rrvec\PM-\rrvec^{\ell}\|_1 &= \sum_{v_i\in \V}\sum_{v_j\in \V}{\rrvec^{\ell}_j\cdot \PM_{j,i}} - \sum_{v_j\in \V}\rrvec^{\ell}_j \\
& =  \sum_{v_j\in \V}\rrvec^{\ell}_j\sum_{v_i\in \V}{\PM_{j,i}} - \sum_{v_j\in \V}\rrvec^{\ell}_j\\
& = \sum_{v_j\in \V}\rrvec^{\ell}_j\sum_{v_i\in \N(v_j)}{\frac{1}{d(v_j)}} -\sum_{v_j\in \V}\rrvec^{\ell}_j = 0,
\end{align*}
\end{small}
implying that $\|\rvec^{\ell+1}\|_1 + \|\pvec^{\ell+1}\|_1$ in Eq. \eqref{eq:r-q-f-ell} equals $0$, namely, Eq.~\eqref{eq:r-q-f} still holds.

As per Line 6, in each $\ell$-th iteration, Algo.~\ref{alg:fwd} converts $(1-\alpha)$ fraction of $\rrvec^{\ell}$ into $\pvec^{\ell}$, i.e., at least $(1-\alpha)\cdot \epsilon\cdot d(v_i)$ out of each non-zero entries in $\rrvec^{\ell}$ is passed to $\pvec^{\ell}$. After $L$ iterations, we obtain $\pvec^{L}$. Recall that by Eq. \eqref{eq:r-q-f}, $\|\pvec^{L}\|_1\le \|\fvec\|_1$. We obtain
\begin{equation}\label{eq:L-gamma-q-f}
\sum_{\ell=1}^{L}{\sum_{i\in \mathsf{supp}(\rrvec^{\ell})}{(1-\alpha)\cdot \epsilon\cdot d(v_i) }}\le \|\pvec^{L}\|_1 \le \|\fvec\|_1,
\end{equation}
which leads to $\sum_{\ell=1}^{L}{\sum_{i\in \mathsf{supp}(\rrvec^{\ell})}{d(v_i)}}\le \frac{\|\fvec\|_1}{(1-\alpha)\epsilon}$.

Notice that in Line 6, each non-zero entry in $\rrvec^{\ell}$ will result in $d(v_i)$ operations in the matrix-vector multiplication $\rrvec^{\ell}\PM$. Thus, the total cost of Line 6 for $L$ iterations is
$\sum_{\ell=1}^{L}{\sum_{i\in \mathsf{supp}(\rrvec^{\ell})}{d(v_i)}}$, which is bounded by $\frac{\|\fvec\|_1}{(1-\alpha)\epsilon}$ as aforementioned.
In addition, in each iteration, Lines 5 and 7 also merely involve operations on the non-zero entries in $\rrvec^{\ell}$. Their total cost for $L$ iterations can then be bounded by $\frac{\|\fvec\|_1}{(1-\alpha)\epsilon}$ as well. As for Line 3, in the first iteration, we need to inspect every entry in $\rvec^{1}$, and hence, the time cost is $\|\rvec^{1}\|_0=\left|\mathsf{supp}(\fvec)\right|$. In any subsequent $\ell$-th iteration, we solely need to inspect the entries in $\rvec^{\ell}$ affected by Line 7 in the previous iteration, which is also bounded by the non-zero entries in $\rrvec^{\ell-1}$. Hence, the overall time complexity of Algo.~\ref{alg:fwd} is $O\left(\max\left\{\left|\mathsf{supp}(\fvec)\right|,\frac{\|\fvec\|_1}{(1-\alpha)\epsilon}\right\}\right)$. The theorem is proved.
\end{proof}

\subsubsection{Proof of Theorem~\ref{lem:acc-adaptive}}
\begin{proof}
Similar to the proof of Theorem~\ref{lem:LP}, we let $\bvec^{\ell}$ be the remaining residual in the $\ell$-th iteration in Algo.~\ref{alg:iter-fwd}. By Lines 6 and Lines 8-11, Eq.~\eqref{eq:q-bound} still holds.
Notice that $\bvec^{\ell}$ is $\mathbf{0}$ when $\ell$-th iteration runs Lines 5-6. As such, $\bvec^{\ell}$ always satisfies $\forall{v_i\in \V}$, $\frac{\bvec^{\ell}_i}{d(v_i)}<\epsilon$ and then Eq.~\eqref{eq:f-q-eps} follows as in the proof of Theorem~\ref{lem:LP}.

Next, we analyze the time complexity of Algo.~\ref{alg:iter-fwd}.
Notice that according to Line 4 in Algo.~\eqref{alg:iter-fwd}, the total cost $C_{tot}$ entailed by the non-greedy operations (Lines 4-6) is bounded by $O\textstyle \left(\frac{\|\fvec\|_1}{(1-\alpha)\epsilon}\right)$. As for greedy operations (Lines 8-11), they will be conducted as in Algo.~\eqref{alg:fwd}. Note that Algo.~\ref{alg:iter-fwd} also terminates when $\rrvec$ is $\mathbf{0}$. Therefore, the total amount of greedy operations in \adiff is at most that needed in \gdiff (see Theorem~\ref{lem:LP}). In sum, the overall complexity of Algo.~\ref{alg:iter-fwd} is $\textstyle O\left(\max\left\{\left|\mathsf{supp}(\fvec)\right|,\frac{\|\fvec\|_1}{(1-\alpha)\epsilon}\right\}\right)$.
\end{proof}

\subsubsection{Proof of Lemma \ref{lem:q-size-adaptive}}
\begin{proof}
We assume Algo.~\ref{alg:iter-fwd} conducts Lines 3-11 for $L$ iterations. and we refer to the vectors $\pvec$, $\rvec$, and $\rrvec$ in each $\ell$-th iteration as in the proof of Theorem \ref{lem:LP}. Further, we assume that in $\ell_1$-th, $\ell_2$-th, $\ldots$, and $\ell_T$-th iterations, \adiff executes Lines 8-11.
Based on Inequality~\eqref{eq:L-gamma-q-f} and Eq.~\eqref{eq:r-q-f}, we can get
\begin{small}
\begin{align*}
\sum_{j=1}^{T}{\sum_{i\in \mathsf{supp}(\rrvec^{\ell_j})}{d(v_i) }} \le \frac{\|\pvec^{\ell_T}\|_1}{(1-\alpha) \epsilon} \le \frac{\|\fvec\|_1-\|\rvec^{\ell_T}\|_1}{(1-\alpha) \epsilon}.
\end{align*}
\end{small}
Let $\mathcal{L}=\{1,2,\ldots,L\}$ and $\mathcal{T}=\{\ell_1,\ell_2,\ldots,\ell_T\}$. As for the $\mathcal{L}\setminus \mathcal{T}$ iterations, Algo.~\eqref{alg:iter-fwd} conducts Lines 5-6. Notice that in such cases, we have $C_{tot} + \mathsf{vol}(\rvec)< \frac{\|\fvec\|_1}{(1-\alpha) \epsilon}$, meaning that
\begin{small}
\begin{align*}
\sum_{j\in \mathcal{L}\setminus \mathcal{T}}{\sum_{i\in \mathsf{supp}(\rvec^{j})}{d(v_i) }} \le \frac{\|\fvec\|_1}{(1-\alpha) \epsilon}.
\end{align*}
\end{small}
According to Lines 5-6 and Lines 8-11, all the non-zero elements in final $\pvec$ are from non-zero entries in $\rrvec^{j}\ \forall{j\in \mathcal{T}}$ and $\rvec^{j}\ \forall{j\in \mathcal{L}\setminus \mathcal{T}}$. Then,
\begin{small}
\begin{align*}
& \mathsf{vol}(\pvec) =\sum_{i\in \mathsf{supp}(\pvec)}{d(v_i)} \\
& \le \sum_{j=1}^{T}{\sum_{i\in \mathsf{supp}(\rrvec^{\ell_j})}{d(v_i) }} + \sum_{j\in \mathcal{L}\setminus \mathcal{T}}{\sum_{i\in \mathsf{supp}(\rvec^{j})}{d(v_i) }}\\
& \le \frac{2\|\fvec\|_1-\|\rvec^{\ell_T}\|_1}{(1-\alpha) \epsilon} = \frac{\beta\|\fvec\|_1}{(1-\alpha) \epsilon},
\end{align*}
\end{small}
where $\beta$ stands for a constant in the range $[1,2]$ since $0\le \|\rvec^{\ell_T}\|_1\le \|\fvec\|_1$.
Similarly, we obtain
\begin{small}
\begin{align*}
|\mathsf{supp}(\pvec)|\le \sum_{i\in \mathsf{supp}(\pvec)}{d(v_i)}=\mathsf{vol}(\pvec) \le \frac{\beta\|\fvec\|_1}{(1-\alpha) \epsilon}.
\end{align*}
\end{small}
When $\sigma\ge 1$, \adiff only executes Lines 8-11, and thus, $\mathcal{L}\setminus\mathcal{T}$ is an empty set. The above inequalities become
\begin{small}
\begin{equation*}
|\mathsf{supp}(\pvec)|\le \mathsf{vol}(\pvec) \le \frac{\|\fvec\|_1}{(1-\alpha) \epsilon}, 
\end{equation*}
\end{small}
which completes the proof.
\end{proof}

\subsubsection{Proof of Lemma~\ref{lem:SVD}}
\begin{proof}
We first need the following theorem.
\begin{theorem}[\bf Eckart–Young Theorem \cite{gloub1996matrix}]\label{lem:eym}
Suppose that $\MM_{k}\in\mathbb{R}^{n\times k}$ is the rank-$k$ approximation to $\MM\in\mathbb{R}^{n\times n}$ obtained by exact SVD, then
$\min_{rank(\widehat{\MM})\le k}{\|\MM-\widehat{\MM}\|_2}=\|\MM-\MM_{k}\|_2=\lambda_{k+1}$,
where $\lambda_{k+1}$ stands for the $(k+1)$-th largest singular value of $\MM$.
\end{theorem}
Suppose that $\UM\boldsymbol{\Lambda}\VM^{\top}$ is the exact $k$-SVD of $\XM$, by Theorem \ref{lem:eym}, we have $\|\UM\boldsymbol{\Lambda}\VM^{\top}-\XM\|_2\le \lambda_{k+1}$,
where $\lambda_{k+1}$ is the $(k+1)$-th largest singular value of $\XM$. 
Let $\widehat{\UM}\widehat{\boldsymbol{\Lambda}}\widehat{\VM}^{\top}$ be the $k$-SVD of $\XM\XM^{\top}$. Similarly, from Theorem \ref{lem:eym}, we get $\|\widehat{\UM}\widehat{\boldsymbol{\Lambda}}\widehat{\VM}^{\top}-\XM\XM^{\top}\|_2\le \widehat{\lambda}_{k+1}$,
where $\widehat{\lambda}_{k+1}$ is the $(k+1)$-th largest singular value of $\XM\XM^{\top}$.

According to~\cite{strang2022introduction}, columns in $\UM$ are the eigenvectors of matrix $\XM\XM^{\top}$ and the squared singular values of $\XM$ are the eigenvalues of $\XM\XM^{\top}$. Given that singular values are non-negative, all the eigenvalues of $\XM\XM^{\top}$ are also non-negative. $\boldsymbol{\Lambda}^2$ and $\UM$ contain the $k$-largest eigenvalues and corresponding eigenvectors of $\XM\XM^{\top}$, and ${\lambda_{k+1}}^2=\widehat{\lambda}_{k+1}$.  Further, by Theorem 4.1 in \cite{zhang2018arbitrary}, it can be verified that $\boldsymbol{\Lambda}^2$ and $\UM$ are the top-$k$ singular values and left/right singular vectors of $\XM\XM^{\top}$, respectively, and ${\lambda_{k+1}}^2$ is the $(k+1)$-th largest singular value of $\XM\XM^{\top}$. Consequently, 
$\|\UM\boldsymbol{\Lambda}^2\UM^{\top}-\XM\XM^{\top}\|_2 \le {\lambda_{k+1}}^2$ and the proof is done.
\end{proof}

\subsubsection{Proof of Theorem~\ref{lem:yy-exp-cos}}
\begin{proof}
Recall that for $v_i\in \V$, vector $\xvec^{(i)}$ is $L_2$ normalized, i.e., $\|\xvec^{(i)}\|_2=1$. Thus, we can derive $\|\xvec^{(i)}-\xvec^{(j)}\|^2_2=2(1-\cos{(\xvec^{(i)},\xvec^{(j)})}=2(1-\xvec^{(i)}\cdot \xvec^{(j)} ) \in [0,4]$. On its basis, $f(v_i,v_j)$ in Eq.~\eqref{eq:exp-cos} can be transformed as follows:
\begin{small}
\begin{align}
& f(v_i,v_j) = \exp{\left(\frac{\xvec^{(i)}\cdot \xvec^{(j)}}{\delta}\right)} = \exp{\left(\frac{1-\frac{1}{2}\|\xvec^{(i)}-\xvec^{(j)}\|^2_2}{\delta}\right)} \notag \\
& = \exp{\left(\frac{1}{\delta}-\frac{\|\xvec^{(i)}-\xvec^{(j)}\|^2_2}{2\delta}\right)}\notag \\
& = \exp\left(\frac{1}{\delta}\right)\cdot \exp{\left(-\frac{\|\xvec^{(i)}-\xvec^{(j)}\|^2_2}{2\delta}\right)}.\label{eq:f-exp-xx}
\end{align}
\end{small}

According to Theorem 1 in \cite{yu2016orthogonal} and the mathematical form of $\widehat{\YM}=\frac{1}{\delta}\XM\boldsymbol{\Sigma}\QM=\frac{1}{\delta}{\UM{\boldsymbol{\Lambda}}\boldsymbol{\Sigma}\QM}$ via Lines 6-9, we have 
\begin{small}
\begin{align*}
& \mathbb{E}\left[ \boldsymbol{K}\cdot \boldsymbol{K}^{\top}\right] = \exp{\left(-\frac{\|\xvec^{(i)}-\xvec^{(j)}\|^2_2}{2\delta}\right)},
\end{align*}
\end{small}
where $\boldsymbol{K}=\frac{1}{\sqrt{d}}\cdot sin(\yhvec^{(i)}) \mathbin\Vert cos(\yhvec^{(j)})$.
Plugging the definitions of $f(v_i,v_j)$ in Eq.~\eqref{eq:f-exp-xx} and $\YM$ in Eq.~\eqref{eq:Y-ecos} into the above equation proves the theorem.
\end{proof}

\subsubsection{Proof of Lemma~\ref{lem:Z-cost}}
\begin{proof}
According to ~\cite{halko2011finding}, the invocation of $k$-SVD over $\XM$ runs in $O(ndk+nk^2)$ time. Note that the number of iterations in $k$-SVD is regarded as a constant and thus is omitted since it is set to small integers, e.g., 7, in practice, and $k$ is less than $d$. By Eq.~\eqref{eq:y-mean-z}, the computations of ${\yvec}^{\ast}$ and $\zvec^{(i)}\ \forall{v_i\in \V}$ at Lines 10-11 need $O(nk)$ time. Thus, when $f(\cdot,\cdot)$ is the cosine similarity function, the construction of TNAM $\ZM$ requires $O(ndk)$ time.

In comparison, when $f(\cdot,\cdot)$ is the exponential cosine similarity function, the QR decomposition at Line 7 and constructing $\YM$ at Line 9 take $O(k^3)$ and $O(nk^2)$ time, respectively. In turn, the overall time complexity of Algo.~\ref{alg:Z} is bounded by $O(ndk)$, which can be reduced to $O(nd)$ since $k$ is regarded as a constant.
\end{proof}

\subsubsection{Proof of Theorem \ref{lem:main-acc}}
\begin{proof}
Let $\rhovec^{\circ}$ be the vector returned by Algo.~\ref{alg:iter-fwd} invoked at Line 5 in Algo.~\ref{alg:main}. According to Theorem~\ref{lem:acc-adaptive}, $\forall{v_j\in \V}$, 
\begin{equation*}
\sum_{j\in \mathsf{supp}(\pivec^{\prime})}{\sum_{v_i\in \V}{\pivec^{\prime}_i\cdot s(v_i,v_j)}\cdot d(v_j) \cdot \pi(v_j,v_t)}-\rhovec^{\circ}_t \ge 0\ \text{and}
\end{equation*}
\begin{equation*}
\sum_{j\in \mathsf{supp}(\pivec^{\prime})}{\sum_{v_i\in \V}{\pivec^{\prime}_i\cdot s(v_i,v_j)}\cdot d(v_j) \cdot \pi(v_j,v_t)}-\rhovec^{\circ}_t \le \epsilon \cdot d(v_t),
\end{equation*}
yielding $$0\le \sum_{j\in \mathsf{supp}(\pivec^{\prime})}{\sum_{v_i\in \V}{\pivec^{\prime}_i\cdot s(v_i,v_j)}\cdot \frac{d(v_j)}{d(v_t)} \cdot \pi(v_j,v_t)}-\frac{\rhovec^{\circ}_t}{d(v_t)} \le {\epsilon}.$$
By the fact of $\pi(v_t,v_j)=\frac{d(v_j)}{d(v_t)}\cdot \pi(v_j,v_t)$ (Lemma 1 in ~\cite{lofgren2015bidirectional}) and $\rhovec^{\prime}_t=\frac{\rhovec^{\circ}_t}{d(v_t)}$ (Line 6), we have
\begin{align*}
0\le \sum_{j\in \mathsf{supp}(\pivec^{\prime})}{\sum_{v_i\in \V}{\pivec^{\prime}_i\cdot s(v_i,v_j)}\cdot \pi(v_t,v_j)}-\rhovec^{\prime}_t \le {\epsilon}.
\end{align*}
Further, we obtain
\begin{align*}
\rhovec^{\prime}_t &\le \sum_{j\in \mathsf{supp}(\pivec^{\prime})}{\sum_{v_i\in \V}{\pivec^{\prime}_i\cdot s(v_i,v_j)}\cdot \pi(v_t,v_j)} \\
\rhovec^{\prime}_t &\ge \sum_{j\in \mathsf{supp}(\pivec^{\prime})}{\sum_{v_i\in \V}{\pivec^{\prime}_i\cdot s(v_i,v_j)}\cdot \pi(v_t,v_j)} - {\epsilon} .
\end{align*}
Notice that $\pivec$ obtained at Line 2 in Algo.~\ref{alg:main} satisfies $0 \le \pi(v_s,v_i)-\pivec^{\prime}_i \le \epsilon\cdot d(v_i)\ \forall{v_i\in \V}$ according to Theorem~\ref{lem:acc-adaptive}. We then derive
\begin{equation*}
\rhovec^{\prime}_t \le \sum_{j\in \mathsf{supp}(\pivec^{\prime})}{\sum_{v_i\in \V}{\pi(v_s,v_i)\cdot s(v_i,v_j)}\cdot \pi(v_t,v_j)}\le \rhovec_t
\end{equation*}
and
\begin{align*}
& \rhovec^{\prime}_t \\
& \ge \sum_{j\in \mathsf{supp}(\pivec^{\prime})}{\sum_{v_i\in \V}{\left(\pi(v_s,v_i)-\epsilon  d(v_i)\right)\cdot s(v_i,v_j)}\cdot \pi(v_t,v_j)} - {\epsilon},
\end{align*}
where the latter leads to
\begin{align*}
& \rhovec_t - \rhovec^{\prime}_t \le \epsilon + \sum_{j\in \mathsf{supp}(\pivec^{\prime})}{\sum_{v_i\in \V}{\epsilon d(v_i) \cdot s(v_i,v_j)}\cdot \pi(v_t,v_j)} \\
&\quad + \sum_{j\in \{1,\ldots,n\} \setminus \mathsf{supp}(\pivec^{\prime})}{\sum_{v_i\in \V}{ \pi(v_s,v_i)\cdot s(v_i,v_j)}\cdot \pi(v_t,v_j)}.
\end{align*}
Recall that $0 \le \pi(v_s,v_i)-\pivec^{\prime}_i \le \epsilon\cdot d(v_i)\ \forall{v_i\in \V}$. Thus, $\forall{i} \notin \mathsf{supp}(\pivec^{\prime})$, $\pi(v_s,v_i)\le \pivec^{\prime}_i + \epsilon\cdot d(v_i) = \epsilon\cdot d(v_i)$. Accordingly,
\begin{align*}
\rhovec_t - \rhovec^{\prime}_t & \le \epsilon + \sum_{v_j\in \V}{\sum_{v_i\in \V}{\epsilon d(v_i) \cdot s(v_i,v_j)}\cdot \pi(v_t,v_j)} \\
& \le \epsilon + \sum_{v_i\in \V}{\epsilon d(v_i)\cdot \sum_{v_j\in \V}{s(v_i,v_j)\cdot \pi(v_t,v_j)}} \\
&\textstyle \le \left(1+\sum_{v_i\in \V}{d(v_i)\cdot \max_{v_j\in \V}{s(v_i,v_j)}}\right)\cdot \epsilon,
\end{align*}
which proves the theorem.
\end{proof}

\subsubsection{Proof of Lemma~\ref{lem:GNN-sol}}
\begin{proof}
By setting its derivative w.r.t. $\HM$ to zero and , we obtain the optimal $\HM$ as:
\begin{align}
& \frac{\partial{\{(1-\alpha)\cdot\|\HM - \HM^{\circ}\|^2_F + \alpha \cdot trace(\HM^{\top}(\IM-\NAM)\HM)\}}}{\partial{\HM}}=0 \notag\\
& \Longrightarrow (1-\alpha)\cdot(\HM - \HM^{\circ}) + \alpha (\IM-\NAM)\HM = 0 \notag\\
& \Longrightarrow \HM = (1-\alpha)\cdot \left(\IM-\alpha\NAM\right)^{-1} \HM^{\circ}. \label{eq:Z-derivative}
\end{align}
By the property of the Neumann series, we have 
$(\IM - \alpha \NAM)^{-1} = \sum_{\ell=0}^{\infty}{\alpha^t \NAM^{\ell}}$. Plugging it into Eq.~\eqref{eq:Z-derivative} completes the proof.
\end{proof}

%% file: tex/additional-experiments.tex
\subsection{Additional Experiments}\label{sec:add-exp}

\subsubsection{Parameter Analysis}\label{sec:param}
In this set of experiments, we empirically investigate the impact of three key parameters in \algoc and \algoe: the restart factor $\alpha$, parameter $\sigma$, and the dimension $k$ of TNAM vectors. For each of them, we run \algoc and \algoe over {\em Cora}, {\em PubMed}, {\em BlogCalaog}, {\em Flickr}, and {\em ArXiv}, respectively, by varying the parameter with others fixed.

\input{tex/figs/vary_params}

\stitle{Varying $\alpha$}
Figs.~\ref{fig:vary-alpha} and \ref{fig:vary-alpha-e} show the precision achieved by \algoc and \algoe on five datasets, respectively, when $\alpha$ is varied from $0.0$ to $0.9$ with step size $0.1$. It can be clearly observed that both \algoc and \algoe present nearly identical behaviors on all datasets when varying $\alpha$. That is, the precision scores increase conspicuously with $\alpha$ increasing. The only exception is on {\em BlogCL}, where the best result is attained when $\alpha=0.8$. Recall that in Algo.~\ref{alg:iter-fwd}, when $\alpha$ is small, \adiff will convert substantial residuals into reserves of nearby nodes and distribute only a few to far-reaching neighbors, yielding local clusters with diminutive size, and hence, sub-par result quality. 

\stitle{Varying $\sigma$}
Next, we study the parameter $\sigma$ for balancing greedy and non-greedy operations in \adiff. Recall that a large $\sigma$ indicates running more greedy operations in \adiff, which degrades to \gdiff when $\sigma=1$.
Figs.~\ref{fig:vary-sigma} and \ref{fig:vary-sigma-e} plot the precision scores when increasing $\sigma$ from 0.0 to $1.0$ in \algoc and \algoe, respectively, on five datasets.
We can observe that both \algoc and \algoe (i) are not sensitive to $\sigma$ on {\em Cora} and {\em PubMed}, (ii) undergo a significant performance downturn when $\sigma$ is beyond $0.1$ on {\em BlogCatlog} and {\em Flickr}, and (iii) on {\em ArXiv}, see a considerable uptick in precision when $\sigma$ rises from $0$ to $0.2$ and invariant results afterward. \algoc and \algoe favor a small $\sigma$ on {\em BlogCatlog} and {\em Flickr} datasets with high average degrees (${m}/{n}>60$) because greedy operations are sensitive to high-degree nodes and tend to return small local clusters on such graphs (i.e., fewer non-zero entries in $\pvec$), as analyzed in Section~\ref{sec:gdiff}.

\stitle{Varying $k$}
Figs.~\ref{fig:vary-k} and \ref{fig:vary-k-e} show the precision values of \algoc and \algoe when varying the dimension $k$ of TNAM vectors $\ZM$ in $\{8, 16, 32, 64, 128, d\}$.
On citation networks {\em Cora}, {\em PubMed}, and {\em ArXiv}, we can see that the performance of both \algoc and \algoe remains stable when $k$ is increased from $8$ to $d$, except a slight drop on {\em Cora} when $k=8$.
In comparison, on social networks {\em BlogCL} and {\em Flickr}, a remarkable improvement and reduction in performance can be observed when $k$ increases from $8$ to $32$ and from $32$ to $d$, respectively.
The reason is that both {\em BlogCL} and {\em Flickr} have numerous distinct attributes in $\XM$ ($d=8,189$ and $d=12,047$), which embody substantial noisy information. Meanwhile, our $k$-SVD essentially denoises attribute data by extracting $k$ key components.
The observations manifest the effectiveness of our TNAM construction technique (Section~\ref{sec:construct-z}) in capturing the attribute similarity of node pairs with a small dimension $k$, e.g., $16$ or $32$.

\subsubsection{Ablation Study}
To study the effectiveness of the SNAS in our \ppr definition (Eq.~\eqref{eq:csim}), our graph diffusion algorithm \adiff in Section~\ref{sec:adiff}, as well as the $k$-SVD in Section~\ref{sec:construct-z}, we create three ablated versions for \algoc and \algoe, respectively. Particularly, the variants of \algoc and \algoe without \adiff are implemented using \gdiff as the diffusion component.

Table~\ref{tbl:ablation} reports the best precision scores attained by each method on 8 datasets. We can see remarkable performance decreases in both \algoc and \algoe after disabling any of these three ingredients, especially the SNAS. The only exception is on {\em Amazon2M}, where \algoc achieves a better result when SNAS is removed from \ppr, whereas \algoe exhibits a radically different phenomenon. This indicates that the exponential cosine similarity is more robust in modelling the similarity of node attributes. From Table~\ref{tbl:ablation}, \adiff is another key component affecting the resulting quality, which accords with our analysis of \gdiff's deficiencies in Section~\ref{sec:adiff}. Consistent with the observations from Figs.~\ref{fig:vary-k} and ~\ref{fig:vary-k-e}, the $k$-SVD in Algo.~\ref{alg:Z} improves the performance of \algo due to its denoising ability.

\input{tex/figs/ablation}

\input{tex/figs/Cond-WCSS}

\subsubsection{Conductance and WCSS}
This set of experiments studies the average {\em conductance}~\cite{lovasz1993random} and the average {\em within-cluster sum of squares} (WCSS)~\cite{macqueen1967some} of ground-truth and the local clusters output by \algoc, \algoe, and 17 competitors. Given a local cluster $C_s$, conductance only measures the connectivity between nodes in $C_s$ and nodes outside $\C_s$, while WCSS merely evaluates the variance of attribute vectors of nodes in $C_s$. Intuitively, lower conductance and WCSS indicate a higher clustering quality.
The conductance and WCSS values of all methods on the eight datasets are reported in Table~\ref{tbl:node-clustering-cond-WCSS}.
We highlight the top 3 best results (with the smallest differences from the ground truth) on each dataset in blue, with darker shades indicating higher quality.
From Table~\ref{tbl:node-clustering-cond-WCSS}, we can see that none of the evaluated methods perform best on all datasets.
But notably, \algoc and \algoe obtain the top $3$ conductance or WCSS results on eight and seven datasets, respectively, whereas the best competitor \textsf{PR-Nibble} is ranked in the top $3$ on six datasets, meaning that local clusters by \algo achieve a good balance of structure cohesiveness and attribute homogeneity.
Another observation is that compared to WCSS, conductance values by different methods vary markedly. This is because nodes in a graph have divergent degrees, and hence, a small change in a cluster can lead to a large difference in conductance.

\input{tex/figs/scalability}

\subsubsection{Scalability Evaluation}\label{sec:scale}
This section experimentally evaluates the scalability of \algoc and \algoe on four large graphs {\em ArXiv}, {\em Yelp}, {\em Reddit}, and {\em Amazon2M} by varying diffusion threshold $\epsilon$ and dimension $k$. 

Fig.~\ref{fig:scale-c-eps-time} and \ref{fig:scale-e-eps-time} depict the running times of \algoc and \algoe on the four datasets when decreasing $\epsilon$ from $1$ to $10^{-8}$, respectively. Specifically, 
the average runtime of both \algoc and \algoe increases by roughly an order of magnitude when there is a tenfold decrease in $\epsilon$, which is consistent with our complexity analysis of \algo in Section \ref{sec:main-algo}.

Although \algoc and \algoe are local algorithms, i.e., their time complexities are independent
of $n$ and $m$, we can observe that under the same $\epsilon$ settings, their empirical times vary on datasets with varied sizes. Note that this difference is caused by their diverse graph structures that lead to different data locality and memory access patterns. For example, although {\em Reddit} contains more nodes and $99\times$ more edges than {\em ArXiv}, their running times are comparable when $\epsilon\ge 10^{-3}$ and the cost for {\em ArXiv} turns to be markedly higher than that for {\em Reddit} when $10^{-7}\le \epsilon \le 10^{-4}$. The reason is that nodes in {\em ArXiv} are sparsely connected, making the matrix-vector multiplications in Algo.~\ref{alg:iter-fwd} inefficient.
As for {\em Yelp} and {\em Amazon2M}, they encompass significantly more nodes and edges than {\em ArXiv} and {\em Reddit}, and, hence, are more likely to yield cache misses and intensive memory access patterns that result in lower efficiency.

In Fig.~\ref{fig:scale-c-k-time} and \ref{fig:scale-e-k-time}, we display the running times of \algoc and \algoe when increasing $k$ from $8$ to $d$. Notably, the time required by \algoc and \algoe remains stable when varying $k$ in $\{8,16,32,64,128\}$, indicating that the time cost of \algo is dominated by $1/\epsilon$ when $k$ is not large.

\input{tex/figs/without_attributes}
\subsubsection{Quality Evaluation on Graphs without Attributes}
This set of experiments evaluates the performance of \algo (i.e., \algo (w/o SNAS)) against 4 strong LGC baselines (\texttt{PR-Nibble}, \texttt{HK-Relax}, \texttt{CRD}, and \texttt{$p$-Norm FD}) on graphs without node attributes in terms of local clustering quality, using the same evaluation protocol in Section~\ref{sec:quality}.
Table~\ref{tbl:without-attrs-data} lists the statistics of the non-attributed graph datasets of various volumes and types for evaluation~\cite{yang2012defining}.
In \textit{com-DBLP}, the nodes represent authors, and the edges represent co-authorship between published authors. The ground-truth clusters are formed based on the publication venue. \textit{com-Amazon} is a co-purchasing network where products are nodes, and edges represent frequently co-purchased items. The ground-truth clusters are determined by product categories. \textit{com-Orkut} is a social network where users establish friendships and join groups. The groups created by users are considered the ground-truth clusters.

The average precisions of the local clusters produced by \algo and 4 competitors are presented in Table~\ref{tbl:without-attrs}.
It can be observed that \algo consistently delivers the best performance across all three datasets. Specifically, on the medium-sized datasets \textit{com-DBLP} and \textit{com-Amazon}, \algo surpasses the best competitors, \texttt{PR-Nibble} and \texttt{$p$-Norm FD}, by a significant gain of $2.5\%$ and $4.8\%$ in precision, respectively. On the large graph \textit{com-Orkut} with 117.2 million edges, \algo still outperforms the state-of-the-art baseline \texttt{PR-Nibble} with an improvement of $0.2\%$.
The results manifest that our proposed BDD in \algo can accurately capture the affinity between nodes even without node attributes through the bidirectional random walks that consider the node importance from the perspectives of both, as remarked in Section~\ref{sec:BDD-desc}. 

%% file: tex/figs/vary_params.tex
 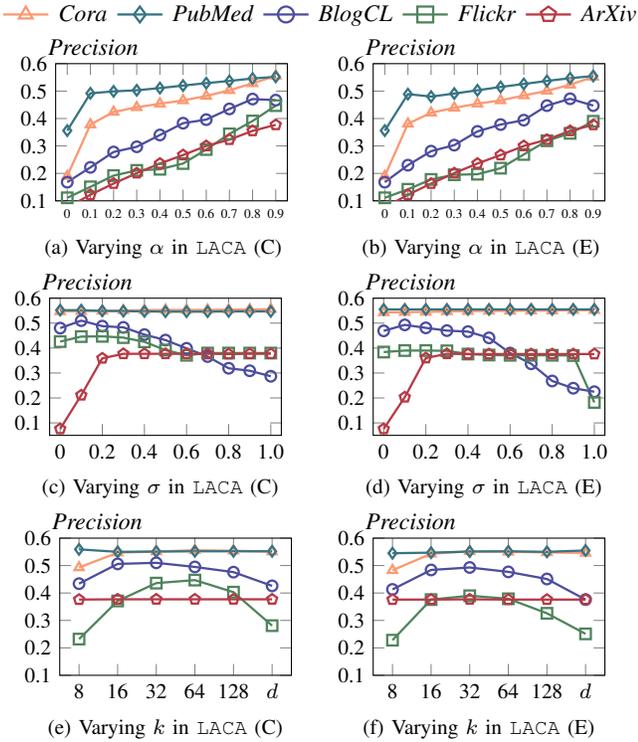
\begin{figure}[!t]
\centering
\begin{small}
\begin{tikzpicture}
    \begin{customlegend}[legend columns=5,
        legend entries={{\em Cora}, {\em PubMed}, {\em BlogCL}, {\em Flickr}, {\em ArXiv}},
        legend style={at={(0.45,1.35)},anchor=north,draw=none,font=\small,column sep=0.05cm}]
    \addlegendimage{line width=0.3mm,mark size=3pt,mark=triangle, color=Cora-color}
    \addlegendimage{line width=0.3mm,mark size=3pt,mark=diamond, color=Pubmed-color}
    \addlegendimage{line width=0.3mm,mark size=3pt,mark=o, color=Blogcatalog-color}
    \addlegendimage{line width=0.3mm,mark size=3pt,mark=square, color=Flickr-color}
    \addlegendimage{line width=0.3mm,mark size=3pt,mark=pentagon, color=Arxiv-color}
    \end{customlegend}
\end{tikzpicture}
\\[-\lineskip]
\vspace{-4mm}
\subfloat[Varying $\alpha$ in \algoc]{
\begin{tikzpicture}[scale=1,every mark/.append style={mark size=2pt}]
    \begin{axis}[
        height=\columnwidth/2.6,
        width=\columnwidth/1.9,
        ylabel={\it Precision},
        xmin=0.5, xmax=10.5,
        ymin=0.10, ymax=0.6,
        xtick={1,2,3,4,5,6,7,8,9,10},
        ytick={0.1,0.2,0.3,0.4,0.5,0.6},
        xticklabel style = {font=\tiny},
        yticklabel style = {font=\footnotesize},
        xticklabels={0,0.1,0.2,0.3,0.4,0.5,0.6,0.7,0.8,0.9},
        yticklabels={0.1,0.2,0.3,0.4,0.5,0.6},
        every axis y label/.style={font=\small,at={(current axis.north west)},right=5mm,above=0mm},
        legend style={fill=none,font=\small,at={(0.02,0.99)},anchor=north west,draw=none},
    ]
    \addplot[line width=0.3mm, mark=triangle, color=Cora-color]  %
        plot coordinates {
(1,	0.190	)
(2,	0.378	)
(3,	0.424	)
(4,	0.441	)
(5,	0.454	)
(6,	0.466	)
(7,	0.482	)
(8,	0.503	)
(9,	0.528	)
(10,	0.556	)
    };

    \addplot[line width=0.3mm, mark=diamond, color=Pubmed-color]  %
        plot coordinates {
(1,	0.356	)
(2,	0.492	)
(3,	0.499	)
(4,	0.503	)
(5,	0.511	)
(6,	0.520	)
(7,	0.529	)
(8,	0.537	)
(9,	0.546	)
(10,	0.552	)
    };

    \addplot[line width=0.3mm, mark=o, color=Blogcatalog-color]  %
        plot coordinates {
(1,	0.168	)
(2,	0.222	)
(3,	0.278	)
(4,	0.297	)
(5,	0.340	)
(6,	0.383	)
(7,	0.396	)
(8,	0.435	)
(9,	0.471	)
(10,	0.467	)
    };

    \addplot[line width=0.3mm, mark=square, color=Flickr-color]  %
        plot coordinates {
(1,	0.111	)
(2,	0.151	)
(3,	0.192	)
(4,	0.211	)
(5,	0.215	)
(6,	0.236	)
(7,	0.287	)
(8,	0.344	)
(9,	0.391	)
(10,	0.447	)
    };

    \addplot[line width=0.3mm, mark=pentagon, color=Arxiv-color]  %
        plot coordinates {
(1,	0.076	)
(2,	0.122	)
(3,	0.163	)
(4,	0.201   )
(5,	0.236	)
(6,	0.267	)
(7,	0.300	)
(8,	0.323	)
(9,	0.355	)
(10,	0.377	)
    };

    \end{axis}
\end{tikzpicture}\hspace{4mm}\label{fig:vary-alpha}%
}
\subfloat[Varying $\alpha$ in \algoe]{
\begin{tikzpicture}[scale=1,every mark/.append style={mark size=2pt}]
    \begin{axis}[
        height=\columnwidth/2.6,
        width=\columnwidth/1.9,
        ylabel={\it Precision},
        xmin=0.5, xmax=10.5,
        ymin=0.10, ymax=0.6,
        xtick={1,2,3,4,5,6,7,8,9,10},
        ytick={0.1,0.2,0.3,0.4,0.5,0.6},
        xticklabel style = {font=\tiny},
        yticklabel style = {font=\footnotesize},
        xticklabels={0,0.1,0.2,0.3,0.4,0.5,0.6,0.7,0.8,0.9},
        yticklabels={0.1,0.2,0.3,0.4,0.5,0.6},
        every axis y label/.style={font=\small,at={(current axis.north west)},right=5mm,above=0mm},
        legend style={fill=none,font=\small,at={(0.02,0.99)},anchor=north west,draw=none},
    ]
    \addplot[line width=0.3mm, mark=triangle, color=Cora-color]  %
        plot coordinates {
(1,	0.190	)
(2,	0.381	)
(3,	0.421	)
(4,	0.439	)
(5,	0.454	)
(6,	0.466	)
(7,	0.484	)
(8,	0.501	)
(9,	0.523	)
(10,	0.552	)
    };

    \addplot[line width=0.3mm, mark=diamond, color=Pubmed-color]  %
        plot coordinates {
(1,	0.356	)
(2,	0.489	)
(3,	0.480	)
(4,	0.491	)
(5,	0.503	)
(6,	0.515	)
(7,	0.526	)
(8,	0.537	)
(9,	0.547	)
(10,	0.555	)
    };

    \addplot[line width=0.3mm, mark=o, color=Blogcatalog-color]  %
        plot coordinates {
(1,	0.168	)
(2,	0.230	)
(3,	0.281	)
(4,	0.303	)
(5,	0.353	)
(6,	0.378	)
(7,	0.394	)
(8,	0.447	)
(9,	0.472	)
(10,	0.447	)
    };

    \addplot[line width=0.3mm, mark=square, color=Flickr-color]  %
        plot coordinates {
(1,	0.111	)
(2,	0.142	)
(3,	0.178	)
(4,	0.195	)
(5,	0.197	)
(6,	0.219	)
(7,	0.269	)
(8,	0.319	)
(9,	0.346	)
(10,	0.390	)
    };

    \addplot[line width=0.3mm, mark=pentagon, color=Arxiv-color]  %
        plot coordinates {
(1,	0.076	)
(2,	0.122	)
(3,	0.163	)
(4,	0.201   )
(5,	0.236	)
(6,	0.267	)
(7,	0.300	)
(8,	0.323	)
(9,	0.355	)
(10,	0.377	)
    };

    \end{axis}
\end{tikzpicture}\hspace{4mm}\label{fig:vary-alpha-e}%
}
\vspace{-2ex}
\subfloat[Varying $\sigma$ in \algoc]{
\begin{tikzpicture}[scale=1,every mark/.append style={mark size=2pt}]
    \begin{axis}[
        height=\columnwidth/2.6,
        width=\columnwidth/1.9,
        ylabel={\it Precision},
        xmin=0.5, xmax=11.5,
        ymin=0.05, ymax=0.6,
        xtick={1,3,5,7,9,11},
        ytick={0.1,0.2,0.3,0.4,0.5,0.6},
        xticklabel style = {font=\footnotesize},
        yticklabel style = {font=\footnotesize},
        xticklabels={0,0.2,0.4,0.6,0.8,1.0},
        yticklabels={0.1,0.2,0.3,0.4,0.5,0.6},
        every axis y label/.style={font=\small,at={(current axis.north west)},right=5mm,above=0mm},
        legend style={fill=none,font=\small,at={(0.02,0.99)},anchor=north west,draw=none},
    ]
    \addplot[line width=0.3mm, mark=triangle, color=Cora-color]  %
        plot coordinates {
(1,	0.547	)
(2,	0.548	)
(3,	0.549	)
(4,	0.551	)
(5,	0.552	)
(6,	0.553	)
(7,	0.553	)
(8,	0.553	)
(9,	0.554	)
(10,	0.555	)
(11,	0.556	)
    };

    \addplot[line width=0.3mm, mark=diamond, color=Pubmed-color]  %
        plot coordinates {
(1,	0.552	)
(2,	0.552	)
(3,	0.549	)
(4,	0.548	)
(5,	0.547	)
(6,	0.546	)
(7,	0.546	)
(8,	0.546	)
(9,	0.547	)
(10,	0.547	)
(11,	0.547	)

    };

    \addplot[line width=0.3mm, mark=o, color=Blogcatalog-color]  %
        plot coordinates {
(1,	0.480	)
(2,	0.510	)
(3,	0.488	)
(4,	0.483	)
(5,	0.454	)
(6,	0.432	)
(7,	0.399	)
(8,	0.366	)
(9,	0.319	)
(10,	0.309	)
(11,	0.287	)

    };

    \addplot[line width=0.3mm, mark=square, color=Flickr-color]  %
        plot coordinates {
(1,	0.426	)
(2,	0.446	)
(3,	0.447	)
(4,	0.442	)
(5,	0.425	)
(6,	0.392	)
(7,	0.371	)
(8,	0.380	)
(9,	0.380	)
(10,	0.380	)
(11,	0.380	)

    };

    \addplot[line width=0.3mm, mark=pentagon, color=Arxiv-color]  %
        plot coordinates {
(1,	0.076	)
(2,	0.211	)
(3,	0.359	)
(4,	0.377   )
(5,	0.377	)
(6,	0.377	)
(7,	0.377	)
(8,	0.377	)
(9,	0.376	)
(10,	0.377	)
(11,	0.377	)

    };

    \end{axis}
\end{tikzpicture}\hspace{4mm}\label{fig:vary-sigma}%
}
\subfloat[Varying $\sigma$ in \algoe]{
\begin{tikzpicture}[scale=1,every mark/.append style={mark size=2pt}]
    \begin{axis}[
        height=\columnwidth/2.6,
        width=\columnwidth/1.9,
        ylabel={\it Precision},
        xmin=0.5, xmax=11.5,
        ymin=0.05, ymax=0.6,
        xtick={1,3,5,7,9,11},
        ytick={0.1,0.2,0.3,0.4,0.5,0.6},
        xticklabel style = {font=\footnotesize},
        yticklabel style = {font=\footnotesize},
        xticklabels={0,0.2,0.4,0.6,0.8,1.0},
        yticklabels={0.1,0.2,0.3,0.4,0.5,0.6},
        every axis y label/.style={font=\small,at={(current axis.north west)},right=5mm,above=0mm},
        legend style={fill=none,font=\small,at={(0.02,0.99)},anchor=north west,draw=none},
    ]
    \addplot[line width=0.3mm, mark=triangle, color=Cora-color]  %
        plot coordinates {
(1,	0.543	)
(2,	0.544	)
(3,	0.545	)
(4,	0.547	)
(5,	0.548	)
(6,	0.549	)
(7,	0.549	)
(8,	0.550	)
(9,	0.550	)
(10,	0.552	)
(11,	0.552	)
    };

    \addplot[line width=0.3mm, mark=diamond, color=Pubmed-color]  %
        plot coordinates {
(1,	0.555	)
(2,	0.555	)
(3,	0.555	)
(4,	0.555	)
(5,	0.555	)
(6,	0.555	)
(7,	0.555	)
(8,	0.555	)
(9,	0.555	)
(10,	0.555	)
(11,	0.555	)

    };

    \addplot[line width=0.3mm, mark=o, color=Blogcatalog-color]  %
        plot coordinates {
(1,	0.469	)
(2,	0.493	)
(3,	0.481	)
(4,	0.470	)
(5,	0.466	)
(6,	0.441	)
(7,	0.380	)
(8,	0.337	)
(9,	0.268	)
(10,	0.239	)
(11,	0.225	)

    };

    \addplot[line width=0.3mm, mark=square, color=Flickr-color]  %
        plot coordinates {
(1,	0.384	)
(2,	0.390	)
(3,	0.390	)
(4,	0.389	)
(5,	0.376	)
(6,	0.372	)
(7,	0.371	)
(8,	0.371	)
(9,	0.371	)
(10,	0.370	)
(11,	0.182	)

    };

    \addplot[line width=0.3mm, mark=pentagon, color=Arxiv-color]  %
        plot coordinates {
(1,	0.076	)
(2,	0.203	)
(3,	0.360	)
(4,	0.377   )
(5,	0.376	)
(6,	0.376	)
(7,	0.376	)
(8,	0.376	)
(9,	0.376	)
(10,	0.376	)
(11,	0.376	)

    };

    \end{axis}
\end{tikzpicture}\hspace{4mm}\label{fig:vary-sigma-e}%
}
\vspace{-2ex}
\subfloat[Varying $k$ in \algoc]{
\begin{tikzpicture}[scale=1,every mark/.append style={mark size=2pt}]
    \begin{axis}[
        height=\columnwidth/2.6,
        width=\columnwidth/1.9,
        ylabel={\it Precision},
        xmin=0.5, xmax=6.5,
        ymin=0.10, ymax=0.6,
        xtick={1,2,3,4,5,6},
        ytick={0.1,0.2,0.3,0.4,0.5,0.6},
        xticklabel style = {font=\footnotesize},
        yticklabel style = {font=\footnotesize},
        xticklabels={8,16,32,64,128,$d$},
        yticklabels={0.1,0.2,0.3,0.4,0.5,0.6},
        every axis y label/.style={font=\small,at={(current axis.north west)},right=5mm,above=0mm},
        legend style={fill=none,font=\small,at={(0.02,0.99)},anchor=north west,draw=none},
    ]
    \addplot[line width=0.3mm, mark=triangle, color=Cora-color]  %
        plot coordinates {
(1,	0.493	)
(2,	0.547	)
(3,	0.551	)
(4,	0.556	)
(5,	0.554	)
(6,	0.551	)
    };

    \addplot[line width=0.3mm, mark=diamond, color=Pubmed-color]  %
        plot coordinates {
(1,	0.559	)
(2,	0.550	)
(3,	0.551	)
(4,	0.552	)
(5,	0.552	)
(6,	0.552	)
    };

    \addplot[line width=0.3mm, mark=o, color=Blogcatalog-color]  %
        plot coordinates {
(1,	0.434	)
(2,	0.506	)
(3,	0.510	)
(4,	0.495	)
(5,	0.476	)
(6,	0.426	)

    };

    \addplot[line width=0.3mm, mark=square, color=Flickr-color]  %
        plot coordinates {
(1,	0.232	)
(2,	0.371	)
(3,	0.436	)
(4,	0.447	)
(5,	0.403	)
(6,	0.281	)

    };

    \addplot[line width=0.3mm, mark=pentagon, color=Arxiv-color]  %
        plot coordinates {
(1,	0.376	)
(2,	0.377	)
(3,	0.377	)
(4,	0.377   )
(5,	0.377	)
(6,	0.377	)

    };

    \end{axis}
\end{tikzpicture}\hspace{4mm}\label{fig:vary-k}%
}
\subfloat[Varying $k$ in \algoe]{
\begin{tikzpicture}[scale=1,every mark/.append style={mark size=2pt}]
    \begin{axis}[
        height=\columnwidth/2.6,
        width=\columnwidth/1.9,
        ylabel={\it Precision},
        xmin=0.5, xmax=6.5,
        ymin=0.10, ymax=0.6,
        xtick={1,2,3,4,5,6},
        ytick={0.1,0.2,0.3,0.4,0.5,0.6},
        xticklabel style = {font=\footnotesize},
        yticklabel style = {font=\footnotesize},
        xticklabels={8,16,32,64,128,$d$},
        yticklabels={0.1,0.2,0.3,0.4,0.5,0.6},
        every axis y label/.style={font=\small,at={(current axis.north west)},right=5mm,above=0mm},
        legend style={fill=none,font=\small,at={(0.02,0.99)},anchor=north west,draw=none},
    ]
    \addplot[line width=0.3mm, mark=triangle, color=Cora-color]  %
        plot coordinates {
(1,	0.483	)
(2,	0.543	)
(3,	0.552	)
(4,	0.550	)
(5,	0.548	)
(6,	0.546	)
    };

    \addplot[line width=0.3mm, mark=diamond, color=Pubmed-color]  %
        plot coordinates {
(1,	0.545	)
(2,	0.547	)
(3,	0.551	)
(4,	0.552	)
(5,	0.550	)
(6,	0.555	)
    };

    \addplot[line width=0.3mm, mark=o, color=Blogcatalog-color]  %
        plot coordinates {
(1,	0.413	)
(2,	0.484	)
(3,	0.493	)
(4,	0.477	)
(5,	0.451	)
(6,	0.376	)

    };

    \addplot[line width=0.3mm, mark=square, color=Flickr-color]  %
        plot coordinates {
(1,	0.228	)
(2,	0.376	)
(3,	0.390	)
(4,	0.379	)
(5,	0.326	)
(6,	0.251	)

    };

    \addplot[line width=0.3mm, mark=pentagon, color=Arxiv-color]  %
        plot coordinates {
(1,	0.376	)
(2,	0.376	)
(3,	0.377	)
(4,	0.376   )
(5,	0.376	)
(6,	0.376	)

    };

    \end{axis}
\end{tikzpicture}\hspace{4mm}\label{fig:vary-k-e}%
}
\end{small}
\caption{Precision when varying parameters.} \label{fig:parameter}
\vspace{-3ex}
\end{figure}

%% file: tex/figs/ablation.tex
\begin{table}[h]
\centering
\renewcommand{\arraystretch}{1.0}
\caption{Ablation study. Darker shades indicate better results.}
\begin{small}
\addtolength{\tabcolsep}{-0.4em}
\resizebox{\columnwidth}{!}{
\begin{tabular}{c|c | c | c| c| c | c| c| c}
\hline
{\bf Method} & \multicolumn{1}{c|}{{\bf {\em Cora}}} & \multicolumn{1}{c|}{{\bf {\em PubMed}}}  & \multicolumn{1}{c|}{{\bf {\em BlogCL}}}  & \multicolumn{1}{c|}{{\bf {\em Flickr}}} & \multicolumn{1}{c|}{{\bf {\em ArXiv}}} & \multicolumn{1}{c|}{{\bf {\em Yelp}}} & \multicolumn{1}{c|}{{\bf {\em Reddit}}} & \multicolumn{1}{c}{{\bf {\em {Amazon2M}}}}  \\
\hline
\algoc      	&	\cellcolor{blue!30}{0.556}	&	\cellcolor{blue!30}{0.552}	&	\cellcolor{blue!30}{0.51}	&	\cellcolor{blue!30}{0.447}	&	\cellcolor{blue!30}{0.377}	&	\cellcolor{blue!30}{0.754}	&	\cellcolor{blue!30}{0.808}	&	\cellcolor{blue!20}{0.465}	\\
w/o $k$-SVD   &	\cellcolor{blue!20}{0.551}	&	\cellcolor{blue!30}{0.552}	&	\cellcolor{blue!10}{0.426}	&	\cellcolor{blue!10}{0.281}	&	\cellcolor{blue!30}{0.377}	&	\cellcolor{blue!30}{0.754}	&	\cellcolor{blue!30}{0.808}	&	\cellcolor{blue!20}{0.465}	\\
w/o \adiff    &	\cellcolor{blue!10}{0.544}	&	\cellcolor{blue!20}{0.551}	&	\cellcolor{blue!20}{0.48}	&	\cellcolor{blue!20}{0.426}	&	0.329	&	\cellcolor{blue!30}{0.754}	&	\cellcolor{blue!10}{0.213}	&	\cellcolor{blue!10}{0.287} \\ 
w/o SNAS       	&	0.486	&	\cellcolor{blue!10}{0.537}	&	0.302	&	0.2	    &	\cellcolor{blue!20}{0.343}	&	\cellcolor{blue!20}{0.687}	&	\cellcolor{blue!20}{0.779}	&	\cellcolor{blue!30}{0.495}	\\ \hline
\algoe      	&	\cellcolor{blue!30}{0.552}	&	\cellcolor{blue!30}{0.555}	&	\cellcolor{blue!30}{0.493}	&	\cellcolor{blue!30}{0.39}	&	\cellcolor{blue!30}{0.377}	&	\cellcolor{blue!30}{0.739}	&	\cellcolor{blue!30}{0.808}	&	\cellcolor{blue!30}{0.521} \\
w/o $k$-SVD   &	\cellcolor{blue!20}{0.546}	&	\cellcolor{blue!20}{0.554}	&	\cellcolor{blue!10}{0.395}	&	\cellcolor{blue!10}{0.251}	&	\cellcolor{blue!20}{0.376}	&	\cellcolor{blue!20}{0.737}	&	\cellcolor{blue!30}{0.808}	&	\cellcolor{blue!20}{0.514}	\\
w/o \adiff     &	\cellcolor{blue!10}{0.54}	&	\cellcolor{blue!10}{0.553}	&	\cellcolor{blue!20}{0.469}	&	\cellcolor{blue!20}{0.384}	&	0.336	&	\cellcolor{blue!10}{0.735}	&	\cellcolor{blue!10}{0.214}	&	0.365 \\ 
w/o SNAS       	&	0.486	&	0.537	&	0.302	&	0.2	    &	\cellcolor{blue!10}{0.343}	&	0.687	&	\cellcolor{blue!20}{0.779}	&	\cellcolor{blue!10}{0.495}	\\ \hline
\end{tabular}
}
\end{small}
\label{tbl:ablation}
\vspace{0ex}
\end{table}

%% file: tex/figs/Cond-WCSS.tex
\begin{table*}[ht]
\centering
\renewcommand{\arraystretch}{1.0}
\caption{Conductance and WCSS.}
\begin{small}
\addtolength{\tabcolsep}{-0.25em}
\resizebox{\textwidth}{!}{
\begin{tabular}{c|cc | cc | cc| cc| cc | cc| cc| cc}
\hline
\multirow{2}{*}{\bf Method} & \multicolumn{2}{c|}{\bf{ {\em Cora}}} & \multicolumn{2}{c|}{\bf{ {\em PubMed}}}  & \multicolumn{2}{c|}{\bf{ {\em BlogCL}}}  & \multicolumn{2}{c|}{\bf{ {\em Flickr}}} & \multicolumn{2}{c|}{\bf{ {\em ArXiv}}} & \multicolumn{2}{c|}{\bf{ {\em Yelp}}} & \multicolumn{2}{c|}{\bf{ {\em Reddit}}} & \multicolumn{2}{c}{\bf{ {\em {Amazon2M}}}}   \\ \cline{2-17}
& Cond. \textdownarrow & WCSS \textdownarrow  & Cond. \textdownarrow & WCSS \textdownarrow  & Cond. \textdownarrow & WCSS \textdownarrow  &  Cond. \textdownarrow & WCSS \textdownarrow & Cond. \textdownarrow & WCSS \textdownarrow  & Cond. \textdownarrow & WCSS \textdownarrow  & Cond. \textdownarrow & WCSS \textdownarrow  &  Cond. \textdownarrow & WCSS \textdownarrow \\ 
\hline
\texttt{Ground-truth} 	&	0.188	&	0.979	&	0.204	&	0.976	&	0.608	&	0.963	&	0.765	&	0.997	&	0.408	&	0.663	&	0.649	&	{0.55}	&	0.226	&	0.594	&	0.173	&	0.981	\\ \hline
\texttt{PR-Nibble}~\cite{andersen2006local}	&	0.337	&	0.966	&	\cellcolor{blue!30}{0.199}	&	\cellcolor{blue!10}{0.974}	&	0.569	&	0.969	&	0.733	&	\cellcolor{blue!30}{0.998}	&	\cellcolor{blue!10}{0.518}	&	0.667	&	0.237	&	\cellcolor{blue!30}{0.55}	&	0.368	&	\cellcolor{blue!30}{0.595}	&	0.369	&	\cellcolor{blue!20}{0.984}	\\
\texttt{APR-Nibble} 	&	0.323	&	0.964	&	\cellcolor{blue!20}{0.196}	&	\cellcolor{blue!10}{0.974}	&	0.671	&	\cellcolor{blue!30}{0.964}	&	0.799	&	0.99	&	\cellcolor{blue!20}{0.345}	&	0.475	&	0.345	&	0.475	&	0.583	&	0.428	&	0.409	&	0.937	\\
\texttt{HK-Relax}~\cite{kloster2014heat} 	&	0.138	&	0.966	&	0.096	&	0.971	&	0.481	&	\cellcolor{blue!10}{0.967}	&	\cellcolor{blue!20}{0.748}	&	\cellcolor{blue!30}{0.998}	&	0.222	&	\cellcolor{blue!30}{0.663}	&	0.13	&	\cellcolor{blue!20}{0.556}	&	\cellcolor{blue!10}{0.196}	&	0.583	&	0.132	&	0.946	\\
\texttt{CRD}~\cite{wang2017capacity} 	&	\cellcolor{blue!20}{0.156}	&	0.942	&	0.173	&	0.947	&	\cellcolor{blue!20}{0.61}	&	\cellcolor{blue!10}{0.967}	&	\cellcolor{blue!10}{0.787}	&	0.974	&	0.28	&	0.644	&	0.139	&	\cellcolor{blue!10}{0.563}	&	0.275	&	0.588	&	\cellcolor{blue!30}{0.168}	&	0.928	\\
\texttt{$p$-Norm FD}~\cite{fountoulakis2020p} 	&	0.131	&	0.954	&	0.178	&	0.958	&	0.709	&	\cellcolor{blue!10}{0.967}	&	0.845	&	\cellcolor{blue!30}{0.996}	&	0.252	&	0.656	&	0.35	&	0.404	&	\cellcolor{blue!30}{0.235}	&	\cellcolor{blue!30}{0.593}	&	0.118	&	  0.971   	\\
\texttt{WFD}~\cite{pmlr-v202-yang23d}	&	0.127	&	0.956	&	0.17	&	0.96	&	0.713	&	\cellcolor{blue!30}{0.964}	&	0.845	&	\cellcolor{blue!30}{0.996}	&	0.251	&	0.657	&	0.339	&	0.405	&	0.468	&	0.508	&	\cellcolor{blue!20}{0.145}	&	  0.971    	\\ \hline
\texttt{Jaccard}~\cite{liben2003link} 	&	0.617	&	0.984	&	0.637	&	0.981	&	0.655	&	0.969	&	0.854	&	\cellcolor{blue!30}{0.998}	&	0.846	&	0.682	&	0.696	&	0.58	&	0.744	&	0.66	&	0.707	&	0.993	\\
\texttt{Adamic-Adar}~\cite{liben2003link} 	&	0.617	&	0.984	&	0.637	&	0.981	&	\cellcolor{blue!30}{0.609}	&	0.969	&	0.584	&	\cellcolor{blue!30}{0.998}	&	0.834	&	0.682	&	0.696	&	0.58	&	0.771	&	0.654	&	0.707	&	0.993	\\
\texttt{Common-Nbrs}~\cite{liben2003link} 	&	0.617	&	0.984	&	0.637	&	0.981	&	\cellcolor{blue!10}{0.594}	&	0.969	&	0.573	&	\cellcolor{blue!30}{0.998}	&	0.834	&	0.682	&	0.696	&	0.58	&	0.777	&	0.654	&	0.707	&	0.993	\\
\texttt{SimRank}~\cite{jeh2002simrank} 	&	0.265	&	\cellcolor{blue!20}{0.98}	&	0.226	&	\cellcolor{blue!10}{0.978}	&	0.723	&	0.968	&	0.954	&	\cellcolor{blue!30}{0.998}	&	-	&	-	&	-	&	-	&	-	&	-	&	-	&	-	\\ \hline
\texttt{SimAttr (C)}~\cite{yin2010unified} 	&	0.714	&	0.975	&	0.469	&	\cellcolor{blue!10}{0.974}	&	0.816	&	\cellcolor{blue!20}{0.965}	&	0.729	&	\cellcolor{blue!30}{0.998}	&	0.811	&	0.683	&	\cellcolor{blue!20}{0.673}	&	0.522	&	0.704	&	0.573	&	0.702	&	0.986	\\
\texttt{SimAttr (E)}~\cite{rahutomo2012semantic}	&	0.714	&	0.975	&	0.469	&	\cellcolor{blue!10}{0.974}	&	0.816	&	\cellcolor{blue!20}{0.965}	&	0.729	&	\cellcolor{blue!30}{0.998}	&	0.811	&	0.683	&	\cellcolor{blue!20}{0.673}	&	0.522	&	0.704	&	0.573	&	0.702	&	0.986	\\
\texttt{AttriRank}~\cite{hsu2017unsupervised} 	&	0.816	&	0.985	&	0.654	&	0.98	&	0.818	&	0.968	&	0.891	&	\cellcolor{blue!30}{0.998}	&	0.925	&	0.684	&	\cellcolor{blue!10}{0.683}	&	0.576	&	0.951	&	0.669	&	0.86	&	0.995	\\ \hline
\texttt{Node2Vec}~\cite{grover2016node2vec} 	&	\cellcolor{blue!30}{0.194}	&	0.984	&	\cellcolor{blue!10}{0.182}	&	0.979	&	0.589	&	0.968	&	0.739	&	\cellcolor{blue!30}{0.998}	&	\cellcolor{blue!30}{0.36}	&	0.67	&	0.407	&	0.579	&	-	&	-	&	-	&	-	\\
\texttt{GraphSAGE}~\cite{hamilton2017inductive} 	&	0.262	&	0.982	&	0.338	&	0.98	&	0.531	&	0.969	&	\cellcolor{blue!30}{0.757}	&	\cellcolor{blue!20}{0.999}	&	-	&	-	&	-	&	-	&	-	&	-	&	-	&	-	\\
\texttt{PANE}~\cite{yang2023pane}   	&	0.465	&	\cellcolor{blue!20}{0.98}	&	0.376	&	\cellcolor{blue!30}{0.976}	&	0.548	&	\cellcolor{blue!10}{0.967}	&	0.715	&	\cellcolor{blue!30}{0.998}	&	0.841	&	0.673	&	0.592	&	0.562	&	0.736	&	0.573	&	0.748	&	0.986	\\
\texttt{CFANE}~\cite{pan2021unsupervised}     	&	0.369	&	\cellcolor{blue!30}{0.979}	&	0.264	&	\cellcolor{blue!20}{0.975}	&	0.754	&	\cellcolor{blue!30}{0.964}	&	0.876	&	0.998	&	-	&	-	&	-	&	-	&	-	&	-	&	-	&	-	\\ \hline
\algoc       	&	\cellcolor{blue!10}{0.227}	&	\cellcolor{blue!10}{0.977}	&	0.106	&	\cellcolor{blue!20}{0.975}	&	0.669	&	\cellcolor{blue!30}{0.962}	&	0.824	&	\cellcolor{blue!30}{0.998}	&	0.248	&	\cellcolor{blue!20}{0.664}	&	\cellcolor{blue!30}{0.642}	&	0.518	&	\cellcolor{blue!20}{0.254}	&	\cellcolor{blue!20}{0.597}	&	0.244	&	\cellcolor{blue!10}{0.985}	\\
\algoe        	&	0.228	&	\cellcolor{blue!10}{0.977}	&	0.108	&	\cellcolor{blue!20}{0.975}	&	0.661	&	\cellcolor{blue!30}{0.962}	&	0.849	&	\cellcolor{blue!10}{0.992}	&	0.247	&	\cellcolor{blue!10}{0.665}	&	0.494	&	0.534	&	\cellcolor{blue!20}{0.254}	&	\cellcolor{blue!10}{0.598}	&	\cellcolor{blue!20}{0.155}	&	\cellcolor{blue!30}{0.981}	\\ \hline
\end{tabular}
}\
\end{small}
\label{tbl:node-clustering-cond-WCSS}
\vspace{0ex}
\end{table*}

%% file: tex/figs/scalability.tex
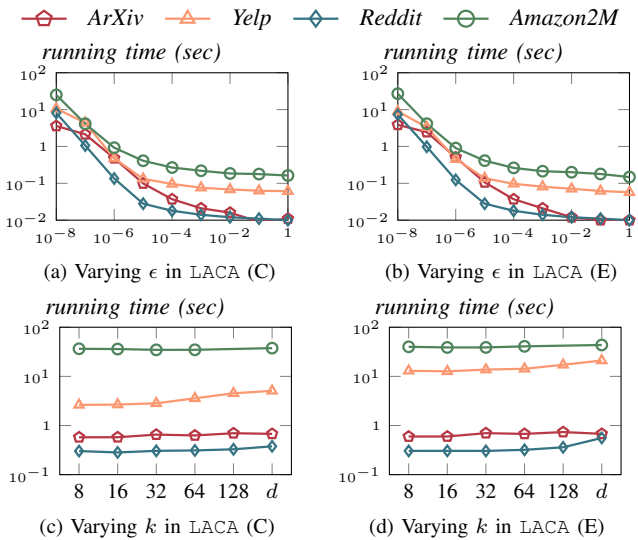
\begin{figure}[ht]
\centering
\begin{small}
\begin{tikzpicture}
    \begin{customlegend}[legend columns=4,
        legend entries={{\em ArXiv}, {\em Yelp}, {\em Reddit}, {\em Amazon2M}},
        legend style={at={(0.45,1.35)},anchor=north,draw=none,font=\small,column sep=0.2cm}]
    \addlegendimage{line width=0.3mm,mark size=3pt,mark=pentagon, color=Arxiv-color}
    \addlegendimage{line width=0.3mm,mark size=3pt,mark=triangle, color=Yelp-color}
    \addlegendimage{line width=0.3mm,mark size=3pt,mark=diamond, color=Reddit-color}
    \addlegendimage{line width=0.3mm,mark size=3pt,mark=o, color=Amazon-color}
    \end{customlegend}
\end{tikzpicture}
\\[-\lineskip]
\vspace{-4mm}
\subfloat[Varying $\epsilon$ in \algoc]{
\begin{tikzpicture}[scale=1,every mark/.append style={mark size=2pt}]
    \begin{axis}[
        height=\columnwidth/2.5,
        width=\columnwidth/1.9,
        ylabel={\it running time (sec)},
        xmin=1e-8, xmax=1,
        ymin=0.01, ymax=100,
        xtick={1e-10, 1e-8, 1e-6, 1e-4, 1e-2, 1},
        ytick={1e-2,1e-1,1,1e1,1e2},
        xticklabel style = {font=\tiny},
        yticklabel style = {font=\tiny},
        xticklabels={$10^{-10}$, $10^{-8}$, $10^{-6}$, $10^{-4}$, $10^{-2}$, $1$},
        yticklabels={$10^{-2}$,$10^{-1}$,$1$,$10^{1}$,$10^{2}$},
        xmode=log,
        ymode=log,
        log basis y={10},
        every axis y label/.style={font=\small,at={(current axis.north west)},right=10mm,above=0mm},
        legend style={fill=none,font=\small,at={(0.02,0.99)},anchor=north west,draw=none},
    ]
    \addplot[line width=0.3mm, mark=pentagon, color=Arxiv-color]  %
        plot coordinates {
(1e-8,	3.601	)
(1e-7,	2.137	)
(1e-6,	0.481	)
(1e-5,	0.099	)
(1e-4,	0.037	)
(1e-3,	0.021	)
(1e-2,	0.016	)
(1e-1,	0.009	)
(1,	0.011	)
    };

    \addplot[line width=0.3mm, mark=triangle, color=Yelp-color]  %
        plot coordinates {
(1e-8,	10.620	)
(1e-7,	4.400	)
(1e-6,	0.475	)
(1e-5,	0.132	)
(1e-4,	0.096	)
(1e-3,	0.076	)
(1e-2,	0.068	)
(1e-1,	0.063	)
(1,	0.061	)

    };

    \addplot[line width=0.3mm, mark=diamond, color=Reddit-color]  %
        plot coordinates {
(1e-8,	8.082	)
(1e-7,	1.061	)
(1e-6,	0.135	)
(1e-5,	0.028	)
(1e-4,	0.018	)
(1e-3,	0.014	)
(1e-2,	0.012	)
(1e-1,	0.011	)
(1,	0.010	)

    };

    \addplot[line width=0.3mm, mark=o, color=Amazon-color]  %
        plot coordinates {
(1e-8,	25.024	)
(1e-7,	4.127	)
(1e-6,	0.925	)
(1e-5,	0.409	)
(1e-4,	0.268	)
(1e-3,	0.219	)
(1e-2,	0.185	)
(1e-1,	0.178	)
(1,	0.163	)

    };

    \end{axis}
\end{tikzpicture}\hspace{4mm}\label{fig:scale-c-eps-time}%
}
\subfloat[Varying $\epsilon$ in \algoe]{
\begin{tikzpicture}[scale=1,every mark/.append style={mark size=2pt}]
    \begin{axis}[
        height=\columnwidth/2.5,
        width=\columnwidth/1.9,
        ylabel={\it running time (sec)},
        xmin=1e-8, xmax=1,
        ymin=0.01, ymax=100,
        xtick={1e-10, 1e-8, 1e-6, 1e-4, 1e-2, 1},
        ytick={1e-2,1e-1,1,1e1,1e2},
        xticklabel style = {font=\tiny},
        yticklabel style = {font=\tiny},
        xticklabels={$10^{-10}$, $10^{-8}$, $10^{-6}$, $10^{-4}$, $10^{-2}$, $1$},
        yticklabels={$10^{-2}$,$10^{-1}$,$1$,$10^{1}$,$10^{2}$},
        xmode=log,
        ymode=log,
        log basis y={10},
        every axis y label/.style={font=\small,at={(current axis.north west)},right=10mm,above=0mm},
        legend style={fill=none,font=\small,at={(0.02,0.99)},anchor=north west,draw=none},
    ]
    \addplot[line width=0.3mm, mark=pentagon, color=Arxiv-color]  %
        plot coordinates {
(1e-8,	3.868	)
(1e-7,	2.451	)
(1e-6,	0.506	)
(1e-5,	0.106	)
(1e-4,	0.037	)
(1e-3,	0.021	)
(1e-2,	0.012	)
(1e-1,	0.010	)
(1,	0.010	)
    };

    \addplot[line width=0.3mm, mark=triangle, color=Yelp-color]  %
        plot coordinates {
(1e-8,	8.902	)
(1e-7,	3.343	)
(1e-6,	0.455	)
(1e-5,	0.137	)
(1e-4,	0.097	)
(1e-3,	0.081	)
(1e-2,	0.071	)
(1e-1,	0.062	)
(1,	0.057	)

    };

    \addplot[line width=0.3mm, mark=diamond, color=Reddit-color]  %
        plot coordinates {
(1e-8,	7.322	)
(1e-7,	0.981	)
(1e-6,	0.124	)
(1e-5,	0.028	)
(1e-4,	0.018	)
(1e-3,	0.014	)
(1e-2,	0.012	)
(1e-1,	0.011	)
(1,	0.010	)

    };

    \addplot[line width=0.3mm, mark=o, color=Amazon-color]  %
        plot coordinates {
(1e-8,	27.052	)
(1e-7,	4.138	)
(1e-6,	0.904	)
(1e-5,	0.412	)
(1e-4,	0.262	)
(1e-3,	0.213	)
(1e-2,	0.200	)
(1e-1,	0.179	)
(1,	0.148	)

    };

    \end{axis}
\end{tikzpicture}\hspace{4mm}\label{fig:scale-e-eps-time}%
}
\vspace{-2ex}
\subfloat[Varying $k$ in \algoc]{
\begin{tikzpicture}[scale=1,every mark/.append style={mark size=2pt}]
    \begin{axis}[
        height=\columnwidth/2.5,
        width=\columnwidth/1.9,
        ylabel={\it running time (sec)},
        xmin=0.5, xmax=6.5,
        ymin=0.1, ymax=100,
        xtick={1,2,3,4,5,6},
        ytick={1e-1,1,1e1,1e2},
        xticklabel style = {font=\footnotesize},
        yticklabel style = {font=\tiny},
        xticklabels={8,16,32,64,128,$d$},
        yticklabels={$10^{-1}$,$1$,$10^{1}$,$10^{2}$},
        ymode=log,
        log basis y={10},
        every axis y label/.style={font=\small,at={(current axis.north west)},right=10mm,above=0mm},
        legend style={fill=none,font=\small,at={(0.02,0.99)},anchor=north west,draw=none},
    ]
    \addplot[line width=0.3mm, mark=pentagon, color=Arxiv-color]  %
        plot coordinates {
(1,	0.576	)
(2,	0.577	)
(3,	0.653	)
(4,	0.628	)
(5,	0.696	)
(6,	0.677	)
    };

    \addplot[line width=0.3mm, mark=triangle, color=Yelp-color]  %
        plot coordinates {
(1,	2.627	)
(2,	2.669	)
(3,	2.832	)
(4,	3.572	)
(5,	4.530	)
(6,	5.075	)
    };

    \addplot[line width=0.3mm, mark=diamond, color=Reddit-color]  %
        plot coordinates {
(1,	0.302	)
(2,	0.283	)
(3,	0.305	)
(4,	0.311	)
(5,	0.328	)
(6,	0.377	)

    };

    \addplot[line width=0.3mm, mark=o, color=Amazon-color]  %
        plot coordinates {
(1,	36.449	)
(2,	36.011	)
(3,	34.653	)
(4,	34.731	)
(6,	37.565	)

    };

    \end{axis}
\end{tikzpicture}\hspace{4mm}\label{fig:scale-c-k-time}%
}
\subfloat[Varying $k$ in \algoe]{
\begin{tikzpicture}[scale=1,every mark/.append style={mark size=2pt}]
    \begin{axis}[
        height=\columnwidth/2.5,
        width=\columnwidth/1.9,
        ylabel={\it running time (sec)},
        xmin=0.5, xmax=6.5,
        ymin=0.1, ymax=100,
        xtick={1,2,3,4,5,6},
        ytick={1e-1,1,1e1,1e2},
        xticklabel style = {font=\footnotesize},
        yticklabel style = {font=\tiny},
        xticklabels={8,16,32,64,128,$d$},
        yticklabels={$10^{-1}$,$1$,$10^{1}$,$10^{2}$},
        ymode=log,
        log basis y={10},
        every axis y label/.style={font=\small,at={(current axis.north west)},right=10mm,above=0mm},
        legend style={fill=none,font=\small,at={(0.02,0.99)},anchor=north west,draw=none},
    ]
    \addplot[line width=0.3mm, mark=pentagon, color=Arxiv-color]  %
        plot coordinates {
(1,	0.597	)
(2,	0.598	)
(3,	0.699	)
(4,	0.675	)
(5,	0.736	)
(6,	0.677	)
    };

    \addplot[line width=0.3mm, mark=triangle, color=Yelp-color]  %
        plot coordinates {
(1,	12.992	)
(2,	12.676	)
(3,	13.732	)
(4,	14.280	)
(5,	17.216	)
(6,	21.041	)
    };

    \addplot[line width=0.3mm, mark=diamond, color=Reddit-color]  %
        plot coordinates {
(1,	0.304	)
(2,	0.305	)
(3,	0.304	)
(4,	0.320	)
(5,	0.360	)
(6,	0.558	)

    };

    \addplot[line width=0.3mm, mark=o, color=Amazon-color]  %
        plot coordinates {
(1,	39.999	)
(2,	38.812	)
(3,	38.839	)
(4,	40.866	)
(6,	43.550	)

    };

    \end{axis}
\end{tikzpicture}\hspace{4mm}\label{fig:scale-e-k-time}%
}
\end{small}
\caption{Efficiency when varying $\epsilon$ and $k$.} \label{fig:scale-time}
\vspace{-1ex}
\end{figure}

%% file: tex/figs/without_attributes.tex
\begin{table}[ht]
\centering
\renewcommand{\arraystretch}{1.0}
\begin{footnotesize}
\caption{Statistics of Datasets without Attributes.}\label{tbl:without-attrs-data}
\begin{tabular}{l|r|r|c}
	\hline
	{\bf Dataset} & \multicolumn{1}{c|}{$\boldsymbol n$ } & \multicolumn{1}{c|}{$\boldsymbol m$ } & {\bf $\overline{|\Y_s|}$} \\
	\hline
    {\em com-DBLP}~\cite{yang2012defining} & 317,080 & 1,049,866 & 1,862 \\
    {\em com-Amazon}~\cite{yang2012defining} & 334,863 & 925,872 &  47 \\
    {\em com-Orkut}~\cite{yang2012defining} & 3,072,441 & 117,185,083 & 621 \\
    \hline
\end{tabular}
\end{footnotesize}
\end{table}

\begin{table}[ht]
\centering
\renewcommand{\arraystretch}{1.0}
\caption{The average precision evaluated with ground-truth. Best is \textbf{bolded} and best baseline \underline{underlined}.}
\begin{small}
\addtolength{\tabcolsep}{-0.4em}
\resizebox{\columnwidth}{!}{
\begin{tabular}{c|c | c | c}
\hline
{\bf Method} & \multicolumn{1}{c|}{{\bf {\em com-DBLP}~\cite{yang2012defining}}} & \multicolumn{1}{c|}{{\bf {\em com-Amazon}~\cite{yang2012defining}}}  & \multicolumn{1}{c}{{\bf {\em com-Orkut}~\cite{yang2012defining}}}  \\
\hline
\texttt{PR-Nibble}~\cite{andersen2006local}      	&	\underline{0.374}	&	0.835	&	\underline{0.251}	\\
\texttt{HK-Relax}~\cite{kloster2014heat}   &	0.305	&	0.865	&	0.233	\\
\texttt{CRD}~\cite{wang2017capacity}    &	0.247	&	0.696	&	0.021	\\ 
\texttt{$p$-Norm FD}~\cite{fountoulakis2020p}       	&	0.331	&	\underline{0.871}	&	0.199	\\
\algo (w/o SNAS)       	&	\textbf{0.399}	&	\textbf{0.919}	&	\textbf{0.253}	\\ \hline
\end{tabular}
}
\end{small}
\label{tbl:without-attrs}
\vspace{0ex}
\end{table}

%% file: tex/alternative-implementation.tex
\subsection{Alternative Implementation of \algo}

\begin{table}[ht]
\centering
\renewcommand{\arraystretch}{1.0}
\caption{The average precision comparing with alternative BDD implementations.}
\begin{small}
\addtolength{\tabcolsep}{-0.25em}
\resizebox{\columnwidth}{!}{
\begin{tabular}{l|c | c | c| c| c | c| c| c}
\hline
\multirow{1}{*}{\bf Method} & \multicolumn{1}{c|}{{\bf {\em Cora}}} & \multicolumn{1}{c|}{{\bf {\em PubMed}}} & \multicolumn{1}{c|}{{\bf {\em BlogCL}}} & \multicolumn{1}{c|}{{\bf {\em Flickr}}} & \multicolumn{1}{c|}{{\bf {\em ArXiv}}} & \multicolumn{1}{c|}{{\bf {\em Yelp}}} & \multicolumn{1}{c|}{{\bf {\em Reddit}}} & \multicolumn{1}{c}{{\bf \em {Amazon2m}}} \\ \cline{2-9}
\hline
\algoc       	&	{0.556}	&	{0.552}	&	{0.51}	&	{0.447}	&	{0.377}	&	{0.754}	&	{0.808}	&	{0.465} \\
\algoc-RS-RS-RS       	& 0.181 & 0.383 & 0.167 & 0.121 & 0.091 & 0.737 & 0.054 & 0.227 \\
\algoc-R-RS-RS       	& 0.224 & 0.446 & 0.184 & 0.152 & 0.209 & 0.737 & 0.194 & 0.223 \\
\algoc-RS-R-RS       	& 0.222 & 0.441 & 0.18 & 0.142 & 0.169 & 0.737 & 0.114 & 0.225 \\
\algoc-RS-RS-R       	& 0.194 & 0.360 & 0.174 & 0.145 & 0.082 & 0.720 & 0.065 & 0.237 \\
\hline
\algoe        	&	{0.552}	&	{0.555}	&	0.493	&	{0.39}	&	{0.377}	&	0.739	&	{0.808}	&	{0.521} \\
\algoe-RS-RS-RS        	& 0.17 & 0.358 & 0.167 & 0.11 & 0.091 & 0.737 & 0.058 & 0.133 \\
\algoe-R-RS-RS        	& 0.179 & 0.364 & 0.178 & 0.113 & 0.208 & 0.737 & 0.191 & 0.396 \\
\algoe-RS-R-RS        	& 0.181 & 0.365 & 0.177 & 0.11 & 0.167 & 0.719 & 0.11 & 0.243  \\
\algoe-RS-RS-R       	& 0.183 & 0.352 & 0.172 & 0.113 & 0.082 & 0.719 & 0.064 & 0.133 \\
\hline
\end{tabular}
}
\end{small}
\label{tbl:alternative-1}
\end{table}

\begin{table}[ht]
\centering
\renewcommand{\arraystretch}{1.0}
\caption{Ablation study on various similarity measures.}
\begin{small}
\addtolength{\tabcolsep}{-0.25em}
\resizebox{\columnwidth}{!}{
\begin{tabular}{l|c | c | c| c| c | c| c| c}
\hline
\multirow{1}{*}{\bf Method} & \multicolumn{1}{c|}{{\bf {\em Cora}}} & \multicolumn{1}{c|}{{\bf {\em PubMed}}} & \multicolumn{1}{c|}{{\bf {\em BlogCL}}} & \multicolumn{1}{c|}{{\bf {\em Flickr}}} & \multicolumn{1}{c|}{{\bf {\em ArXiv}}} & \multicolumn{1}{c|}{{\bf {\em Yelp}}} & \multicolumn{1}{c|}{{\bf {\em Reddit}}} & \multicolumn{1}{c}{{\bf \em {Amazon2m}}} \\ \cline{2-9}
\hline
\algoc       	&	{0.556}	&	{0.552}	&	{0.51}	&	{0.447}	&	{0.377}	&	{0.754}	&	{0.808}	&	{0.465} \\
\algoe        	&	{0.552}	&	{0.555}	&	0.493	&	{0.39}	&	{0.377}	&	0.739	&	{0.808}	&	{0.521} \\
\algo (Jaccard) & {0.524} & {\xmark} & {0.304} & {0.28} & {\xmark} & {\xmark} & {\xmark} & {\xmark} \\
\algo (Pearson) & {0.518} & {0.551} & {0.289} & {0.115} & {-} & {-} & {-} & {-} \\
\hline
\end{tabular}
}
\end{small}
\label{tbl:similarity-metric-1}
\end{table}

\subsubsection{Alternative Implementation of \ppr} \label{sec:bdd-alternative}
To rigorously validate the effectiveness of our BDD, we have implemented the new graph diffusion algorithms for local clustering based on the suggested formulation in the comment and the other three alternative formulations as follows:
\begin{enumerate}[leftmargin=*]
\item RS-RS-RS: Integrating attribute similarity (i.e., SNAS) into all three random walk steps as suggested. Specifically, for each node pair $(v_s,v_t)$, the affinity is defined by $\sum_{v_i,v_j\in V} \rho(v_s,v_i) \cdot \rho(v_i,v_j) \cdot \rho(v_t,v_j)$, where $\rho(v_i,v_j) =
\begin{cases}
\pi(v_i,v_j) \cdot s(v_i,v_j) & \text{if } v_i \text{ is connected to } v_j \text{ via an edge} \\
1 & v_i = v_j
\end{cases}$..
\item R-RS-RS: Integrating attribute similarity (i.e., SNAS) into the second and third random walk steps. For each node pair $(v_s,v_t)$, the affinity is defined by $\sum_{v_i,v_j\in V} \pi(v_s,v_i) \cdot \rho(v_i,v_j) \cdot \rho(v_t,v_j)$.
\item RS-R-RS: Integrating attribute similarity (i.e., SNAS) into the first and third random walk steps. For each node pair $(v_s,v_t)$, the affinity is defined by $\sum_{v_i,v_j\in V} \rho(v_s,v_i) \cdot \pi(v_i,v_j) \cdot \rho(v_t,v_j)$.
\item RS-RS-R: Integrating attribute similarity (i.e., SNAS) into the first and second random walk steps. For each node pair $(v_s,v_t)$, the affinity is defined by $\sum_{v_i,v_j\in V} \rho(v_s,v_i) \cdot \rho(v_i,v_j) \cdot \pi(v_t,v_j)$.
\end{enumerate}
and have compared them against our original BDD.

Table~\ref{tbl:alternative-1} reports the local clustering performance of \algoc and \algoe based on our BDD and the above four alternative definitions. It can be observed that all these four variants undergo severe performance degradation compared to the BDD on most datasets. For instance, on {\em Cora} and {\em Amazon}, \algoc-BDD is able to yield $55.6\%$ and $46.5\%$ in precision, whereas the four alternatives achieve at most $22.4\%$ and $23.7\%$, respectively.
The remarkable superiority of the BDD over the alternatives is due to the fact that these alternatives overly incorporate the attribute similarity (at least two attribute-only transitions) and topological connectivity (three random walk steps) into the random walk diffusion process, rendering the graph traversal rather biased and easier to jump to the nodes that are distant or even disconnected from the seed node $v_s$ via the intermediate nodes with high attribute similarities and long random walks. In turn, it is more likely to produce nodes that are far-reaching from the local cluster around $v_s$.

\subsubsection{Alternative Choices on Similarity Measurements} \label{sec:similarity}
To further demonstrate the superiority of \algoc and \algoe, we have conducted an ablation study that employs the Jaccard and Pearson correlation coefficients as the SNAS in \algoc and \algoe on all datasets. 
Note that the Jaccard coefficient requires the attribute values to be binary and thus are not applicable to datasets with continuous attributes, i.e., {\em PubMed}, {\em ArXiv}, {\em Yelp}, {Reddit}, and {\em Amazon2M} and the Pearson correlation coefficient is unable to report the results on large graphs within three days due to the high complexity ($O(n^2d)$) needed for calculating the similarities of the intermediate node pairs.
As presented in Table~\ref{tbl:similarity-metric-1}, \algoc and \algoe consistently outperform these two variants with considerable gains. For example, on \textit{Flickr}, \algoc can obtain a precision of $44.7\%$, while the precision scores attained by Jaccard and Pearson correlation coefficients are merely $28\%$ and $11.5\%$.

\subsection{Other Related Work}
\subsubsection{Attributed Network Embedding}
Attributed network embedding (ANE) is to embed each node graph $\G$ into a low-dimensional feature vector, preserving both topology and attribute information. The obtained embeddings can be used in many downstream tasks, including graph clustering.
ANE methods can be categorized into two types: factorization-based and learning-based.
Factorization-based methods \cite{yang2015network, huang2017accelerated, yang2019low, yang2023pane} construct and factorize node proximity matrices that integrate graph topology and node attributes to derive low-dimensional vectors. \texttt{TADW} \cite{yang2015network} leverages a second-order adjacency matrix, \texttt{AANE} \cite{huang2017accelerated} matches node representations with attribute proximities, \texttt{PANE} \cite{yang2020scaling,yang2023pane} optimizes forward and backward affinity matrices via random walks and optimizes convergence with greedy initial technique. However, these methods suffer from scalability issues due to the necessity of factorization of an $n \times n$ proximity matrix.
Learning-based methods \cite{zhang2018anrl, gao2018deep, hamilton2017inductive, pan2021unsupervised} are further classified into encoder-decoder and propagation categories. Encoder-decoder methods, including \texttt{DANE} \cite{gao2018deep} and \texttt{ANRL} \cite{zhang2018anrl}, integrate attribute features and graph topology by utilizing multiple autoencoders to minimize input reconstruction loss. Additionally, \texttt{ANRL} develops an attribute-aware model based on the Skip-gram model \cite{mikolov2013distributed}.
\texttt{CFANE} \cite{pan2021unsupervised} integrates propagation-based and encoder-decoder methods, utilizing self-attention to refine its model.
In contrast to our local-based approach, these ANE methods typically require processing nodes across the entire graph.